\documentclass[aps,prb] {revtex4}
\usepackage{epsfig}
\usepackage{subfigure}
\usepackage{booktabs}
\usepackage{comment}
\usepackage{color}

\usepackage{graphicx}
\usepackage{dcolumn}
\usepackage{bm}
\usepackage{mathrsfs}

\begin{document}


\title{ Superconducting Symmetry Phases and Dominant bands in  (Ca-) Intercalated AA- Bilayer Graphene  }
\author{ Rouhollah Gholami}
\affiliation{Physics Department, Faculty of Science Razi University, Kermanshah, Iran\\ Physics Department, Faculty of Science Ilam University, Ilam, Iran}

\author{Rostam Moradian}
 \email{rmoradian@razi.ac.ir}
\affiliation{ Physics Department, Faculty of Science Razi University, Kermanshah, Iran\\
Nanoscience and Nanotechnology Research Center,  Razi University, Kermanshah, Iran}

\author{Sina Moradian}
\affiliation{Department of Electrical and Computer Engineering, University of Central Florida, Orlando FL, USA}
\author{Warren E. Pickett}
\affiliation{Department of Physics UC Davis, One Shields Avenue, Davis CA 95616, USA}
\date{\today}
\begin{abstract}
Built on a realistic multiband tight binding model, mirror symmetry is used to
map a calcium-intercalated 
bilayer graphene Hamiltonian into two independent single layer graphene-like
 Hamiltonians
with renormalized hopping integrals ($\pm 1 $ cone index). The quasiparticles exhibit two types of chirality. Here a quasi-particle consist of two electrons  from opposing layers  where  possess an additional quantum number called ``cone index'' which it can be regarded as  the eigenvalue of mirror symmetry operations 
and  it play  a major  role in describing  the physical behavior of
 of AA-stacked bilayer graphene. To obtain tight binding parameters, both effective 
monolayer graphene Schr\"odinger equations are solved analytically and fitted to 
first principles band structure results. Two  quasi particles (four electrons) can team up to build a Cooper pair with even or odd chirality. 
Treatment of the pairing Hamiltonian 
leads to two decoupled gap equations for each of these effective graphene monolayer
sectors that means pairing of quasi-particles with different cone index is forbidden. The decoupled gap equations are solved analytically to obtain all the possible order 
parameters for superconducting phases. Two nearly ``flat bands'' crossing the Fermi energy, 
each related to the graphene-like structures, are responsible for
two distinct superconductivity gaps that emerge. 
 Depending on how much these  bands are affected by the intercalant and which is 
closer to the Fermi energy, distorted $s$-wave or $d$-wave superconductivity may 
become dominant. Numerical calculations reveal that $d$-wave superconductivity is 
dominant in both sectors.
 For these two dominant 
phases we present the relation of the pairing potential 
to the superconducting critical temperature $T_c$.
Within the range of 0-6 K  which superconductivity
has been observed experimentally, transition from single-gap to dual-gap superconductivity is possible. 
Adopting the two gap viewpoint of superconductivity in C$_6$CaC$_6$, the dominant 
$d$-wave states should  have the same critical temperature.
Around $T_c=2K$ these two relations intersect, otherwise superconductivity has been realized just in the one of these two sectors and disappear in the other one. In this case, the superconducting coexistence with the Dirac electrons is possible.

\end{abstract}


\maketitle

\section{Introduction}
Discovery of new superconducting phases, often at low temperature, has been one of the active achievements in recent decades, sometimes overshadowed 
by the high profile effort to push toward higher temperature  superconductors. 
Remarkable progress in the synthesis or fabrication of nanostructures in the recent years is opening new horizons in engineering the physical properties  of new classes of materials.
 Two dimensional (2D) phases, including superconducting ones,\cite{brun2017,Saito2016} received renewed attention from the 
discovery that monolayer graphene can be peeled off, and subsequently found that it can be synthesized from the gas phase. 
Dirac (charge neutrality)  points, has become of rising interest. 
While graphene, with its massless Dirac points, hosts many unusual quantum properties 
with potential applications, pristine graphene does not support superconductivity 
due to its vanishing density of states (DOS) at the Dirac points. 

Despite that, 
discussion of superconductivity in doped graphene has been profuse with theoretical 
models\cite{Uchoa2007,Nandkishore2012,Nandkishore2014,Black-Schaffer2007,Kiesel2012,Ma2014}.
 With the many pictures  raising various possibilities (such as  chiral d+id pairing\cite{Nandkishore2012,Nandkishore2014,Black-Schaffer2007},  coexistence of both chiral d+id and f-wave\cite{Kiesel2012} and p+ip\cite{Ma2014},) but little of a certain nature. However,  superconductivity in lithium decorated graphene
around 6K was predicted theoretically\cite{Profeta2012} and subsequently reported experimentally.\cite{Ludbrook2015} 
Graphite intercalation compound (GIC) superconductors have 
been known and studied for decades.\cite{Weller-naturephysics2005}
Recent advances in few layer graphene  (FLG)  fabrication methods  have caused  
theoretical attention to their  electronic physics.
   Due to its strong two dimensionality, bilayer graphene (BlG)
has provided an attractive platform for studying 2D electron correlation 
effects.\cite{Nori2016}  
Because of the weak interlayer van der Waals forces, the layers of the graphene 
bilayer can rotate relative to each other and form   
different ordering stakes {\it i.e.} AA, AB (Bernal phase), or even ``twisted bilayer'' 
form. 

Following the indication of a Mott insulating phase in 
 twisted bi-layered graphene\cite{Cao-Mott2018}  and the observation of superconductivity 
 upon doping,\cite{Cao2018} these types of systems have revived interest and may 
realize a new class of superconductor. 
It seems that bilayer graphene can host superconductivity, magnetism,  and other
unusual phases. In addition to  relativistic character of quasi particles in single layer graphene,  interlayer coupling causes the bilayer graphene to exhibit behaviors that are not observed in the  single layer graphene. Most fascinating behaviors of  TBlG are played  
by interlayer hybridization  of  nearby Dirac K points of opposite layers via  
interlayer tunneling in a spatially periodic way\cite{MacDonald2019}. Interlayer coupling causes  quasi particles in the AB-stacked bilayer graphene behave as  massive  
chiral quasi-particle with parabolic dispersion near the Dirac Fermi points.   In this manuscript we will see that quasi-particles in AA-stacked BlG   can exhibit extra  aspects  of  such behaviors.

Beside TBlG, when the electron-electron correlations  effects are taken into consideration 
in  AA and AB-stacking structures, theory suggests a variety of 
instabilities  with potential technological applications, including unusual 
quantum Hall effects, antiferromagnetic phases, and tunable band gap opening   
at the charge neutral points (for a review see [\onlinecite{Nori2016}]), some of 
which have been confirmed experimentally.\cite{Velasco2012}.
Although it is difficult to distinguish AA-stacked BlG  from single layer graphene but Some authors claimed to fabricate AA-BlG experimentally.  
However, it has been received much less attention than more stable  AB-stacked phase.
 Unlike Fermi points in monolayer graphene and AB-stacked, the well nested 
Fermi surface  of pristine AA-BlG 
consists of  small electron and hole pockets of equal area. This feature  
has drastic consequences, tending the system toward electron instabilities 
such as antiferromagnetism at zero doping and bilayer exciton condensation 
when doped.\cite{Nori2016,Rakhmanov2012,Akzyanov2013,Sboychakov2013,SboychakovMetal2013}

 Superconducting instabilities in doped or gated AB-stacking phases  
have been predicted.  An effective two-band Hamiltonian with attractive interactions 
was used in Ref{\cite{ Vucicevic2012}} to investigate the possibility of a time reversal breaking 
$d+id$ phase in moderately doped AB-stacked BlG.  Using a 
weak-coupling renormalization group formalism, the possibility of unconventional 
superconducting orders from repulsive interaction on doped AB-stacked honeycomb 
bilayer in $d$-wave, $f$-wave, and pair density wave channels were discussed in 
by James {\it et al.}\cite{ James2014} Spin triplet $s$-wave pairing could also 
arise.\cite{Hosseini2012}  
  Superconductivity reported in Ca-intercalated bilayer 
graphene represents the thinest limit of graphite intercalation compounds (GICs) \cite{Kanetani2012}, at 
4K\cite{Satoru2016} and around 6.4K in Ca-doped graphene 
laminates\cite{Chapman2016}.
Even so, the superconducting phase of AA-graphene  has been addressed very little
in the literatures very rarely.


 Based on an effective Hamiltonian with an attractive potential between inter- and 
intra-layer near neighbor sublattices 
Alidoust {\it et al. }\cite{Alidoust2019} to study phonon-mediated superconducting 
pairing symmetries that may arise in AA, AB (and AC)-stacking bilayer graphene 
at the charge neutral point and beyond (by varying chemical potential).  They 
claimed that at a finite doping, AB stacking can develop 
singlet and triplet $d$-wave symmetry beside $s$-wave, $p$-wave and $f$-wave that 
can be achieved at the charge neutral point, while the AA-stacked phase, similar 
to the undoped case, is unable to accept $d$-wave pairing.

\begin{figure}
\centerline{\epsfig{file=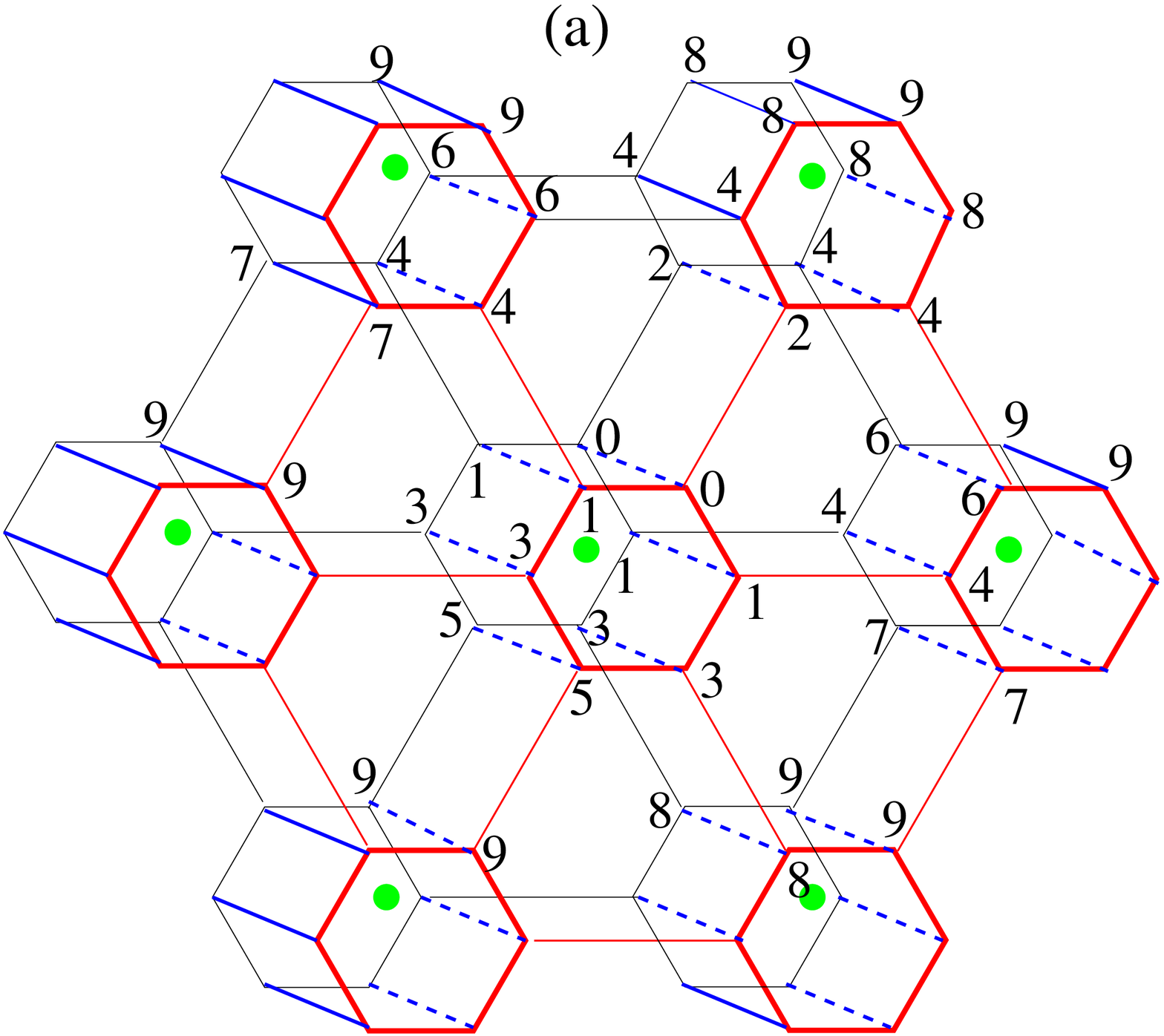,width=6.0cm,angle=0}
\epsfig{file=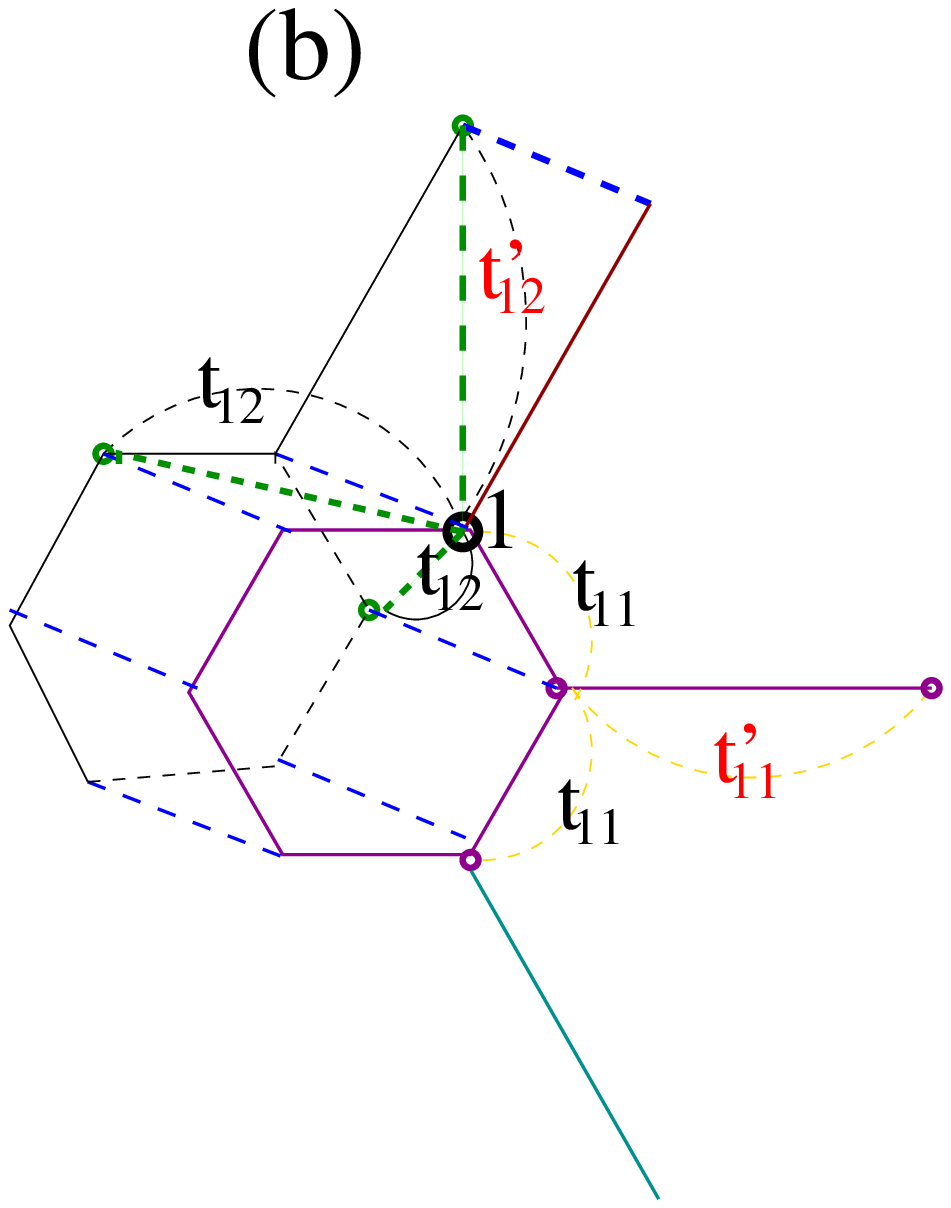,width=4.2cm,angle=0} 
\epsfig{file=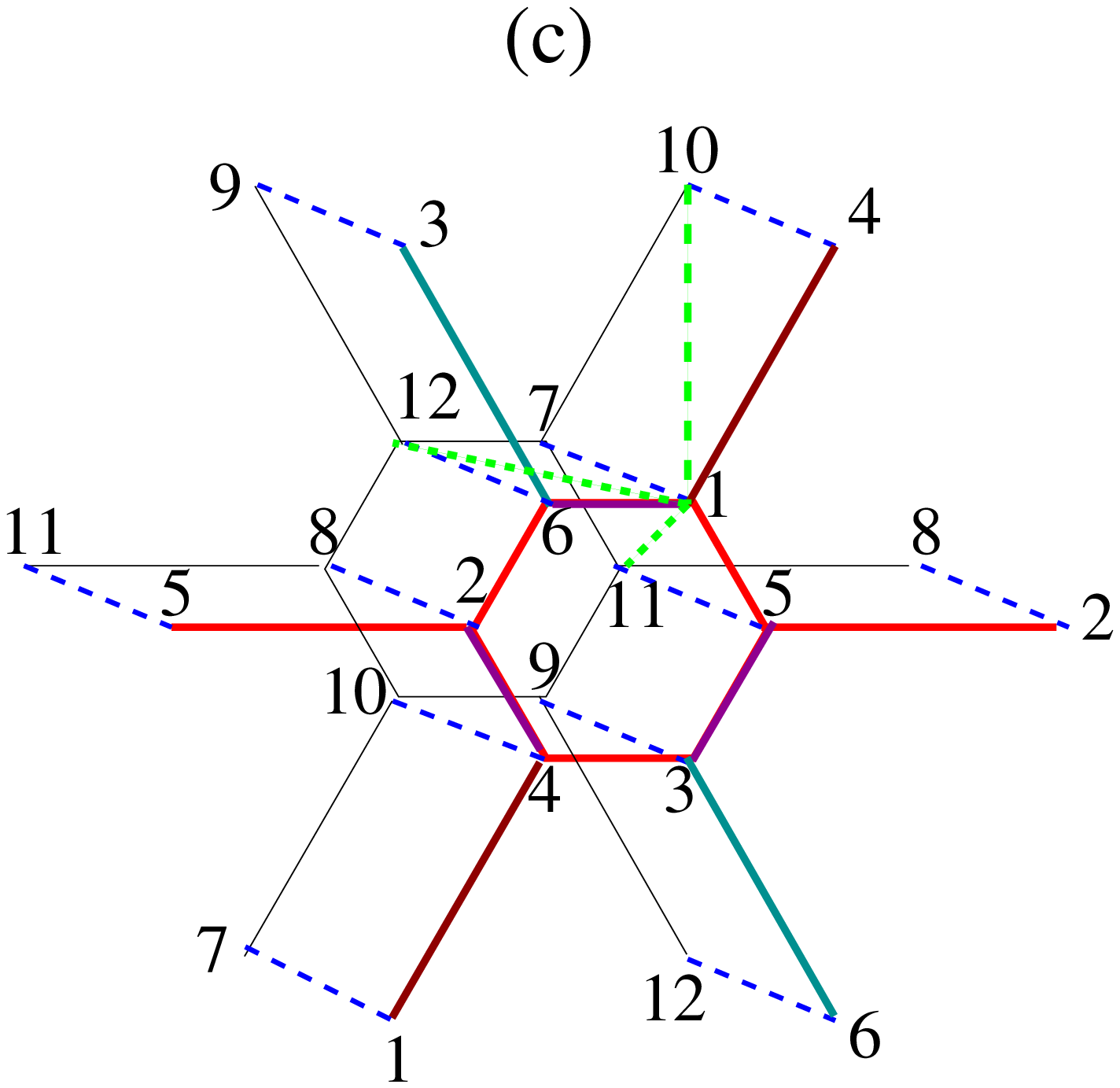,width=5.0cm,angle=0} }
\caption{Structure and notation. 
{\bf(a)} Sketch (exaggerated) of shrunk bilayer graphene where numbers 
indicate C-C first, second, and so on, neighbors of reference carbon 
atom in each layer. 
{\bf(b)} Shows the unit cell of intercalated bilayer graphene. In this Kekul\'e structure the intralayer hopping energies symmetry between first nearest neighbor atoms is broken. Intralayer hopping parameters along hexagonal bonds  are the same and shown by $t_{11}$ while between the long bond  slightly has been changed   given by $t_{11}'$. Similarly interlayer hopping are given by   $t_{12}$ and $t_{12}'$. Symmetry breaking of hopping energies leads to open two unequal gap at the  Dirac points  that folded back to the $\Gamma$  point.
{\bf (c) } Intra-plane superconductor pairing amplitudes
 $(\Sigma_1~\Sigma_2~\Sigma_3)$ are between 1-4, 3-6 and 2-5 subsites 
respectively, $(\Delta_1~\Delta_2~\Delta_3)$ are between 3-5, 2-4 and 1-6 
subsites respectively, 
and $(\Pi_1~\Pi_2~\Pi_3)$ are between 2-6, 1-5 and 3-4 subsites respectively.
 Also, inter-plane superconductor pairing amplitudes
 $(\Sigma_1^{'}~\Sigma_2^{'}~\Sigma_3^{'})$ are between 1-10, 3-12 and 2-11 
subsites respectively, $(\Delta_1^{'}~\Delta_2^{'}~\Delta_3^{'})$ are 
between 3-11, 2-10 and 1-12 subsites respectively, 
and $(\Pi_1^{'}~\Pi_2^{'}~\Pi_3^{'})$ are between 2-12, 1-11 and 3-10 
subsites respectively.  }
 \label{figure:pairing amplitude} 
\end{figure}

Motivated by  experimental observation of superconductivity in  Ca doped bilayer graphene, a more realistic model has been introduced here to obtain all possible superconducting symmetries which can be arise analytically. 
In this manuscript, we follow the notion that Calcium doped bilayer graphene as mentioned in the ref.\cite{Kanetani2012} as the thinest limit of graphite intercalation compounds for which the structure  Consists of  Ca intercalated bilayer AA- stacked graphene (Fig.\ref{figure:pairing amplitude}(a)). However,  recently, another possibility has been raised experimentally \cite{Endo2019}. 

Based on mean-field treatment of an extended Hubbard model,  a realistic tight-binding model has been used  where its parameters are determined by a fitting to DFT band structure. By adding an effective attractive interactions between interlayer and intralayer electrons,  all possible superconductivity pairing symmetry characters of  C$_6$CaC$_6$  has been studied in details ( which can be applied to any related
graphene-like structures such as B$_{3}$N$_{3}$CaB$_{3}$N$_{3}$). We will take advantage of the observation that, mathematically one can use mirror symmetry properties of Bloch coefficients of  intercalated AA -stacked bilayer graphene and interpret their Hamiltonian as two independent single layer pseudo-graphene structures (even and odd sectors) where one of them (even symmetry sector) is decorated with calcium layer. 
This notion leads to the emergence of a topological number
 called cone index ($c=\pm 1$).
Conservation of  cone index during Klein tunneling across an n-p junction  is one of the interesting unique behaviors  in AA-stacked bilayer  where it raises the possibility of ‘cone-tronic’ devices based on AA-BlG.\cite{Sanderson2013} 

In the real space a quasi-particle consist of two electrons  from opposing layers with the same symmetrical position where they possess an additional quantum cone index
beside their chirality nature near the Dirac points. The  cone index concept   is a unique feature in describing quasi-particles behaviors in AA-BlG  with respect to single layer graphene.
Two  quasi particles  (i.e. ``four electrons'') can  team up to build a Cooper pair as one can see in Fig.~\ref{figure:pairing amplitude}(b,c). We will see that only quasiparticles with the same cone index can be paired.

It will be shown that the question of superconducting phases in metal-intercalated bilayer graphene such as C$_6$CaC$_6$ can be decoupled to two independent gap equations corresponding to each of the even and odd sectors which they can be solved analytically (or nearly so)  to obtain all possible  pairing symmetry phases  which can be probed experimentally.

 
  The two gap nature of superconductivity that is one of the unique feature of MgB2 \cite{Putti2006} can be inspected here similarly. We have numerically predicted that in the temperature range of 0-6 K  the phase transition from d-wave single-gap  to dual-gap d-wave superconductivity could be observed. 
Using {\it ab initio}  anisotropic Migdal-Eliashberg theory including 
Coulomb interaction, Margine {\it et al.}\cite{Margine2016} concluded that C$_6$CaC$_6$ 
should support phonon mediated superconductivity with a critical temperature 
$T_c=6-7$K, within the range of observations, and it exhibits two distinct 
superconducting gaps on the electron and hole Fermi surface 
pockets which is in agreement with the result has been obtained in the present manuscript.

The  rest organization of the paper is as follows.
Section II introduces the model Hamiltonian that we study, with Sec. III
setting the stage by obtaining mostly analytic diagonalization of the non-interacting system. In Sec. IV, the treatment of pairing and presentation of superconducting phases is presented.  In Sec. V, analytic insight are complemented by numerical solutions,
followed by a discussion and summary in Sec. VI. Many of the analytic expressions are delegated
to Appendices.  

\section{Model Hamiltonian}
The system we consider, illustrated in Fig.\ref{figure:pairing amplitude}(a),
 consists of $AA$ stacked bilayer graphene intercalated by Ca metal layer in which
 intercalant  atoms  are located on the central symmetry plane of bilayer graphene  
at the center of neighboring carbon hexagons.
The distance between  the graphene
 layers is calculated to be $h= 4.63\text{\AA}$ in the case of  Ca intercalation.
The nearest in-plane Ca-Ca distance is $\xi=4.26\text{\AA}$. 
Charge transfer from Ca to the graphene layers leads to  breaking symmetries 
of hopping amplitudes, and of C-C bond lengths similarly to those of Li decorated 
monolayer graphene[\cite{Rouhollah2018}]. 
The attractive interaction  between metal cations and C atoms after charge transfer 
contracts the Ca-C distance and
reduces the C-C bond lengths in the Ca-centered hexagon to 
$a_{1}=1.419\text{\AA}$. As a result the bond 
length of neighboring C atoms in different hexagons is somewhat larger, at
$a_{2}=1.423\text{\AA}$. Also in this ``shrunken bilayer graphene''\cite{Rouhollah2018} the hopping 
integrals between short-bond inter- and intra-layer carbons are 
respectively $t^{11}_{1}$ and  $t^{12}_{1}$, 
while those between stretched carbon sites will be denoted 
$t^{'11}_{1}$ and  $t^{'12}_{1}$.  
The lattice then becomes a two dimensional hexagonal Bravais lattice with thirteen
atomic sites. The sites of $i$-th cell will be labeled as 
$A_{i1}^{m}, A_{i2}^{m}, A_{i3}^{m}, B_{i1}^{m}, B_{i2}^{m}, B_{i3}^{m}$ 
and $Ca$, where $m$ is layer index and  takes $m=1,2$. 

The Hamiltonian of this system is 
\begin{eqnarray}
 \hat{H}=-\sum_{i\alpha}\sum_{j \beta,\sigma } 
    t_{i\alpha,j\beta}^{\sigma,\sigma}{\hat{c}}_{i\alpha\sigma}^{\dagger}{\hat{c}}_{j\beta\sigma} 
    +\sum_{i\alpha ,\sigma }(\epsilon_{i\alpha}-\mu_{o} ) \hat{n}_{i\alpha\sigma} 
    +\frac{1}{2}\sum_{i\alpha,\sigma}\sum_{j \beta ,\sigma{}'}U _{i\alpha,j \beta}^{\sigma ,\sigma{}'} 
      \hat{n}_{i\alpha\sigma}   \hat{n}_{j\beta\sigma{}'}= {\hat H}_{N} + {\hat H}_{P}.
\label{eq:full-Hamiltonian}
\end{eqnarray}
where $\alpha$ and $\beta$ run over the sublattice orbitals $ A_i^{m} p_z$, 
$B_i^{m} p_z$ and Ca $s$. Here 
${\hat{c}}_{i\alpha\sigma}^{\dagger}$, ${\hat{c}}_{i\alpha\sigma} $ are 
creation and annihilation operators of an electron with spin $\sigma$ on 
subsite $\alpha$ of $i$th lattice site, and
$\hat{n}_{i\sigma}={\hat{c}}^{\dagger}_{i\sigma}{\hat{c}}_{i\sigma} $ 
is the electron number operator.
The chemical potential is $\mu_{0}$
and $t_{i\alpha,j \beta}$ is the hopping integral from $\alpha$ subsite 
of $i$th site to the $\beta$  subsite of $j$th site.
Here $U$ is an effective negative interaction between electrons in the extended (negative $U$) 
Hubbard model that allows the possibility of superconductivity. 
\section{The non-interacting System}
In this section a thirteen band tight binding  model, consisting of twelve 
$p_z$ C orbitals and the Ca $s$ orbital, for  Ca-intercalated bilayer  
graphene is constructed, to be applied to study superconducting states 
of this system within BdG theory.
The non-interacting system Hamiltonian is invariant under mirror symmetry, 
which leads to division of the intercalated AA-BlG band structure into two 
sectors characterized by eigenvalues of mirror operation.  
  Here we take advantage of the mirror symmetry.
We first apply this reduction to the simple case of pristine AA-BlG.

\subsection{Reducible Tight Binding Model for Pristine AA-staked Bilayer Graphene}
\begin{figure}
\centerline{
\epsfig{file=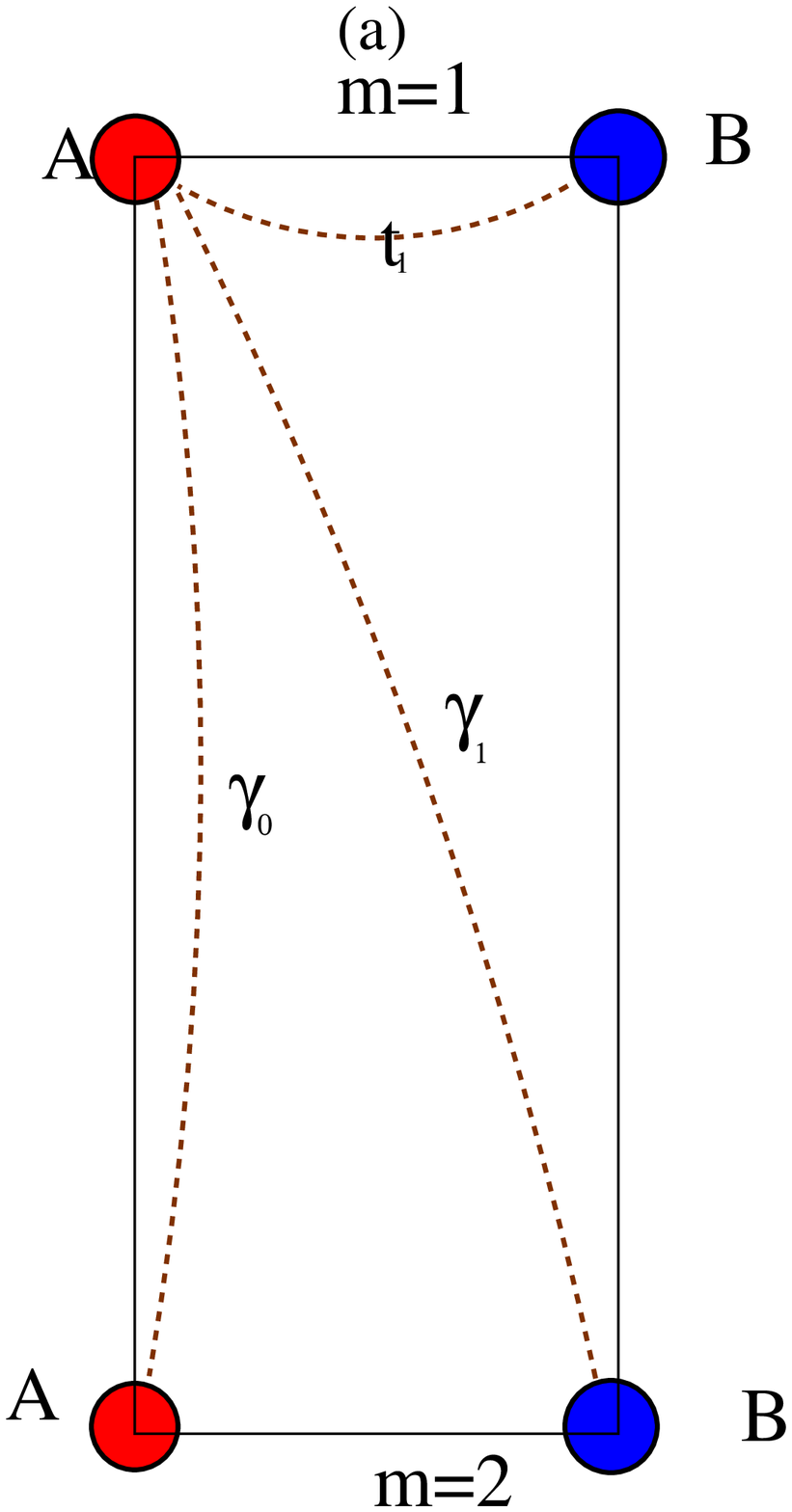,width=3.10cm,angle=0}
\epsfig{file=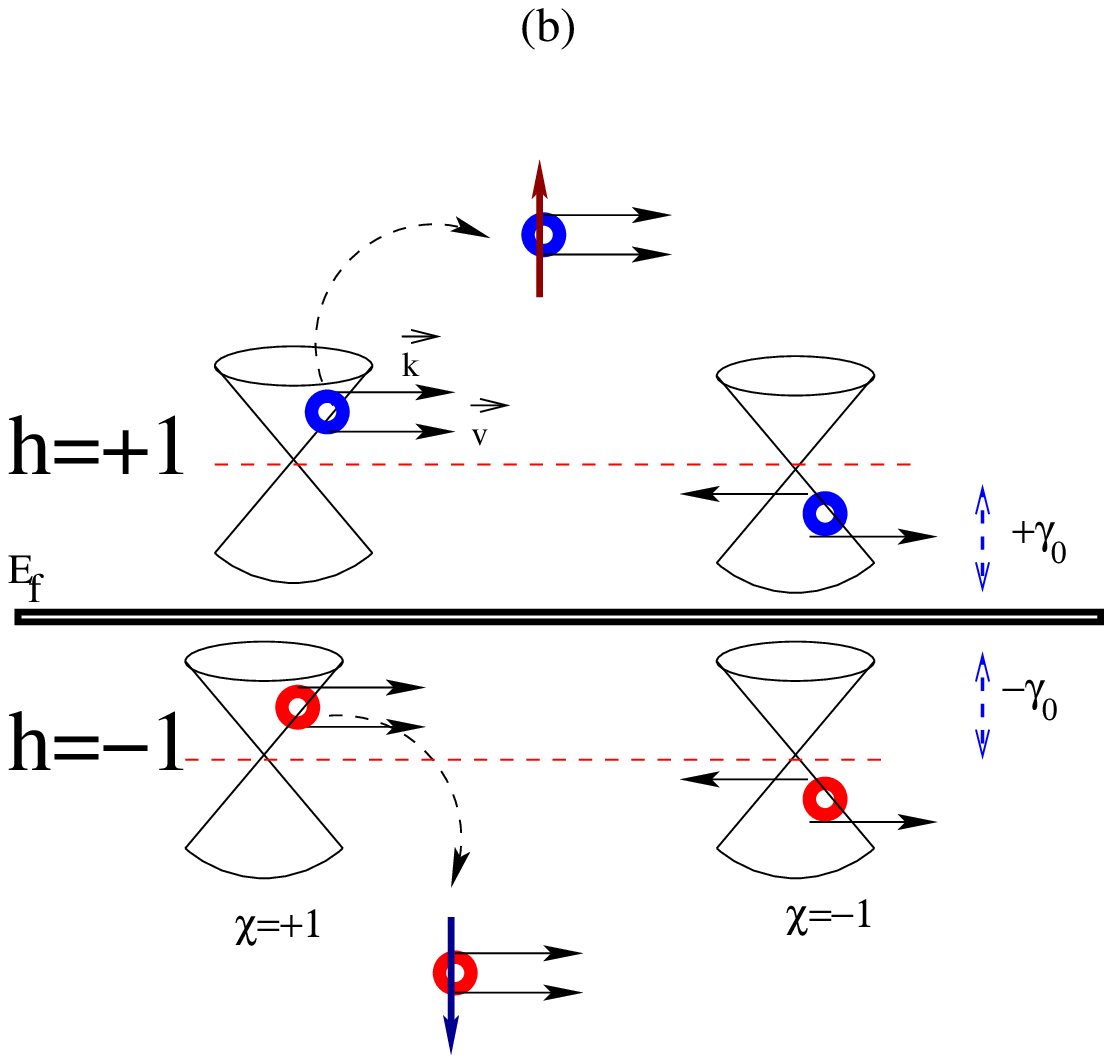,width=6.0cm,angle=0}
\epsfig{file=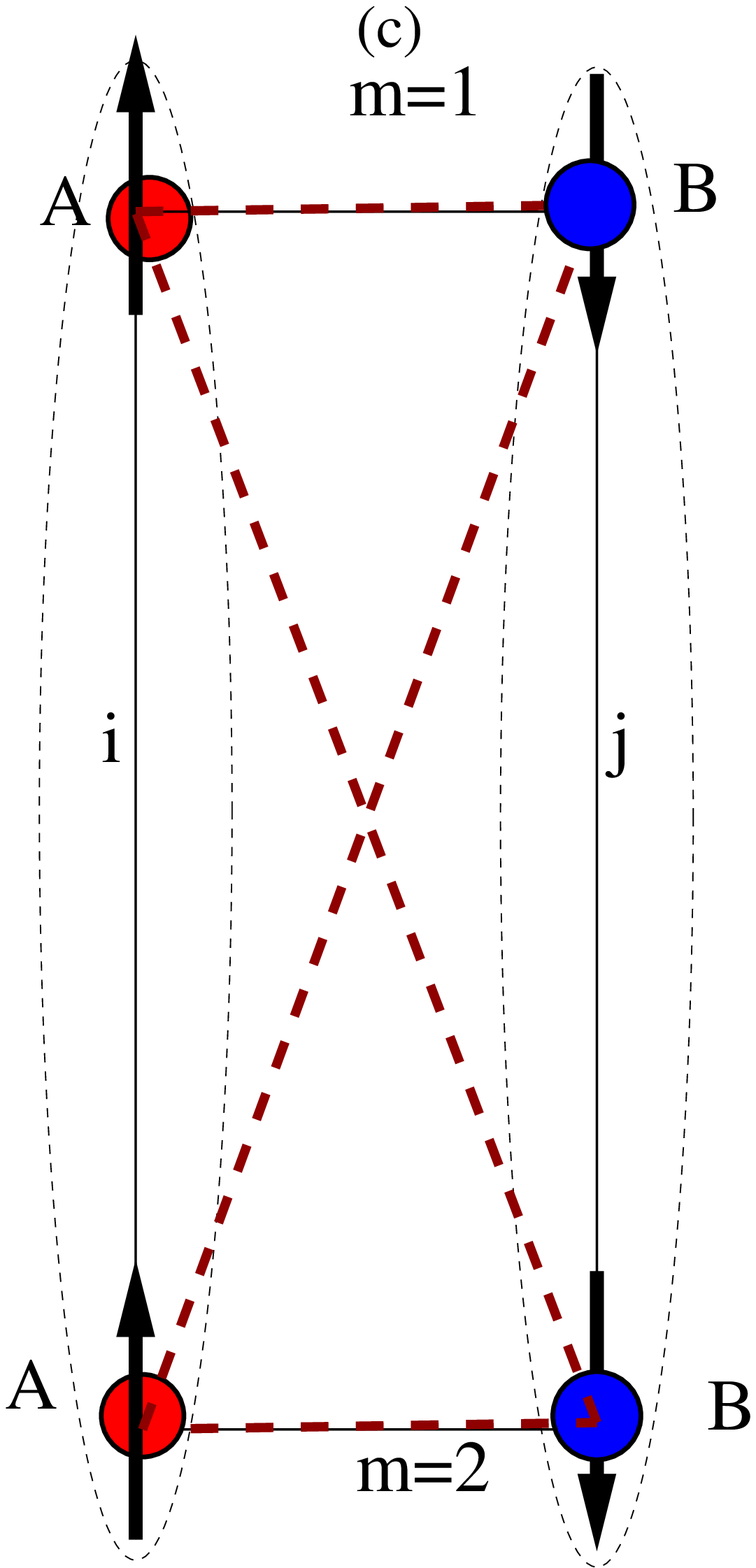,width=3.0cm,angle=0}}
\caption{Structure and notation. (a)Shows AA-stacked BlG unit cell, m= 1,2 indicate layer index ,  $t_1$  and $\gamma_1$  are nearest neighbor intralayer and interlayer  hopping energies respectively and  $\gamma_0$ is direct interlayer hopping. (b) Near the Dirac points K\& K'(K'=-K) the even and odd sector of pristine AA-BlG  are shown by $\gamma_0$-up and down shifted Dirac Cones. Quasi-particles in even sector (odd sector) are shown  by blue (red) circles. $\chi$ shows v-chirality while $h$ is the index of h-chirality  c) shows a Cooper pair. Up and down pseudo-spin (h-Pspin) of irreducible blocks  of  AA-BlG Hamiltonian viz. $H^+$ and $H^-$  are shown by dashed ellipses i \& j. Here a quasi-particle  consists of two  electrons with the same spin which each one  located at similar sub-sites in the opposing layers. These quasi-particles describe by an additional h-chirality index.
Two  quasi particles  (i.e. ``four electrons'') can  team up to build a Cooper pair as one can see in Fig.~\ref{figure:pairing amplitude}(b,c). We will see that only quasiparticles with the same cone index can be paired
 }
 \label{figure:AA-stacked-amplitude} 
\end{figure}
The unit cell of AA-BlG, illustrated in
Fig.~\ref{figure:AA-stacked-amplitude}(a), consists of four atoms 
$A_1$, $B_1$ in the top layer and  
$A_2$, $B_2$ in the bottom layer. The Schr\"odinger equation for this system in 
terms of Bloch coefficients is given by
\begin{eqnarray}
\left[ {\begin{array}{*{20}{c}}
H_{11}&\vline&H_{12}&\\
\hline
H_{21}&\vline&H_{22}&
\end{array}} \right]
\left( {\begin{array}{*{20}{c}}
\chi_1\\
\chi_2\\
\end{array}} \right) 
 = E_{n}\left( {\begin{array}{*{20}{c}}
\chi_1\\
\chi_2\\
\end{array}} \right),~~~\left( {\begin{array}{*{20}{c}}
\chi_1\\
\chi_2\\
\end{array}} \right)=
 \left( {\begin{array}{*{20}{c}}
{\mathscr C}_{A_1}(\vec k)\\
{\mathscr C}_{B_1}(\vec k)\\
\hline
{\mathscr C}_{A_2}(\vec k)\\
{\mathscr C}_{B_2}(\vec k)\\
\end{array}} \right)
\label{eq:H-normal-AA-stacked}
\end{eqnarray}
where  $2\times1$ column matrices $\chi_1$ and $\chi_2$ are components of AA-BlG 
iso-spinors, $H_{11}=H_{22}$ and $H_{12}=H_{21}$ are intralayer and interlayer hopping matrix 
respectively.
The 4-component ''Dirac spinor'' (Dirac representation) is
reducible.
 Using the unitary transformation 
$U=\frac{1}{\sqrt{2}}\left(\begin{array}{*{20}{c}}I_2&~~I_2&\\I_2&-I_2&\end{array}\right)$, 
it decomposes into two irreducible representations, acting only on two 2-component right 
and left hand ``Weyl spinors.'' 
 Similar 
transformations decouple the 4-component iso-spinor of AA-BlG into two 2-component 
chiral iso-spinors. This connection is an pedagogically useful mathematical similarity 
between the Schrödinger equation of AA-BlG  and the Dirac equation,
because it can give a insight into the prediction of quasiparticle behavior in AA-BlG 
in analogy with relativistic particles such as neutrinos. 
Mirror symmetry leads to\begin{eqnarray}
\mathscr{C}_{A_1}(\vec k)=\pm\mathscr{C}_{A_2 }(\vec k) ,~~~~ \mathscr{C}_{B_1}(\vec k)=\pm\mathscr{C}_{B_2 }(\vec k) 
\label{eq:bloch-inversion-symmetry-AA-stacked} 
\end{eqnarray}

Inserting Eq.~\ref{eq:bloch-inversion-symmetry-AA-stacked}  
  into Eq.~\ref{eq:H-normal-AA-stacked} and defining new single layer graphene-like Hamiltonians $H^{\pm}=H_{11}\pm H_{12}$, the four band Schr\"odinger equation 
converts into two decoupled single layer graphene two band Schr\"odinger equations 
of the form
 \begin{eqnarray}
H^+\chi_R=E_n^+\chi_R,~~~~H^-\chi_L=E_n^-\chi_l,
\label{eq:H-normal-AA-stacked-decouple}
\end{eqnarray}
wherein  2-component iso-spinors  (has been shown in Fig.~\ref{figure:AA-stacked-amplitude}c) are given by
 \begin{eqnarray}
\chi_R=\frac{1}{\sqrt{2}}(\chi_1+\chi_2)=\left( {\begin{array}{*{20}{c}}
{\mathscr C}_{A}^{+}(\vec k)\\
{\mathscr C}_{B}^{+}(\vec k)\\
\end{array}} \right),~~~
\chi_L=\frac{1}{\sqrt{2}}(\chi_1-\chi_2)=\left( {\begin{array}{*{20}{c}}
{\mathscr C}_{A}^{-}(\vec k)\\
{\mathscr C}_{B}^{-}(\vec k)\\
\end{array}} \right).
\end{eqnarray}
 The $\pm$ sign appearing in the even and odd sector Hamiltonians  are the $h=\pm 1$ eigenvalues of the mirror symmetry operator $\hat S_h=\left(\begin{array}{*{20}{c}}0&~~I_2&\\I_2&0&\end{array}\right)$ which is  the same as fifth Dirac gamma matrix $\hat\gamma^5$ 
that reflects the right ($\chi_R$) or left hand ($\chi_L$) chirality of quasiparticles in the relativistic quantum field theory.  This additional $\pm$ topological 
index called  by Sanderson and Ang (AS) as ``cone index.''\cite{Sanderson2013}.
SA have shown that quasiparticles in AA-BlG are not only chiral but are also characterized 
by  ``cone index,''.  
So we will refer to the cone index as h-chirality index and to usual  helicity appears again as v-chirality. According to this notion, one can describe Dirac cones in AA-BlG shown in the Fig. \ref{figure:AA-stacked-amplitude}(b) with two kind of chirality with respect to   
asymmetric in such a way that the structure and its vertical 
 ({\bf``v-chirality''}) and horizontal ({\bf``h-chirality''}) mirror image are not superimposable.
This chirality is a general aspect of AA-BlG quasi-particles
that holds for general hoppings and over the entire Brillouin zone. 
 SA show that electron transport across a barrier must conserve the cone index, a 
consequence of the Klein tunneling behavior in AA-stacked BlG. In the following sections we will extend  the consequence of the cone index notion to superconductivity pairing.
 
A quasi-particle  in the even sector (odd sector) consists of two electron (2$e$ charge) 
from opposing layers with the same spatial symmetry which possess $h=+1$ ($-1$) cone 
index and on-site energy of $\varepsilon^+=+\gamma_0$ ($-\gamma_0$) respectively. 
Hopping of a quasi-particle from A subsite to nearest neighbor B subsite in the 
even-sector (odd-sector) changes the energy by $t^+=t_1+\gamma_1$  ($t^-=t_1-\gamma_1$). 
 This decoupling is more than just a simple mathematical diagonalization and symmetry
characterization.
One can interpret the Hamiltonian of AA-BlG as a single layer honeycomb lattice 
Hamiltonian with two types of charge carriers described by
 $$H=-\sum_{i,j}\sum_{\sigma=\pm}t_{i,j}^{\sigma}a_{i\sigma}^{\dag}b_{j\sigma}+
\sum_{i\sigma}\varepsilon^{\sigma}(a_{i\sigma}^{\dag}a_{i\sigma} + b_{i\sigma}^{\dag}b_{i\sigma})+h.c$$
 where $a_{i\sigma}^{\pm\dag}$ ($b_{i\sigma}^{\pm\dag}$) is the creation operator of a quasiparticle in the A(B) subsite of the $i$th site with $\sigma=\pm  1$ wherein $\sigma$  represents h-chiral Pseudo-spin (h-Pspin). 
 where $a_{i\sigma}^{\pm\dag}$ ($b_{i\sigma}^{\pm\dag}$) is the creation operator 
of a quasiparticle in the A(B) subsite of the $i$th site with $\sigma=\pm  1$ 
wherein $\sigma$  represents $h$-chiral Pseudo-spin ($h$-Pspin). 


In the intralayer and interlayer nearest neighbor hopping approximation,
viz. $t_1$, $\gamma_0$ and $\gamma_1$ the system Hamiltonian  is given by 
\begin{eqnarray}
\left[ {\begin{array}{*{20}{c}}
H_{11}&\vline&H_{12}&\\
\hline
H_{12}&\vline&H_{11}&
\end{array}} \right]
\approx
-\left[ {\begin{array}{*{20}{c}}
{\mu}&{t_1}f(\vec k)&\vline& {{\gamma_0}}&{\gamma_1}f(\vec k)&\\
{{t_1}f^{*}(\vec k)}&\mu&\vline& {{\gamma_1}f^*(\vec k)}&{\gamma_0}&\\
\hline
{{\gamma_0}}&{\gamma_1}f(\vec k)&\vline& \mu&{t_1}f(\vec k)&\\
{{\gamma_1}f^*(\vec k)}&{\gamma_0}&\vline& {t_1}f^*(\vec k)&\mu&\\
\end{array}} \right];~~
H^{\pm}=H_{11}\pm H_{12}=-\left[ {\begin{array}{*{20}{c}}
&{\mu\pm\gamma_0}&{t^{\pm}}f(\vec k)&\\
&{{t^{\pm}}f^{*}(\vec k)}&{\mu\pm\gamma_0}&\\
\end{array}} \right]
\label{eq:normal-AA-stacked Hamiltonian}
\end{eqnarray}
 Here $t^{\pm}=t_1\pm \gamma_1$ and
$t_1$ ($\gamma_1$) is the nearest neighbor intralayer(interlayer) hopping between 
A and B sublattices and the direct interlayer hopping (i.e. $A_1$ to $A_2$ or 
$B_1$ to $B_2$) is given by $\gamma_0$. Here $\mu$ is chemical potential, also 
$f(\vec k)=\sum_{i=1}^{3}e^{i\vec{k}.\vec{\delta_i}}$ 
wherein $\vec{\delta_i}$ vectors connect A subsite to three in-plane nearest neighbor  B subsites.

The decoupled Schr\"odinger equation Eq.~\ref{eq:H-normal-AA-stacked-decouple}
has eigenvalues
\begin{eqnarray}
E_{n}^{\pm}(\vec k)=\mu+\eta_{v-c}(t^{\pm}|f(\vec k)|\pm\gamma_0)
\end{eqnarray}
where  $\eta_{v-c}=\pm 1$. The four bands of AA-BlG separates into two independent 
up-down shifted  single layer  graphene bands 
where they are referred to as the even and odd sector, respectively.

Near the Dirac points, the dispersion energies 
  $E_n^{\pm}(\vec {k})=\hbar v_f^{\pm}|\vec {k}|\pm\gamma_0$ 
of odd and even sectors are shown by two up-down shifted Dirac cones in 
Fig.~\ref{figure:AA-stacked-amplitude}(b) where
  $v_f^{\pm}=\frac{\sqrt{3}t^{\pm}a}{2\hbar}$ is Fermi velocity of $h=\pm 1$ quasi particles. 
Generalizing the tight binding model to include further neighbor hopping terms 
can highlight some hidden aspects of the AA-stacked Dirac cone quasi-particles. 
For second neighbor interlayer hopping, $\gamma_1$ taken into regard one can 
distinguish quasi-particles with the same chirality (v-chirality) and different  
cone index (h-chirality) from their velocities which could be inspected experimentally.
The Fermi velocity of Dirac cone with $h=-1$ chirality decreases as 
interlayer hopping increases, while the velocity of $h=+1$ quasi-particle increases:
$$
v_f^{+}-v_f^{-}=\frac{\sqrt{3}a}{\hbar}\gamma_1.
$$ 
In the strong inter layer coupling $t_1\to (-)\gamma_1$  Fermi velocity $v_f^-(v_f^+)\to 0$  and odd(even) sector bandwidth tends to zero. 
As shown in the next subsection, interlayer coupling  $\gamma_1$ may be considerable,  
so that inequality could be considerable. 

\subsection{Analytic Tight Binding Model for Intercalated Bilayer Graphene}
\begin{figure}
\centerline{\epsfig{file=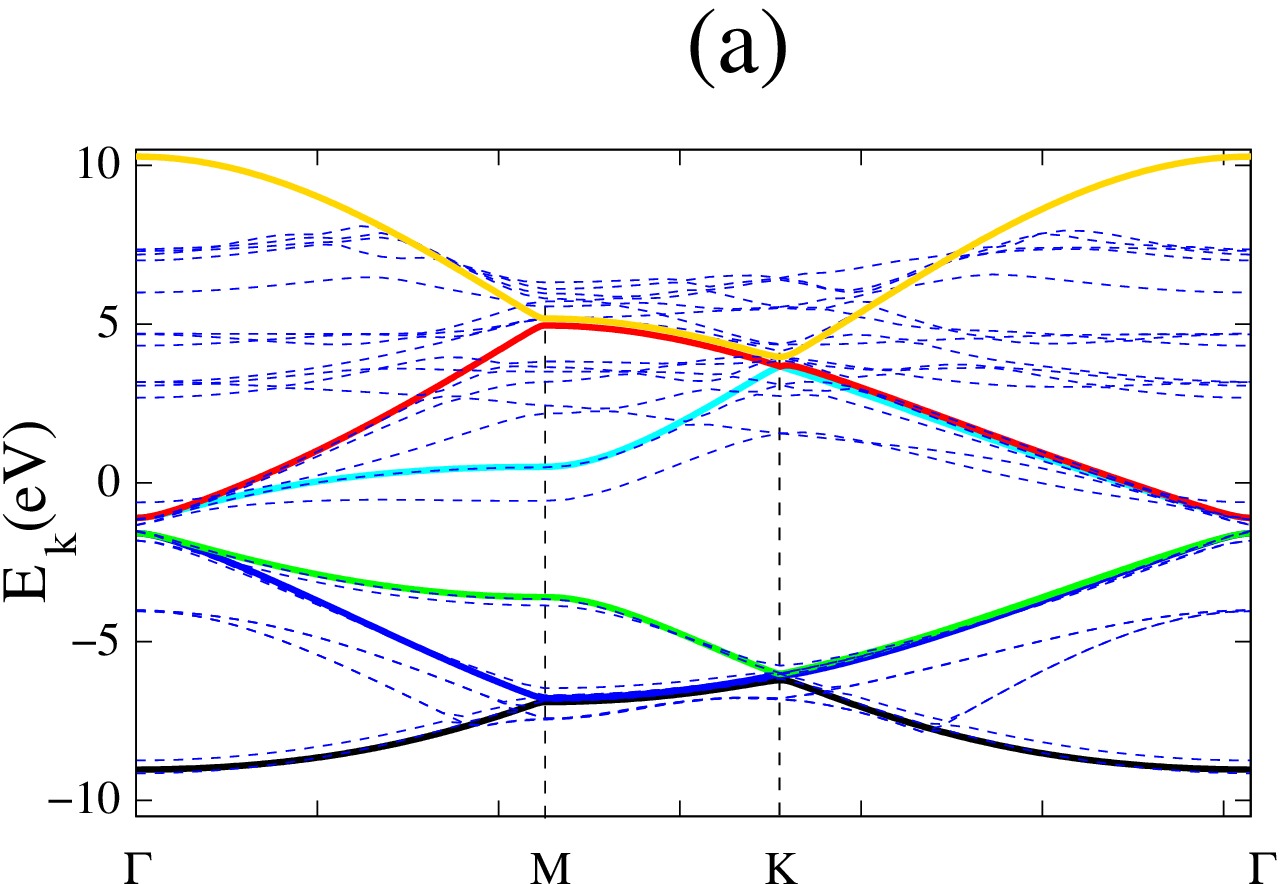,width=7.5cm,angle=0}\hspace{0.5cm}\epsfig{file=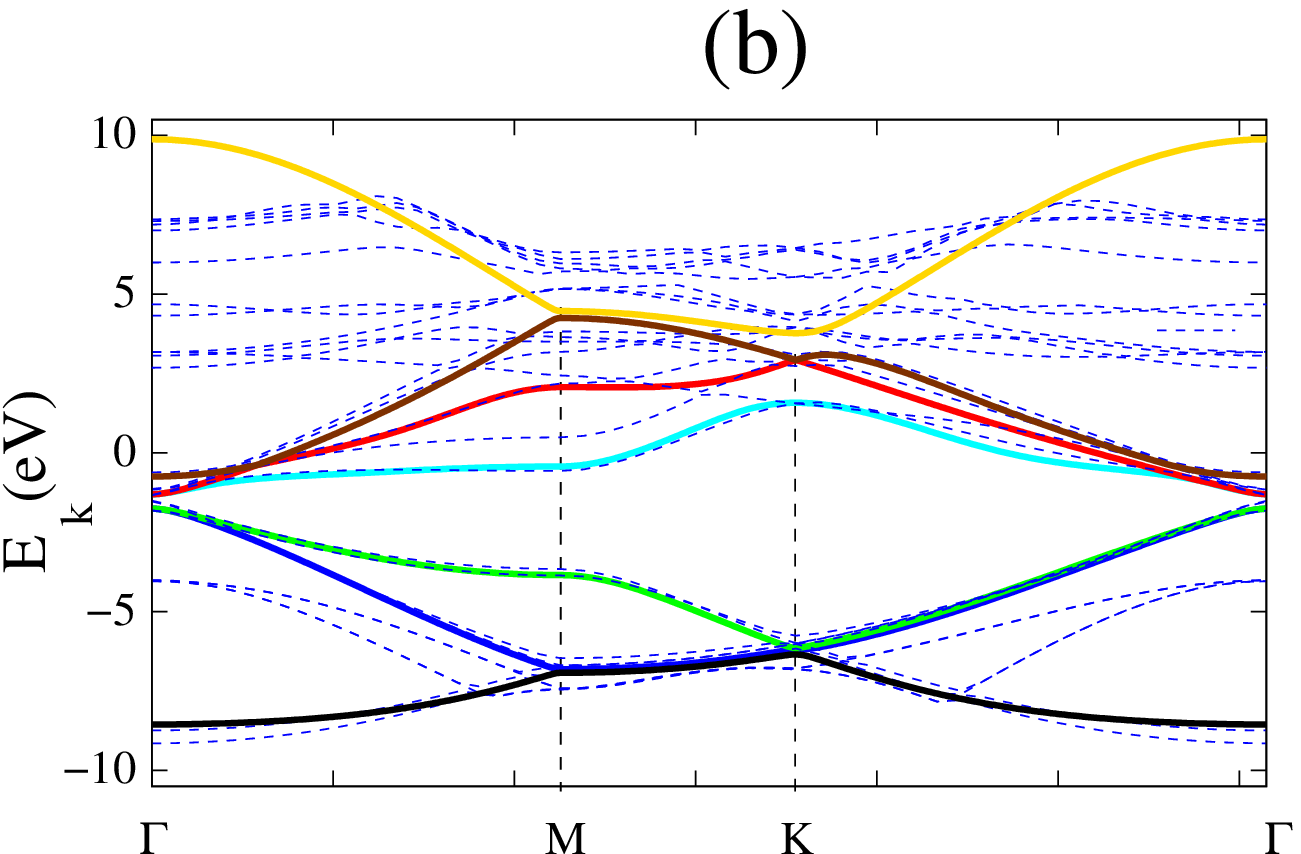,width=7.5cm,angle=0}}
\caption{Band structure of C$_6$CaC$_6$. 
Left panel:  bands emphasizing the six pseudo-graphene layer bands  ($H^{-}(\vec{k})$). 
Right panel: emphasis on the seven pseudo-graphene layer bands  ($H^{+}(\vec{k})$). 
Thin dashed lines indicate the DFT bands, while the fitted bands are shown in color. 
 }
 \label{figure:fiting-band structure of graphene} 
\end{figure}

In this subsection we will generalize the previous procedures to include the case  of experimentally  observed structures such as Li or Ca intercalated bilayer graphene. We will follow the notion it believe that these structure are intercalated AA-stacked bilayer graphene as has been shown in the Fig.~\ref{figure:pairing amplitude}. 
From the beginning, the Hamiltonian is generalized to incorporate 
several broken symmetries, including the on-site energies, 
hopping integrals, and bonds lengths (geometry). Due to this generalization,
 it can be used to  
obtain analytic dispersion energies of not only C$_6$CaC$_6$, but also related
graphene-like structures such as B$_{3}$N$_{3}$CaB$_{3}$N$_{3}$.
The Hamiltonian of such  a non-interacting system is 
\begin{equation}
 \hat{H}_{N}=-\sum_{i\alpha}\sum_{j \beta,\sigma } t_{i\alpha,j\beta}^{\sigma,\sigma}c_{i\alpha\sigma}^{\dagger}c_{j\beta\sigma} +\sum_{i\alpha ,\sigma }(\epsilon_{i\alpha}-\mu_{o} ) \hat{n}_{i\alpha\sigma}.
\label{eq:H-normal}
\end{equation}
where $\alpha$ and $\beta$ run over sublattice orbitals $ A_i^{m}$, $B_i^{m}$ and the 
intercalated atom (e.g. $Ca$) orbital. 
The Schr\"odinger equation for this system in terms of Bloch coefficients 
in $\vec{k}$ space becomes
\begin{eqnarray}
\sum^{12}_{\beta=0} \varepsilon_{\alpha\beta}({\vec k}) \mathscr{C}_{\beta }+(\epsilon_{\alpha}-\mu_{o})\mathscr{C}_{\alpha } = E({\vec k})\mathscr{C}_{\alpha } \;\;\mbox{where}\;\; \varepsilon_{\alpha\beta}({\vec k})=-\frac{1}{N}\sum_{ij} e^{i{\vec k}.({\vec r}_{i\alpha}-{\vec r}_{j\beta})}t^{\sigma\sigma}_{i\alpha j\beta}.
\label{eq:schro-bloch-state-Fourier}
\end{eqnarray}
Here the $\beta$=0, 1, 2, ..., 12 subscripts refer respectively to 
intercalant Ca, $A_1^{1}$, $A_2^{1}$, $A_3^{1}$, 
$B_1^{1}$, $B_2^{1}$, $B_3^{1}$, $A_1^{2}$, $A_2^{2}$, $A_3^{2}$, 
$B_1^{2}$, $B_2^{2}$ \& $B_3^{2}$ and N is the number of unit cells. 
Here $\epsilon_{A_{i}}=\epsilon_{A}$ and $\epsilon_{B_{i}}=\epsilon_{B}$. 

Mirror symmetry of this system result in the relations 
\begin{eqnarray}
\mathscr{C}_{\alpha}(\vec k)=\pm\mathscr{C}_{\alpha+6 }(\vec k) ,~~~~ \alpha=1,2,...,6
\label{eq:bloch-inversion-symmetry} 
\end{eqnarray}
which reflects the mirror symmetry through the Ca plane that separates even and
odd states (i.e. $h=\pm 1$ h-chirality). 
 By inserting Eq.~\ref{eq:bloch-inversion-symmetry} into
Eq.~\ref{eq:schro-bloch-state-Fourier}, with more detail given in Appendix A, 
one obtains two independent  Schr\"odinger equations corresponding to 
 eigenvectors ${\Psi _i^{-}}(\vec k)^T =(0,~~~
   \boldsymbol{\mathcal C}_i(\vec k), ~-\boldsymbol{\mathcal C}_i(\vec k))$, 
              ${\Psi _j^{+}}(\vec k)^T =({\mathscr C}_0(\vec k),~~
   \boldsymbol{\mathcal C}_j(\vec k), ~+\boldsymbol{\mathcal C}_j(\vec k))$ 
  respectively. For the odd eigensystem, the Schr\"odinger 
   Eq.\ref{eq:k-schr-bigraphen1} reduces to following $6\times 6$ 
  matrix eigenvalue problem
\begin{eqnarray}
H^-(\vec k)\boldsymbol{\mathcal C}_{n}(\vec k)=\left( H_{11}-H_{12}\right)\boldsymbol{\mathcal C}_{n}(\vec k)=E_n^{-}\boldsymbol{\mathcal C}_{n}(\vec k),~~~n=1,2,...,6.
\label{eq:k-schr-bigraphen-first6-solutions} 
\end{eqnarray}
The Schr\"odinger  Eq.~\ref{eq:k-schr-bigraphen-first6-solutions} can be solved
analytically, with the six eigenvalues presented in  
Appendix I, Eq.~\ref{eq:full-shrun-eigen}. These expressions are unaffected by the
intercalant layer due to the separation of even and odd mirror symmetries,
but the presence of Ca will renormalize parameters.
   For the even mirror sector, the Schr\"odinger equation 
Eq.\ref{eq:k-schr-bigraphen1} reduces to the following $7\times 7$ 
matrix eigenvalue problem
\begin{eqnarray}
H^{+}_{c}(\vec k)\left( {\begin{array}{*{20}{c}}
{\mathscr C}_0(\vec k)\\
{\sqrt{2}{\boldsymbol{\mathcal C} _n} {\vec k} )}
\end{array}} \right) =\left[ {\begin{array}{*{20}{c}}
{{h_0}(\vec k)}&\vline& {\sqrt 2 {h_{01}}(\vec k)}\\
\hline
{\sqrt 2 {h_{10}}(\vec k)}&\vline& H^+(\vec k)
\end{array}} \right]\left( {\begin{array}{*{20}{c}}
{\mathscr C}_0(\vec k)\\
\sqrt{2}{{\boldsymbol{\mathcal C} _n}( {\vec k} )}
\end{array}} \right) = E_{n}^{+}\left( {\begin{array}{*{20}{c}}
{\mathscr C}_0(\vec k)\\
{\sqrt{2}{\boldsymbol{\mathcal C} _n}( {\vec k} )}
\end{array}} \right),
\label{eq:k-schr-bigraphen-next7-solutions} 
\end{eqnarray}
where $n =7 ,...,13$. The other seven bands of the 
Schr\"odinger equation Eq.~\ref{eq:k-schr-bigraphen1} can be obtained 
from solving the new Schr\"odinger  
Eq.~\ref{eq:k-schr-bigraphen-next7-solutions}.
 The k-dependent part of corresponding matrix components of $H_{11}$ and 
$H_{12}$ are identical in form, differing only in hopping parameters, 
hence  $H^{\pm}(\vec{k})$ can be considered as a shrunken graphene 
monolayer Hamiltonian  
with renormalized hopping parameters. It follows that, Similar to the pristine bilayer graphene (see \cite{Nori2016}), intercalated bilayer 
graphene can be interpreted as two independent pseudo-graphene monolayers 
where one of them is dressed by  a modified hopping Ca layer. 
The thirteen bands of bilayer graphene divide into two groups, 
six bands group (odd symmetries) corresponding to $H^-$ Hamiltonian and 
seven bands group  (even symmetries) which are eigenvalues of $H_c^+$ matrix.  

 Mathematically many of the results 
obtained in ref.~[\onlinecite{Rouhollah2018}] can be generalized to 
these graphene like structures 
but with renormalized hopping parameters. For general $\vec{k}$, except  
at $\Gamma$, it is challenging to obtain an exact analytical solution of 
the  Schr\"odinger equations of Eq.~\ref{eq:k-schr-bigraphen-next7-solutions}. 
At the graphene Dirac points which have become folded back to the 
this supercell $\Gamma$ point, 
symmetry breaking of nearest neighbor intra- and interlayer hopping 
parameters i.e. $t_{11},~{t'}_{11}$  and  $t_{12},~{t'}_{12}$ (Fig.~\ref{figure:pairing amplitude}b) results in two unequal small gaps with different centers, 
corresponding to six (odd-sector) and seven (even-sector) band pseudo-graphene Hamiltonians, given by
 \begin{eqnarray}
  E_{g}^{+}=2|t_1^+-t_1'^{+}|,~~~~ E_{g}^{-}=2|t_1^{-}-t_1'^{-}|
  \label{eq:gap-gamma}
 \end{eqnarray}
where $ t_1^{\pm}=t_{11}\pm t_{12}$ and $ t_1'^{\pm}=t'_{11}\pm t'_{12}$. For the use of graphene as field effect transistors, it is necessary to create an tunable gap.  Tunable and sizable band gap can be constructed in single layer\cite{Farjam2009} by decoration and in the bilayer graphe by intercalation as can be seen from Eq.~\ref{eq:gap-gamma}.

 The effect of symmetry braking of the the inter-layer coupling parameter i.e. $\Delta \gamma_1=t_{12}-{t'}_{12}$ leads to inequality of even and odd sector gaps.  Knowing the size of these energy gaps, one can find the difference in the first nearest neighbor intra and inter-layer hopping parameters symmetry breaking  (suppose $\Delta t>0$,  $\Delta \gamma_1>0$ and  $\Delta t>\Delta \gamma_1$),
 \begin{eqnarray}
  E_{g}^{+}- E_{g}^{-}=4\Delta \gamma_1,~~~~~ E_{g}^{+}+E_{g}^{-}=4\Delta t
  \label{eq:diff-gap-gamma}
 \end{eqnarray}
  where $\Delta t=t_{11}-{t'}_{11}$.  These two gaps are characteristic of AA-IBlG.
  In the case of Li- intercalated BlG, experimental ARPES spectra (Fig.~4 Ref.{{\cite{Caffrey2016}}}) shows two distinct gaps of wide $E_{g}^-=0.20eV$ and $E_{g}^+=0.46eV$.  In this reference  the authors has equated the ratio of these two gaps as the ratio of the interlayer skew coupling parameters ($\gamma_2$ and $\gamma_2 '$  in their notation).   Equation \ref{eq:diff-gap-gamma}  slightly correct the discussion  that has been stated in the  [Sec. III, sub-sec.C of Ref.{\cite{Caffrey2016}}] about this ratio. 
  
  Eq.~\ref{eq:k-schr-bigraphen-next7-solutions} results in the event 
that intercalant layer hopping parameters to graphene sheets are negligible, 
as in the case of Li-decorated graphene where Li atoms fully 
ionize and the Li-associate band lies above the Fermi energy, so Li-C 
hopping effects are negligible. In this particular case the odd (-) and the even (+) nontrivial 
 eigenvalues of $H^{\pm}$ matrix, are given by
\begin{eqnarray}
E_{sh;m,l}^{\pm}(t_{i},\vec{\xi_{i}},\vec{k}) = \mu_{m}^{\pm}(\vec k) - {\mu _{o}} + \frac{1}{2}\left[ {{\varepsilon _A} + {\varepsilon _B} + {{( - 1)}^l}\sqrt {{{\left( {{\varepsilon _A} - {\varepsilon _B}} \right)}^2} + 4{w_m^{\pm}}(t_{i},\vec{\xi_{i}},\vec{k})}   } \right],~~~m=1,2,3;~l=1,2 
\label{eq:full-shrun-eigen-text} 
\end{eqnarray}
 wherein $\mu_{m}^{\pm}(\vec k)$, $w_m^{\pm}(t_{i},\vec{\xi_{i}},\vec{k})$ are defined in Eqs.~\ref{eq:diag-eigenvalues} and \ref{eq:cubic-w-solutions} respectively.  Details for obtaining these results 
are presented in Appendix A and ref.~[\onlinecite{Rouhollah2018}].

Now further notation is established. Similar to the  case of pristine AA-BlG investigated in the previous subsection,
this transformation recast the  noninteracting Hamiltonian 
Eq.~\ref{eq:H-normal} as the direct sum of two
 single layer pseudo-graphene structures with  renormalized hopping 
integrals  in the  right and left hand chiral representation of the form
\begin{eqnarray}
\hat{H}_{N}=\hat{H}^{+}_{N}\oplus\hat{H}^{-}_{N}=
\sum_{ij\sigma}\sum^{6}_{\alpha,\beta=0}{t^{+}_{i\alpha\sigma,j\beta\sigma}}\hat{c}^{+\dag}_{i\alpha\sigma}\hat{c}^{+}_{j\beta\sigma}
\oplus\sum_{ij\sigma}\sum^{6}_{\alpha,\beta=1}{t^{-}_{i\alpha\sigma,j\beta\sigma}}\hat{c}^{-\dag}_{i\alpha\sigma}\hat{c}^{-}_{j\beta\sigma}
\end{eqnarray} 
wherein, as illustrated in the Fig.\ref{figure:AA-stacked-amplitude}(c), we introduced quasiparticle creation operators and hopping integrals in real space, 
\begin{eqnarray}
\hat c^{(\pm) \dagger}_{i\alpha\sigma}=\frac{1}{\sqrt 2}(\hat c^{\dagger}_{i\alpha\sigma}\pm\hat c^{\dagger}_{i,\alpha+6,\sigma}),~~t^{\pm}_{i\alpha\sigma,j\beta\sigma}=t^{inter}_{i\alpha\sigma, j\beta\sigma}\pm t^{intra}_{i\alpha\sigma, j\beta+6\sigma}, ~~~\alpha,\beta=1,...,6.
\label{eq:pair-creation-operators}
\end{eqnarray}
The creation operator $\hat c^{(\pm)\dagger}_{i\alpha\sigma}$ creates an quasi-particle with $h=\pm 1$ h-chirality  and spin ($\sigma$) at $i\alpha$ atomic subsite of each of these two  graphene-like structures, participate directly in the formation of superconducting phases.
 
The separation of thirteen bands of intercalated bilayer graphene into 
groups of six  bands with odd-symmetry and seven bands  with even symmetry  has strong 
advantages.  The tight-binding model band structures can be fit to DFT band 
structure results with greater accuracy and simplicity, for example. 
Especially when the pairing interactions are introduced, this transformation
reduces the speed of numerical calculations significantly and provides 
additional insight into physical properties of bilayer graphene.
\subsection{Fit to DFT band structures}
This formalism has been applied and divided into two separated effective 
single layer (shrunken)- pseudo-graphene models. 
DFT calculation is used to obtain the electronic structure data.  Gnu-plot of DFT data has been shown with blue thin dashed lines in the background of Fig.~\ref{figure:fiting-band structure of graphene}(a),(b).
The six odd bands and seven even bands were fit 
to DFT bands with results shown  with color lines in the Fig.~\ref{figure:fiting-band structure of graphene}.  The main problem that emerged here is that odd and even sector bands are not separated  in the DFT data. But by inspecting the DFT band structure  
and  to be careful in analytical calculations,  knowing that odd sector does not affected by Ca-C coupling the odd and even sector flat bands  can easily be distinguished. Emerging of two distinct gaps in the  Dirac point of both sectors is the other guide to perform fitting. 
The reduced fitting parameters are given in 
Tables \ref{table:shrunk-hoppping-6} and \ref{table:shrunk-hoppping-7}.
  
We follow the model presented in ref.~[\onlinecite{Rouhollah2018}]. As illustrated 
in Fig. \ref{figure:pairing amplitude}(a) 
on each of bilayer graphene sheets,
the $A_1$ sublattice site of the central unit cell is chosen as the origin labeled by $0$,  
and the $B_1$ site in 
the adjacent hexagon is considered as the second C atom neighbor. While just slightly longer 
than the nearest neighbor atoms $B_2$ and  $B_3$ in the same hexagon,  
this neighbor is labeled by $n=2$, and so on the further neighbors are labeled. 
In  Fig.~\ref{figure:pairing amplitude} (a), the big  
hexagon included up to nine intra-plane neighbors but for the pristine graphene 
it is surrounded by five neighbors. 
C-C hopping from $0$-subsite to intra-plane $n$th neighbor($t_{i0jn}^{intra}$)
plus (minus)  hopping from $0$-subsite to 
inter-plane $n$th neighbor ($t_{i0jn}^{inter}$) has  been shown by $t^{{\pm}CC}_{0n}$.
 In-plane Ca-Ca hoppings $t^{Ca-Ca}_{0m}$ are included up to $m=4$ neighbors. 
 Modified Ca to C hopping integrals in Eq. \ref{eq:k-schr-bigraphen-next7-solutions}, 
which are defined as $\sqrt{2}$ times the 
hopping from central Ca to $m$th neighbor 
 C atoms,  are denoted by $t^{CaC}_{0m}$ and obtained up to $m=5$ neighbors. 
 
 The six odd bands and seven even bands specified by Eqs.~\ref{eq:k-schr-bigraphen-first6-solutions}
and \ref{eq:k-schr-bigraphen-next7-solutions} respectively, have reduced hopping integrals given by
$t^{\pm}_{im\sigma,jn\sigma}=t^{inter}_{im, jn}\pm t^{intra}_{im,jn}$, 
DFT calculated bands were fitted to tight binding odd bands
of Eq.~\ref{eq:full-shrun-eigen-text}, with results  presented in 
Fig.~\ref{figure:fiting-band structure of graphene}(a) and Table I. 
The even bands which are solutions of 
Eq.~\ref{eq:k-schr-bigraphen-next7-solutions}  are obtained numerically  
and fitted to the DFT bands. The results are illustrated in 
Fig.~\ref{figure:fiting-band structure of graphene}(b) and Table II.

There are two 
flat bands with $d$-wave Bloch character: one in each of the odd and even sectors. 
The opposite signs of nearest neighbor inter- and intra-layer 
hopping amplitudes, $t_1^{11}$ and  $t_1^{12}$, leads to reduced  bandwidths
 of the even states (+ sign) in Eq.~\ref{eq:k-schr-bigraphen-next7-solutions}. 
Larger interlayer hopping $t_1^{12}$ leads to smaller
 bandwidth ($H^{\pm}=H_{11}\pm H_{12}$), while 
for the other six odd-symmetry bands, the bandwidth can increase as can be seen in 
Fig.~\ref{figure:fiting-band structure of graphene}. On the other hand, the bandwidth 
of the even sector flat band is reduced, again due to the calcium to carbon hopping 
while the odd sector is not affected by Ca-C hoppings.  For this reason, the 
flat band belonging to the even bands group plays a major role in 
superconductivity. 
\begin{table}[ht]
\caption{The C-C hopping parameters (in eV) for the six odd symmetry bands of  
C$_6$CaC$_6$ are denoted by $t^{(-)CC}_{0n}$ where 
the index $n$ indicates $n$-th neighbour.} 
\begin{tabular}{|c|c|c c|c c| c c|c c|c |} 
\toprule [2pt]
$n$ & $0$ & $1$ & $2$ & $3$ & $4$ & $5$ & $6$ & $7$ & $8$ & $9$\\ [1ex] 
\toprule [1pt]
$t_{0n}^{(-)CC}$ & $\epsilon_{c}^{-}=-1.00$ & $t_1^{-}=3.04$ & $t_1^{'-}=0.92t_1^{-}$ & $t_2^{-}=-0.23$ & $t_2^{'-}=0.92t_2 $ & $t_3^{-}=-0.29$&$t_3^{'-}\approx t_3$ & $t_4^{-}=-0.02$ & $ t_4^{'-}\approx t_4^{-}$ & $t_5^{-}=-0.05$\\ [1ex]
 \bottomrule[1pt]
\end{tabular}
\label{table:shrunk-hoppping-6} 
\end{table}
\begin{table}[ht]
\caption{The C-C hopping parameters (in eV) for the seven even-symmetry bands of C$_6$CaC$_6$  
are denoted by $t^{(+)CC}_{0n}$ where the index $n$ indicates the $n$-th neighbour.  
In the intercalant plane,  Ca-Ca hopping parameters 
are denoted by $t_{0m}^{CaCa}$ where $m$ is $m$-th Ca neighbor of central Ca. 
The modified Ca-C hopping parameter is  $t_{0m}^{CaC}$.} 
\begin{tabular}{|c|c|c c|c c| c c|c c|c |} 
\toprule [2pt]
$n$ & $0$ & $1$ & $2$ & $3$ & $4$ & $5$ & $6$ & $7$ & $8$ & $9$\\ [1ex] 
\toprule [1pt]
$t_{0n}^{(+)CC}$ & $\epsilon_{c}^{+}=-0.60$ & $t_1^{+}=2.94$ & $t_1^{'+}=0.92t_1$ & $t_2^{+}=-0.24$ & $t_2^{'+}=0.92t_2 $ & $t_3^{+}=0.27$&$t_3^{'+}\approx t_3^{+}$ & $t_4^{+}=-0.02$ & $ t_4^{+}\approx t_4^{+}$ & $t_5^{+}=-0.08$\\ [1ex]
\toprule [2pt]
$m$ & $0$ & $1$ & $2$ & $3$ & $4$ & &5&&& \\ [1ex] 
 \bottomrule[1pt]
$t_{0m}^{CaCa}$ &$\epsilon_{Ca}=1.12$&$-0.35$&$0.06$& $0.06$ &$-0.02 $&$0.00$& & &&\\ [1ex]
 \bottomrule[1pt]
$t_{0m}^{CaC}$ &$-$&$0.17$&$-0.14$& $0.08$ &$-0.07$ &$-0.05$&&  && \\ [1ex]
 \bottomrule[1pt]
\end{tabular}
\label{table:shrunk-hoppping-7} 
\end{table}

\section{Superconducting Pairing and States}
\subsection{ Bogoliubov-de Gennes Transformation}
We treat the thirteen band Hubbard model in mean field approximation to 
investigate superconductivity in intercalated bilayer graphene. 
Singlet pairing is considered and, as illustrated in 
Fig.~ \ref{figure:pairing amplitude}, pairing interactions are pictured 
in real space as interactions between 
nearest neighbors on inter- and intra-layer carbon atoms.
This superconducting Hamiltonian can be 
transformed, as for the non-interacting case, to the direct sum of 
two independent superconducting Hamiltonian 
corresponding to odd and even symmetries pseudo-graphene structures 
\begin{eqnarray}
{\hat H}_{su}=\sum_{\vec k} \Lambda ^{\dag }(\vec k)\left( {\begin{array}{*{20}{c}}
{{H_{su}^{+}}(\vec k)}&0\\
0&{  H_{su}^{-}( \vec k)}
\end{array}} \right)\Lambda (\vec k)
\label{eq:trans-Hsu-matrix-text}
\end{eqnarray}
where
$ \Lambda^{\dagger} (\vec k)=\left([\hat c^{\dagger}_{0\uparrow}(\vec k)\hat c^{+\dagger}_{1\uparrow}(\vec k)...\hat c^{+\dagger}_{6\uparrow}(\vec k)\;\hat c_{0\downarrow}(-\vec k)\hat c_{1\downarrow}^{+}(-\vec k)...\hat c_{6\downarrow}^{+}(-\vec k)]~~~[\hat c^{-\dagger}_{1\uparrow}(\vec k)\;\hat c^{-\dagger}_{2\uparrow}(\vec k)...\hat c^{-\dagger}_{6\uparrow}(\vec k)\;\hat c_{1\downarrow}^{-}(-\vec k)\;\hat c_{2\downarrow}^{-}(-\vec k)...\hat c_{6\downarrow}^{-}(-\vec k)]\right)$,
 in which $\hat c^{(\pm) \dagger}_{m\sigma}(\vec k)=
  \frac{1}{\sqrt 2}(\hat c^{\dagger}_{m\sigma}(\vec k)
  \pm\hat c^{\dagger}_{m+6,\sigma}(\vec k))$ 
  and ${H_{su}^{+}}$ and  ${H_{su}^{-}}$ 
  are Hamiltonians of even and odd symmetry pseudo-graphene 
structures respectively; for more information see Appendix C. Decoupling of 
these Hamiltonians means there is no effective pairing between an 
electron in the even sector with one in the odd sector. 
Using the fact that the gap is small on the electronic scale, 
applying perturbation up to second order gives  
quasiparticle energies from
Eq.\ref{eq:trans-Hsu-matrix-text} 
(see ref. [\onlinecite{Rouhollah2018}]) as
\begin{eqnarray}
E_{m,s}^{Q+} (\vec k)= s\left( E_{m}^{+}(\vec k)   + \sum_{n = 1}^7 \frac{{\left| \Delta_{mn}^{+}(\vec k) \right|}^2}{E_{m}^{+}(\vec k) + E_{n}^{+}(\vec k) }  \right),~~~\Delta _{mn}^{+}(\vec k) = \sum_{\alpha  = 1}^9 \Omega _{mn}^{+\alpha} (\vec k)\Delta^{\alpha }_{+} ~~    s =  \pm 1,~~~m=1,2,...7
\label{eq:quasi-su1-spectrum}
\end{eqnarray}
\begin{eqnarray}
E_{m,s}^{Q-} (\vec k)= s\left( E_{m}^{-}(\vec k)   + \sum_{n = 8}^{13} \frac{{\left| \Delta^{-}_{mn}(\vec k) \right|}^2}{E_{m}^{-}(\vec k) + E_{n}^{-}(\vec k) }  \right),~~~\Delta _{mn}^{-}(\vec k) = \sum_{\alpha  = 1}^9 \Omega _{mn}^{-\alpha} (\vec k)\Delta^{\alpha }_{-} ~~    s =  \pm 1,~~~m=8,9,...13
\label{eq:quasi-su2-spectrum}
\end{eqnarray}
Here $(\Delta^{1}_{\pm}~~\Delta^{2}_{\pm}~~\Delta^{3}_{\pm})=[(g_1\Sigma_1\pm g_{1}^{'}\Sigma_{1}^{'})~~(g_1\Sigma_2\pm g_{1}^{'}\Sigma_{2}^{'})~~(g_1\Sigma_3\pm g_{1}^{'}\Sigma_{3}^{'})]$, $(\Delta^{4}_{\pm}~~\Delta^{5}_{\pm}~~\Delta^{6}_{\pm})=[(g_0\Delta_1\pm g_{0}^{'}\Delta_{1}^{'})~~(g_0\Delta_2\pm g_{0}^{'}\Delta_{2}^{'})~~(g_0\Delta_3\pm g_{0}^{'}\Delta_{3}^{'})]$ and $(\Delta^{7}_{\pm}~~\Delta^{8}_{\pm}~~\Delta^{9}_{\pm})=[(g_0\Pi_1\pm g_{0}^{'}\Pi_{1}^{'})~~(g_0\Pi_2\pm g_{0}^{'}\Pi_{2}^{'})~~(g_0\Pi_3\pm g_{0}^{'}\Pi_{3}^{'})]$, where $<ij>$ subscript has been dropped for brevity.  Also 
\begin{eqnarray}
\Omega^{1\pm}_{mn}(\vec k) &=& \mathscr{C}^{*}_{1 } ({E_{m}^{\pm}})\mathscr{C}_{4 }({E_{n}^{\pm}})e^{i\vec k.{{\vec \tau }_1}} + 
\mathscr{C}^{*}_{4} ({E_{m}^{\pm}})\mathscr{C}_{1 }({E_{n}^{\pm}})e^{-i\vec k.{\vec \tau }_{1 }}\nonumber\\
\Omega^{2\pm}_{mn}(\vec k) &=& \mathscr{C}^{*}_{3 } (E_{m}^{\pm})\mathscr{C}_{6 }({E_{n}^{\pm}}){e^{i\vec k.{{\vec \tau }_2}}} + \mathscr{C}^{*}_{6 } ({E_{m}^{\pm}})\mathscr{C}_{3 }({E_{n}^{\pm}})e^{ - i\vec k.{{\vec \tau }_{2}}}\nonumber\\
\Omega^{3\pm}_{mn}(\vec k) &=& \mathscr{C}^{*}_{2 } ({E_{m}^{\pm}}){\mathscr{C}_{5 }}({E_{n}^{\pm}}){e^{i\vec k.{{\vec \tau }_3}}} + \mathscr{C}^{*}_{5 } ({E_{m}^{\pm}}){\mathscr{C}_{2 }}({E_{n}^{\pm}}){e^{ - i\vec k.{{\vec \tau }_3}}}\nonumber \\
\Omega^{4\pm} _{mn}(\vec k) &=& \mathscr{C}^{*}_{2} ({E_{m}^{\pm}}){\mathscr{C}_{6 }}({E_{n}^{\pm}}){e^{i\vec k.{{\vec \delta }_1}}} + \mathscr{C}^{*}_{6 } ({E_{m}^{\pm}}){\mathscr{C}_{2 }}({E_{n}^{\pm}}){e^{ - i\vec k.{{\vec \delta }_{1}}}} \nonumber\\
\Omega^{5\pm}_{mn}(\vec k) &=& \mathscr{C}^{*}_{1 } ({E_{m}^{\pm}}){\mathscr{C}_{5 }}({E_{n}^{\pm}}){e^{i\vec k.{{\vec \delta }_2}}} + \mathscr{C}^{*}_{5 } ({E_{m}^{\pm}}){\mathscr{C}_{1 }}({E_{n}^{\pm}}){e^{ - i\vec k.{{\vec \delta }_2}}}\nonumber\\
\Omega^{6\pm}_{mn}(\vec k) &=& \mathscr{C}^{*}_{3 } ({E_{m}^{\pm}}){\mathscr{C}_{4 }}({E_{n}^{\pm}}){e^{i\vec k.{{\vec \delta }_3}}} + \mathscr{C}^{*}_{4} ({E_{m}^{\pm}}){\mathscr{C}_{3 }}({E_{n}^{\pm}}){e^{ - i\vec k.{{\vec \delta }_3}}}\nonumber\\
\Omega^{7\pm} _{mn}(\vec k) &=& \mathscr{C}^{*}_{3} ({E_{m}^{\pm}}){\mathscr{C}_{5}}({E_{n}^{\pm}}){e^{i\vec k.{{\vec \delta }_1}}} + \mathscr{C}^{*}_{5 } ({E_{m}^{\pm}}){\mathscr{C}_{3 }}({E_{n}^{\pm}}){e^{ - i\vec k.{{\vec \delta }_1}}}\nonumber\\
\Omega^{8\pm} _{mn}(\vec k) &=& \mathscr{C}^{*}_{2 } ({E_{m}^{\pm}}){\mathscr{C}_{4 }}({E_{n}^{\pm}}){e^{i\vec k.{{\vec \delta }_2}}} + \mathscr{C}^{*}_{4} ({E_{m}^{\pm}}){\mathscr{C}_{2}}({E_{n}^{\pm}}){e^{ - i\vec k.{{\vec \delta }_2}}}\nonumber\\
\Omega^{9\pm} _{mn}(\vec k) &=& \mathscr{C}^{*}_{1 } ({E_{m}^{\pm}}){\mathscr{C}_{6 }}({E_{n}^{\pm}}){e^{i\vec k.{{\vec \delta }_3}}} + \mathscr{C}^{*}_{6 } ({E_{m}^{\pm}}){\mathscr{C}_{1 }}({E_{n}^{\pm}}){e^{ - i\vec k.{{\vec \delta }_3}}}.
\label{eq:band-orderparameter-coef}
\end{eqnarray}
Here  ${\mathscr{C}_{m }}({E_{n}^{-}})$ is the $m$th component of $n$th column eigenvector of $H^{-}$  and ${\mathscr{C}_{m }}({E_{n}^{+}})$ is  $(m+1)$th component  of $n$th eigenvector of $H_{c}^{+}(\vec k)$. Band order parameters $\Delta _{mn}^{+}(\vec k)$  are defined such that first electron is in the $m$th band and second electron is in the $n$th band of $H_{c}^{+}(\vec k)$, also  $\Delta _{mn}^{-}(\vec k)$  is defined such that first electron is in the $m$th band and second electron is in the $n$th band of $H^{-}(\vec k)$.
Note that an electron in the $m$th band of  $H_{c}^{+}(\vec k)$  and an electron in $n$th band of $H^{-}(\vec k)$ cannot be paired; {\it i.e} for this case $\Delta_{mn}^{\pm}(\vec k)=0$.

The Bogoliubov-de Gennes transformation used in Eq.~\ref{eq:trans-Hsu-matrix-text} shows 
 that pairing amplitudes should be $\Delta_{\pm}^{\alpha}=<\hat{c}^{\pm}_{\alpha,i}\hat{c}^{\pm}_{\alpha,j}>$
 which implies that all inter- and intra-layer pairing amplitudes in real space are equal, 
$g_0=g_0^{'}$ and $g_1=g_1^{'}$.
 This restriction makes the matrix gap equations hermitian and implies that band order 
parameters, $\Delta_{mn}^{\pm}(\vec{k})$ can be interpreted physically as pairing of electrons in different bands with pairing interaction $g_0^{\pm}$. In this limit $\Delta_{mn}^{\pm}(\vec{k})$ is equal to the product of band Green function and $g_{0}$, 
\begin{eqnarray}
\Delta_{mn}^{\pm}(\vec k)=g_{0}^{\pm}\langle \hat{d}^{\pm\uparrow}_{m}(\vec k)\hat{d}^{\pm\downarrow}_{n}(\vec k)\rangle
\label{eq:order-green}
\end{eqnarray}
where $\hat{d}^{\pm\sigma}_{i}(\vec k)=\sum^{7}_{m=1}{\mathscr{C}}^{\pm*}_{m} 
  (E_{i}(\vec k)) 
  \hat{c}^{\sigma}_{m}(\vec k)$ annihilates an electron with spin $\sigma$ in the 
$i$th even or odd bands with energy $E_{i}^{\pm}(\vec k)$.

 \subsection{Two Gap Superconducting Pairings and States}
 The linearized gap equation can be decoupled  by minimizing free energy with respect to 
nearest neighbor  pairing, or equivalently with respect to  $\Delta_{\pm}^{\alpha}$,
for more detail see Appendix C. Minimization of free energy 
   with respect to $\Delta_{\pm}^{\alpha}$ gives
\begin{eqnarray}
\left[ \begin{array}{*{20}{c}}
A^{\pm}&B^{\pm}&B^{\pm}\\
B^{\pm}&C^{\pm}&D^{\pm}\\
B^{\pm}&D^{\pm}&C^{\pm}
\end{array} \right]\left( \begin{array}{*{20}{c}}
\Sigma_{i}\pm \Sigma_{i}^{'}\\
\Pi_{i}\pm \Pi_{i}^{'}\\
 \Delta_{i}\pm \Delta_{i}^{'}
\end{array} \right) =-\frac{1}{g_{0}^{\pm}}\left( \begin{array}{*{20}{c}}
\Sigma_{i}\pm \Sigma_{i}^{'}\\
 \Pi_{i}\pm\Pi_{i}^{'}\\
\Delta_{i}\pm \Delta_{i}^{'}
\end{array} \right)
\label{eq:matrix-form-gap-eq-six-seven-bands-text}
\end{eqnarray}
in which $A$, $B$, $C$ and $D$ matrices have been introduced as
\begin{eqnarray}
A^{\pm} = \left[ \begin{array}{*{20}{c}}
\Gamma^{\pm} _{11}&\Gamma^{\pm} _{12}&\Gamma^{\pm} _{12}\\
\Gamma^{\pm} _{12}&\Gamma^{\pm} _{11}&\Gamma^{\pm} _{12}\\
\Gamma^{\pm} _{12}&\Gamma^{\pm} _{12}&\Gamma^{\pm} _{11}
\end{array} \right],\;    C^{\pm}= \left[ \begin{array}{*{20}{c}}
\Gamma^{\pm} _{44}&\Gamma^{\pm} _{45}&\Gamma^{\pm} _{45}\\
\Gamma^{\pm} _{45}&\Gamma^{\pm} _{44}&\Gamma^{\pm} _{45}\\
\Gamma^{\pm} _{45}&\Gamma^{\pm} _{45}&\Gamma^{\pm} _{44}
\end{array} \right],\;\;
B^{\pm} = \left[ \begin{array}{*{20}{c}}
\Gamma_{14}&\Gamma^{\pm} _{15}&\Gamma^{\pm} _{15}\\
\Gamma^{\pm} _{15}&\Gamma^{\pm} _{14}&\Gamma^{\pm} _{15}\\
\Gamma^{\pm} _{15}&\Gamma^{\pm} _{15}&\Gamma^{\pm} _{14}
\end{array} \right],\;\;    D^{\pm} = \left[ \begin{array}{*{20}{c}}
\Gamma^{\pm} _{47}&\Gamma^{\pm} _{48}&\Gamma^{\pm} _{48}\\
\Gamma^{\pm} _{48}&\Gamma^{\pm} _{47}&\Gamma^{\pm} _{48}\\
\Gamma^{\pm} _{48}&\Gamma^{\pm} _{48}&\Gamma^{\pm} _{47}
\end{array} \right].
\label{eq:matrix-gap-elements}
 \end{eqnarray}
wherein, $\Gamma$ matrix elements are given by
   \begin{equation}
\Gamma _{\beta \alpha }^{\pm} = \frac{1}{N}\sum_{\vec k} 
   \sum_{i }\sum_{j } \frac{\tanh (\frac{{ E_i^{\pm}}}{ 2 k_{B}T})}{E_{j}^{\pm}(\vec k) 
  + E_{i}^{\pm}(\vec k) } \left( \Omega _{ij}^{\pm\alpha} (\vec k)\Omega _{ji}^{\pm \beta }(\vec k) 
  + \Omega _{ji}^{\pm\alpha} (\vec k)\Omega _{ij}^{\pm \beta }(\vec k) \right).  
\label{eq:gap-Equation-Hc-text}
\end{equation}
Equation \ref{eq:matrix-form-gap-eq-six-seven-bands-text}
can be interpreted as two independent gap equations for odd (minus sign) and 
even (plus sign) pseudo-graphene systems. 
The impact is that superconductivity can be established independently in two distinct sectors
this system. In the next section we numerically inspect which of these pseudo-graphene sectors, 
odd or even, play major rules of superconductivity.

Gholami {\it et al.}\cite{Rouhollah2018} solved such gap equations for Li-decorated
single layer graphene. The gap equations in Eq.~\ref{eq:matrix-form-gap-eq-six-seven-bands}
 are the same form so they can be solved similarly. The $A$, $B$, $C$, and $D$ matrices 
have identical structures, 
hence they share eigenvectors: 
$V_{s}^{T}=(1\;\;1\;\;1)$, $V_{d_{xy}}^{T}=(1\;\;-1\;\;0)$, and 
$V_{d_{x^2-y^2}}^{T}=(1\;\;1\;\;-2)$, where the latter two are degenerate. Their eigenvalues,
in obvious notation, are 
\begin{eqnarray}
{{a}_s^{\pm}}& =&{\Gamma^{\pm} _{11}} + 2{\Gamma^{\pm} _{12}}\;,\;\;  
{{b}_s^{\pm}}={\Gamma^{\pm} _{14}} + 2{\Gamma^{\pm} _{15}}\;,\;\; 
{{c}_s^{\pm}}={\Gamma^{\pm} _{44}} + 2{\Gamma^{\pm} _{45}}\;,\;\; 
{{d}_s^{\pm}}= {\Gamma^{\pm} _{47}} + 2{\Gamma^{\pm} _{48}}\nonumber\\
{a_d^{\pm}} &=&{\Gamma^{\pm} _{11}} - \;\;{\Gamma^{\pm} _{12}}\;,\;\; {b_d^{\pm}} ={\Gamma^{\pm} _{14}} - \;\;{\Gamma^{\pm} _{15}}
\;,\;\;  {c_d^{\pm}} ={\Gamma^{\pm} _{44}} - \;{\Gamma^{\pm} _{45}}\;,\;\;   {d_d^{\pm}} ={\Gamma^{\pm} _{47}} -\; {\Gamma^{\pm} _{48}}.
\label{eq:gapequation-matrix-unhermit-elements}
\end{eqnarray}

Similar to decorated single layer graphene \cite{Rouhollah2018}, for each of  the gap equations given by
 Eq.~\ref{eq:matrix-form-gap-eq-six-seven-bands} there are nine independent solutions. 
The first three superconducting states with island (localized) character  can be expressed
in compact form as
\begin{eqnarray}
[\Psi_{\Delta^{\pm}_{sy}}^{0}]^{T}=[0,\quad V_{{sy}},\quad -V_{sy}], 
   ~~~J_{sy}^{0\pm}= c_{sy}^{\pm} - d_{sy}^{\pm}
\label{eq:gap-general-function-island-text}
\end{eqnarray}
 where  $V_{sy}$ refers to one of the $V_s$, $V_{d_{xy}}$ or $V_{d_{x^2-y^2}}$-wave symmetries.  Pairing in these phases cannot propagate.
The other six superconducting states of 
Eq.~\ref{eq:matrix-form-gap-eq-six-seven-bands}
 have the explicit form
\begin{eqnarray}
[\Psi_{\Delta^{\pm}_{sy}}^{l}]^{T}=[\alpha_{sy}^{l,\pm} V_{sy}\quad, V_{sy}\quad V_{sy}]
\label{eq:gap-general-function}
\end{eqnarray}
where
\begin{eqnarray} 
\alpha _{sy}^{l,\pm} =  \frac{J_{sy}^{l,\pm} - c_{sy}^{\pm} - d_{sy}^{\pm}}{b_{sy}^{\pm}},~~
J_{sy}^{l,\pm} =\frac{1}{2}\left({ {a_{sy}^{\pm}} + {c_{sy}^{\pm}} + {d_{sy}^{\pm}} 
             +(-1)^l \sqrt {8(b_{sy}^{\pm})^2 + {{\left[ {{c_{sy}^{\pm}} + {d_{sy}^{\pm}} 
             -  {a_{sy}^{\pm}}} \right]}^2}} } \right),~~l=1,2.
\label{eq:quad-eigenvalue}
\end{eqnarray}
Here $c_{sy}=c_{s}^{\pm}, c_d^{\pm} $, $b_{sy}=b_{s}^{\pm}, b_d^{\pm}$, $d_{sy}=d_{s}, d_d^{\pm} $ and 
$J_{sy}^{l,\pm}=-\frac{1}{g_0^{\pm}}$ for each symmetry, and $+$ superscript refer to the even 
sector and $-$ superscripts to the odd sector. In each of above categories, ${d_{{x^2} - {y^2}}}$ 
and ${d_{xy}}$ phases are degenerate. 	Similar to decorated single layer graphene, only three of 
solutions for which $l=2$ are physically reachable in the framework of mean field theory. 
In the limit of pristine bilayer  graphene, these three states convert to usual s-wave and d-wave symmetries.
In the sec. V, for odd and even sectors we illustrate from numerical solutions which of these three phases are 
dominant.
\subsection{Flat band(s) Superconductivity: Strong interlayer coupling}
To make a rough estimate and  provide mathematical insight into the physics, one can diagonalize normal state Hamiltonian of pristine bilayer graphene in the mini-Brillouin zone of C$_6$CaC$_6$.
As shown in Fig. \ref{figure:fiting-band structure of graphene}, two conduction bands
corresponding to odd and even sector are weakly dispersive near the Fermi energy along $\Gamma \to M$, which seems to play a major role in the formation of superconducting Cooper's pairs.

In the case of pristine bilayer graphene, odd (-) and even (+) so  called  flat bands are the minimum of ($E_{\alpha,2}^{\pm}(\vec{k}),E_{\beta,2}^{\pm}(\vec{k})$) along different high symmetry paths that are given by Eqs.~\ref{eq:alpha-branch} and \ref{eq:beta-branch}, where their Bloch wave function, viz. Eq.~\ref{eq:pristine-eigenvectors}, are similar to those of ref. \cite{Rouhollah2018}. They have linear combination of $d_{x^2-y^2}$ and $d_{xy}$ character and are responsible for superconducting pairing $d_{x^2-y^2}$ and $d_{xy}$. 

One can ask: what is so special about these flat bands?
To address this question, we return to the matrix gap equation of Eq.~\ref{eq:gap-Equation-Hc-text}.
The right hand side contains the product of a form factor given by
$\Omega_{ni}^\alpha (\vec k)\Omega_{ni}^{*\beta}(\vec k)
                           +\Omega_{ni}^\beta   (\vec k)\Omega_{ni}^{*\alpha}(\vec k)$
 and the thermal occupation factor over the energy denominator i.e.
 $\frac{\tanh (\frac{{ E_n}}{ 2 k_{B}T})}{E_{n}(\vec k)  + E_{i}(\vec k) }$.
The form factor is a function of the Bloch wave coefficients of normal state Hamiltonian.
Using Eqs.~\ref{eq:pristine-eigenvectors},
\ref{eq:pristine-Bloch phase}  one can investigate that in the limited case of pristine bilayer  graphene at the nearest neighbor approximation, these Bloch wave  coefficients are the same for both  sectors and this is almost for the next neighbor approximation. As such,
it is independent of chemical potential $\mu$.
Thus the form factor is the same for the both odd and even sector of band structures. 
Since $\frac{tanh(x)}{x}\to 1$ as $x\to 0$, when one of the conduction odd or even flat bands and so their corresponding valance bands becomes completely flat at the Fermi level then 
$\frac{\tanh(\frac{{\beta E_i}}{2})}{E_{i}(\vec k)+E_{j}(\vec k)}\to \frac{\beta}{4}$,
  where $E_{i}(\vec k)$ or $E_{j}(\vec k)$ are one of these flat bands. 
 In this case the dominant contribution comes from these mutual conduction and valence 
flat bands, and one can show that all gap equation block matrix elements in 
Eq. \ref{eq:matrix-gap-elements} are equal to $A^{\pm}.$ In this event, depending on 
whether the flat bands belong to the odd or even sector, one can use 
Eqs.~\ref{eq:band-orderparameter-coef}, \ref{eq:gap-Equation-Hc-text}, and
\ref{eq:pristine-eigenvectors} to show that 
 \begin{eqnarray}
 \Gamma_{11}^{\pm}\to \frac{\beta_c}{9},~~~~\Gamma_{12}^{\pm}\to -\frac{\beta_c}{36},~~(\beta_c=\frac{1}{k_{B}T_c}).
\end{eqnarray}  
Cooper pair interaction potentials $g_0$ of d-wave symmetry, {\it i.e.} 
$g_0^d$ and s-wave symmetry $g_0^s$,  are given by
 \begin{eqnarray}
 g_{0}^{d}&=&\frac{1}{3(\Gamma_{11}^{\pm}-\Gamma_{12}^{\pm})}=\frac{12}{5}k_BT_c\nonumber\\
  g_{0}^{s}&=&\frac{1}{3(\Gamma_{11}^{\pm}+2\Gamma_{12}^{\pm})}=6k_BT_c
\end{eqnarray}
In this case $\Gamma_{12}^{\pm}<0$ and $g_0^d$ is less than  $g_0^s$, so 
$d$-wave symmetry is dominant, with an extraordinary decrease in pairing potential interaction 
proportional to the critical temperature. This ``ultra'' decrease of pairing interaction can 
explain the importance of the flat bands in the formation of Cooper pairs in twisted bilayer graphene. 
 Here, another  point that can be deduced from mathematical calculations is that in the limit of strong interlayer hopping when 
inter-layer hoppings tends to minus (plus) of intra-layer hoppings, as one can see from 
Appendix D, all of the six bands of even (odd) sector become flat while the other sector 
bandwidths increases. Then one can show that the gap matrix elements are
\begin{eqnarray}
\Gamma_{i,j}\to\beta_c \delta_{ij};~~~~g_0=k_BT_c
\end{eqnarray}
 and so all possible superconducting symmetries are degenerate 
with pairing potential $g_0=k_BT_c$. 
 \section{Numerical Results}
\subsection{General features}
To know in the variety of doping regimes which of the pairing symmetries 
(distorted $s$-wave or $d$-wave) are dominant, and also to 
inspect in the which sectors of the band structure Cooper pairs with the lowest
pairing potential can constructed,  
   superconducting gap equations of odd and even sectors {\it i.e.} 
Eq.~\ref{eq:matrix-form-gap-eq-six-seven-bands-text} are solved numerically. The result
is shown in Fig.~\ref{figure:u-mu}.

 Similar to Li intercalated single layer graphene (ref. [\onlinecite{Rouhollah2018}]), 
at moderate doping,  $d$-wave superconductivity  always dominates in both sectors  
of the C$_6$CaC$_6$ band system. Distorted $s$-wave symmetry only survives at high doping levels.
 For odd and even sector flat bands, the density 
of states peaks at the $M$ critical point. This point, for the odd flat band, is located 
about 0.5 eV above the Fermi level, and about 0.44 eV below the Fermi level for the case of the 
even flat band. Exploration of each of these flat bands can be  engineered 
by applying a gate voltage on the bilayer via a change in the chemical potential $\mu$. 
As illustrated in Fig.~\ref{figure:u-mu},  
around $\mu =0$ at $T_c=1K$ {\it i.e.} C$_6$CaC$_6$ is not affected by gate voltage. 
Dominant $d$-wave symmetry pairing phases are degenerate, so pairing can arise in both 
odd and even flat bands. This occurs when the averaged density of states of both bands are 
the same near the Fermi level. When the odd flat band is dominant near Fermi level due to 
electron doping,  $d$-wave pairing dominates in this band, with the minimum of the pairing
interaction energy ($g_0^{-}=0.35$) corresponding to the $M$ point near $\mu=0.5 eV$. 

\subsection{Hole doping}
\begin{figure}
\centerline{\epsfig{file=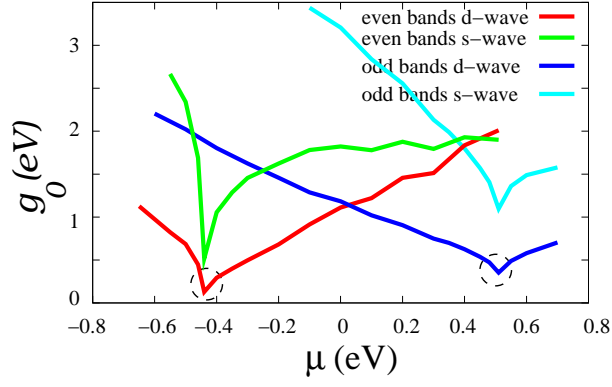 ,width=8.0cm,angle=0}}
\caption{Pairing interaction potential $g_0$ versus  chemical potential $\mu$. 
Both odd (six bands) and even (seven bands) symmetry solutions are shown for
$T_c=1K$. Their corresponding $d$-wave phases are degenerate around $\mu=0$ and pairing
separately contributes to 
a phase with two differing gaps. For both odd and even flat bands, $d$-wave symmetries 
are dominant at moderate doping, while at high doping a phase transition from $d$-wave 
to distorted $s$-wave occurs. This transition can be seen for the even flat band near 
$\mu=0.5 eV$. Hole doping causes the $d$-wave symmetry pairing to prevail between electrons 
in the even flat band case, with a minimum of pairing potential interaction $g_0^{+}=0.12 eV$ 
where there occurs a critical M point at -0.44 eV shown by a  dashed circle. Vice versa, 
electron doping leads to $d$-wave symmetry pairing between electrons in the odd flat band,
with  a minimum of pairing potential interaction at $g_0^{+}=0.12 eV$ that occurs at the 
critical M point at -0.44 eV shown by a  dashed circle.}
 \label{figure:u-mu} 
\end{figure}

Hole doping by gating, by Ca$\rightarrow$Na substitution, or by Ca deficiency, 
leads to the situation where the even flat band dominates. Dominant $d$-wave
symmetry occurs in this band with the lowest pairing energy ($g_0^{+}=0.12$), again at  
the critical $M$ point at $\mu=-0.44 eV$. From Fig.~\ref{figure:u-mu} it can be seen 
that when the even flat band is dominant (hole doping), the pairing potential $g_0$ of 
$d$-wave symmetry emerging from this band is less than the case that dominant $d$-wave 
symmetry  occurs in the odd flat band (by electron doping). For example, with a factor of
one-third at their critical $M$ point {\it i.e.} $g_0^{+}=\frac{1}{3}g_0^{-}$, that means 
that when the even flat band reaches near the Fermi level, reduction of bandwidth due to both 
interlayer C-C interaction($H_{12}$) and C-Ca layer interactions lead to a sharp increase 
in the density of states. While both C-C and Ca-C interlayer interaction decrease the pairing 
potential $g_0$, one can numerically inspect that reduction of the bandwidth due to graphene 
interlayer interaction, more affected the energy of pairing in the even flat band than 
Ca-C layers interaction.
 \begin{figure}
\centerline{\epsfig{file=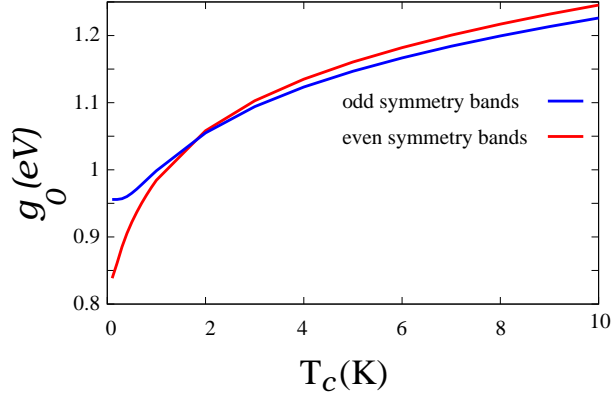 ,width=8.0cm,angle=0}}
\caption{Critical temperature $T_c $ versus pairing potential $g_0$. Shown are the 
dominant $d$-wave solutions of the odd (six bands) and even (seven bands) symmetry
 superconducting gap equations. At $T_c=2K$ the phases are degenerate, and pairing
separately contributes to a phase with two gaps.}
 \label{figure:twogap} 
\end{figure}

In the case of Ca intercalated bilayer graphene, for a given $T_c$ the pairing 
interaction potential  $g_0$
(proportional to  superconducting gap energy, $|\Delta|^2$) for dominant $d$-wave phases 
of the odd and even superconducting gaps are illustrated in Fig. \ref{figure:twogap}. 
It is evident that for a given critical temperature $T_c$,  superconductivity can be 
single gap or two gap and  dominant superconducting pairings can occur  between electrons 
in the odd $H^{-}$, or even $H^{+}_c$, sector.

\section{Discussion and Summary}

 Discovery of new superconducting phases, often at low temperature, has been one of the active achievements in recent decades. Superconductivity  in  the lithium-coated single  layer graphene with respect to  to the case of calcium-decorated single layer is more  capable  \cite{Ludbrook2015} and also reported experimentally, whereas this situation is opposing  in the  bilayer graphene. 
     Superconductivity has been reported in the Ca-intercalated bilayer graphene   around $T_c=4K$  while Li-intercalated bilayer graphene is not superconductor \cite{Satoru2016,Chapman2016}.
      Experimental fabrication of Li and Ca-intercalated bilayer graphene has been reported in the ref.\cite{Caffrey2016} and  ref.\cite{Kanetani2012} respectively. Li and Ca atoms are suggested to intercalate between graphene layers with an ordered structures similar to that of bulk GICs like LiC$_6$  i.e. two graphene layers are AA-stacking.      
      That  it demonstrates the  vitally important role  of intercalant inter-band in  the formation of  superconductivity Cooper pairs.  

Most theoretical microscopic models of pristine honeycomb bilayer superconductivity concentrate on the more stable AB stacking of bilayer graphene.\cite{Vucicevic2012,James2014}. To our knowledge, there are few studies that focus on pristine AA stacked  bilayer graphene, \cite{Alidoust2019} and no  analogous studies that concentrate on intercalated bilayer graphene.
Based on {\it ab initio} calculations of
electron-phonon coupling, anisotropic Migdal-Eliashberg theory has been applied by some authors \cite{Margine2016,Durajski2019}to give strong evidence that this system is a phonon mediated  
two-gap superconductor with predicted $T_c$ around 7K.
Recently unconventional superconductivity up to $T_c=1.7$K has been reported 
in gated twisted bilayer graphene where the layers are rotated relative to each other 
by a magic angle of $1.1 ^{\circ}$. Superconductivity in this low doping regime of band filling 
cannot be addressed within the framework of conventional electron-phonon coupling based on
Migdal's adiabatic approximation.  This discovery has opened speculation that this
superconducting behavior may shed light on other systems in which superconductivity arises
from an insulating phase.\cite{Cao2018}
This development also highlights studies such as ours, which does not rely on the
mechanism, but instead on more general pairing concepts and the specific electronic structure. 
 
 Many theorists have suggested that exotic  superconductivity gaps  arise in some materials are related to a peculiarity of the normal-state band structure. In this kind  of issue  angle-resolved photoemission spectroscopy (ARPES)  extensively has been applied to analysis of the normal state band structure. To determine structural and electronic properties of material, tight binding model in addition to DFT calculation has been used to interpret experimental  results achieved from ARPES. 
Following this point of view, 
extended Hubbard  model has been  used here to address superconductivity  of Ca- intercalated bilayer graphene.

The main results  are achieved in two 
 steps: first, for the normal-state (non-interacting part) a more realistic effective tight-binding model with two decoupled 
 symmetry sectors is derived, with the parameters determined by a 
 fitting to the DFT band structure; second, the dominant 
superconducting pairing channels are discussed based on a mean-field 
 treatment of a Hubbard model obtained by adding (effective) attractive 
 interactions between the electrons. The summary, results and comparisons are presented below.
   
 \subsection{Non interacting Part: Normal state}
In the first part of this manuscript we have taken the advantages of mirror symmetry operation on the AA-stacked BlG through the central plane and generalized it to include intercalated bilayer graphene.
\subsubsection{AA-stacked pristine bilayer graphene: Two kinds of quasiparticles}
The honeycomb lattice structure  makes  quasi-particles in the single layer graphene behave as a massless Dirac particles at low energies  that provide a proper platform to examine the  characteristic
 effects of QED, such as the Klein paradox and Zitterbewegung, which were never
observed in particle physics.
In addition to relativistic nature of quasiparticles in single layer graphene,  they exhibit extra  aspects  of  such behaviors in AA-stacked BlG. Interlayer coupling causes the bilayer graphene to exhibit properties that are not observed in the  single layer graphene. In QED, the 4-component ''Dirac spinor'' (Dirac representation) 
 decomposes into two irreducible representations, acting only on two 2-component right 
and left hand ``Weyl spinors.'' There is an pedagogically useful mathematical similarity 
between the Schrödinger equation of AA-BlG in these two representations  and the Dirac equation. The non-interacting AA-stacked Hamiltonian is invariant under mirror symmetry, 
leads to division of the  AA-BlG band structure into two
even and odd  sectors characterized by eigenvalues of mirror operation $h=\pm 1$ (analogous to two decouple Weyl equations for massless relativistic chiral particles).   Each of these sectors  describes a graphene-like structure i.e. $H^+$ and $H^-$. 
In this notion as has been shown in the Fig.~\ref{figure:AA-stacked-amplitude}c, up and down pseudo-spin (h-Pspin) of irreducible blocks  of  AA-BlG Hamiltonian viz. $H^+$ and $H^-$, consists of two  electrons with the same spin which each one  located at similar sub-sites in the opposing layers. These quasi-particles describe by an additional  index has been called ``cone index''. Here we refer to this index as h-chirality index.
 According to this notion, one can describe Dirac cones in AA-BlG shown in the Fig. \ref{figure:AA-stacked-amplitude}(b) with two kind of chirality with respect to   
asymmetric in such a way that the structure and its vertical 
 ({\bf``v-chirality''}) and horizontal ({\bf``h-chirality''}) mirror image are not superimposable.
 This chirality (h-chirality) is a general aspect of AA-BlG quasi-particles
that holds for general hoppings and over the entire Brillouin zone and it is unrelated to the helicity operation. This is in contrast to the famous graphene chirality (helicity) that occurs just at low energies near the Dirac cones.
 
Physically,  AA-stacked BlG can be interpreted as a ``single layer honeycomb lattice''  that 
instead of  $1e^-$ charge carriers, there are two types of fermionic quasi-particles 
with $2e^-$ charge,  moving 
through it 
that differ in  a quantum number called cone index (h-chirality). Quasi-particles with 
different $h$-chirality  don't interact but move independently. Also $(\pm 1)$ $h$-chiral 
quasi-particles have $\pm\gamma_0$ on-site energies (similar to the positive and negative energy of the particles and the anti-particles in the QED).
Hopping of quasi-particles with (+1) $h$-chirality constructs the even sector of band 
structure  while  the odd sector made by (-1) $h$-chiral quasiparticles. 
Near the Dirac cone points, quasi-particles with ($\pm 1$) $h$-chirality are moving 
with Fermi velocities  $v_f^{\pm}$. 
One can 
distinguish quasi-particles with the same chirality (v-chirality) and different  
cone index (h-chirality) from their velocities which in the case of strong interlayer coupling  could be observed experimentally.
 
\subsubsection{Intercalated AA-staked bilayer graphene: Mirror symmetry operation  advantage}
Based on mean-field treatment of an extended Hubbard model,  a realistic 
thirteen band tight binding  model, has been constructed to include the case  of experimentally  observed structures such as Ca intercalated bilayer graphene,  where its parameters are determined by a fitting to DFT band structure. We followed the notion that Calcium doped bilayer graphene as the thinest limit of graphite intercalation compounds  (Fig.\ref{figure:pairing amplitude}(a)). 

In  our previous work, the effects of Li-decoration on structural and electronic band structure of single layer graphene has been demonstrated in details and also symmetry character of the band-branches illustrated as well as  the possible superconducting phases of lithium decorated single layer graphene LiC$_6$ were obtained analytically and analyzed.\cite{Rouhollah2018}. 
The Brillouin zone (BZ) of these structure is one third of that of graphene, with the Dirac points folded back to the $\Gamma$ point. In this mini-BZ, the two $\pi$ bands of (pristine) graphene folds to six branches and  their different symmetries (d+id, s,...) are also separated as illustrated in Fig. 2 of 
ref.~[\onlinecite{Rouhollah2018}]. 
Generalization of these results to include intercalated bilayer graphene (IBlG) 
has strong advantages and provides 
additional insight into its physical properties. This
is possible through 
decoupling of  normal and superconducting Hamiltonians of IBlG into two independent  corresponding single layer pseudo-graphene Hamiltonians, coupled only by a common chemical potential.

Similar to the pristine AA-staked BlG, accounting for  symmetries of Bloch wave coefficients, the $13\times13$ Hamiltonian 
of IBlG converts, by mirror symmetry, to  two decouple
sectors: an $7\times7$ even symmetry sector $H_{c}^{+}$ with involvement of the intercalant (coated single layer
 pseudo-graphene)
and the $6\times 6$ odd sector $H^{-}$,
for which the intercalant provides only renormalized hopping amplitudes and break down symmetry of hopping integrals (six bands shrunken single layer  pseudo-graphene). Therefore, all previous discussions about $2e^{-}$ charge, h-Pspin and chirality  of  quasiparticles  in pristine AA-stacked BlG are extended to include IBlG. 

Periodic perturbation  of graphene layers potential  due to ordered   Intercalant atoms causes hopping integrals symmetry to break and so two distinct  gaps of size $E_{g}^{\pm}=2|t_{11}^{\pm}-{t'}_{11}^{\pm}|$   open at the Dirac point (folded to $\Gamma$ point) of each of the even and odd sector pseudo- graphene structures. These two gaps are characteristic of AA-IBlG.
 Knowing the size of these energy gaps, one can find the difference in the first nearest neighbor intra and inter-layer hopping parameters symmetry breaking  i.e. ($\Delta t=t_{11}-{t'}_{11}$) and ($\Delta \gamma=t_{12}-{t'}_{12}$). 
 In the case of Li- intercalated BlG, experimental ARPES spectra (Fig.~4 Ref.{{\cite{Caffrey2016}}}) shows two distinct gaps of wide $E_{g}^-=0.20eV$ and $E_{g}^+=0.46eV$.  We slightly correct the discussion  that has been stated in the  [Sec. III, sub-sec.C of Ref.{\cite{Caffrey2016}}] about the relation between these two gaps and symmetry breaking interlayer coupling parameters. 
 
 In the case of Li- intercalated BlG, Li-s orbital is fully ionized and Li-C hybridization is negligible, so the odd and even sector band structure are similar to the band structure of  pristine shrunken- graphene $C_6$. The difference of course is due to $\pm$ sign that appears in the even and odd sectors between intra and inter layer hopping terms which  leads to  different bandwidth.  The even and odd sectors  Schr\"odinger equation solved analytically (or nearly so).  From the beginning, the Hamiltonian is generalized to incorporate 
several broken symmetries, including the on-site energies, 
hopping integrals, and bonds lengths (geometry). Due to this generalization,
 it can be used to  
obtain analytic dispersion energies of not only C$_6$CaC$_6$, but also related
graphene-like structures such as B$_{3}$N$_{3}$CaB$_{3}$N$_{3}$.
\subsubsection{Tight Binding Parametrization of Ca- intercalated bilayer graphene  from DFT }
Dividing the thirteen bands into seven even-symmetry bands and six odd-symmetry bands, considerably  facilitates tight Binding Parametrization  from DFT. We used up to nine neighbor approximation  tight binding  model considering the symmetry breaking of bond length and hopping integral parameters across different direction of hexagons. The main problem with DFT data is that the odd and even sector data bands are not separated.     But by inspecting the DFT band structure  
and  to be careful in analytical calculations (knowing that odd sector does not affected by Ca-C coupling directly and so remain graphen like),  the odd and even sector DFT flat bands  can easily be distinguished. Emerging of two distinct gaps in the  Dirac point of both sectors is the other guide to perform fitting. 
The reduced fitting parameters are given in 
Tables \ref{table:shrunk-hoppping-6} and \ref{table:shrunk-hoppping-7} with results are shown in the Fig.~\ref{figure:fiting-band structure of graphene}.
 
 \subsection{Interacting Part: Superconductivity}
SA show that electron transport across a barrier must conserve the cone index, a 
consequence of the Klein tunneling behavior in AA-stacked BlG. Here it is discussed  that Cone index footprint can be traced also in the formation of superconducting Cooper pairs.
Due to this index, the salient differences  are emerged between the Cooper pairs in single layer graphene and AA-stacked bilayer graphene.  The two types of  even and odd symmetry superconductivity are predictable in AA-stacked bilayer graphene. 
 \subsubsection{Odd/ Even Superconducting  Gap equations and Symmetry Phases}
   Similar to the normal state Hamiltonian, the superconducting Hamiltonian is also block diagonalized into two  sectors. Each of these two blocks represents the superconducting Hamiltonian of the even and odd single layer pseudo-graphene structures.   
  The impact is that superconductivity in AA-stacked BlG can be established independently in two distinct band structure sectors. 
  Two  quasi particles  (i.e. ``four electrons'') just with the same
h-chirality (i.e. cone index)  can  team up to build a Cooper pair (Fig.~\ref{figure:AA-stacked-amplitude}c). 
In the other words, pairing  in bilayer graphene arises between quasiparticles inside of the   coated  single layer
pseudo-graphene $H^{+}_{c}$ band structure (even sector-superconductivity) 
 or   uncoated shrunken  single layer pseudo-graphene $H^{-}$ (odd sector-superconductivity) separately; even-odd pairing  is impossible without further symmetry breaking.
 
 Two distinct  superconductivity gap equations corresponding to 
$H_{c}^{+}$ and $H^{-}$ single layer pseudo-graphene structures emerged from minimization of 
the free energy. 
 These behaviors show that general aspects of superconductivity 
in the (Li-)decorated single layer graphene\cite{Rouhollah2018} and (Ca-) intercalated bilayer 
graphene are similar, and their behaviors are different primarily in the probability 
of two gap superconductivity in the bilayer structures. A difference of course is that 
interlayer interaction becomes a key factor; the Cooper pairs in AA-stacked BlG instead of $2e^-$ charge have $4e^-$ charge  and  additionally  have an right or left hand h-chirality index. This theoretical prediction requires empirical inspect.

These two even and odd superconducting gap equations were solved analytically to 
obtain the relations between the superconducting pairing potential and resulting ordered phases.
The two sets of gap equations have solutions similar to those obtained in our previous work, for decorated 
single layer graphene.\cite{Rouhollah2018} 
Seven hybridized  orbitals in pseudo- coated single layer graphene support nine possible bond pairing amplitudes.
There are nine superconducting phases with $p_x$, $p_y$, $f$, $s^{\pm}$, $d_{xy}^{\pm}$ and $d_{x^2-y^2}^{\pm}$ atomic orbital-like symmetries 
corresponding to each of these even(+)/odd(-) gap equations. Only three of them are physically 
reachable, denoted by $\Psi^{{\pm}}_{2,s}$, $\Psi^{{\pm}}_{2,{d{xy}}}$, and 
$\Psi^{{\pm}}_{2,{d_{x^2-y^2}}}$.
 These symmetries almost preserve properties from a two band model of pristine graphene (Fig.~4 Ref.\cite{Rouhollah2018}).  The $d$-wave solutions are degenerate and so it  can support chiral $d_{x^2-y^2}+id_{xy}$ superconductivity in each of these sectors.
These three 
phases are distorted by intercalation.
In fact, the significant  difference which appears  between  two bands pristine $C_2$   pairing symmetries  and shrunken  graphene $C_6$(decorated graphene) is a skewness factor  i.e. $\alpha_{sy}^{l,\pm}\neq 1$ in front of the self consistent gaps solutions.
In each of these even or odd sectors, one band  is weakly dispersive near the Fermi energy along $\Gamma \to M$ where its Bloch wave function has linear combination of $d_{x^2-y^2}$ and $d_{xy}$ character, and is responsible for $d_{x^2-y^2}$ and $d_{xy}$ pairing with lowest pairing energy in our model(see Ref.\cite{Rouhollah2018}). Because of the high density of states of carriers in this band, d-wave superconductivity is more robust against disorder than s-wave.
\subsubsection{  Dominant Bands: Possibility of Two Gap  superconductivity }
  Superconductivity could be established in the odd or even sector of intercalated AA-stacked  or simultaneously in both sectors. Even/odd sectors are coupled just via chemical potential.   Two nearly ``flat bands'' with d-wave symmetry Bloch character, crossing the Fermi energy, 
each related to the graphene-like structures, are responsible for
two distinct d-wave superconductivity gaps that could be emerged. The distorted s-wave superconductivity is constructed between quasi-particles in upper bands of both sector that have s-wave symmetry character.
  At moderate doping, distorted d-wave superconductivity is dominant in both sectors while in high doping, distorted s-wave becomes preferable. Superconductivity with different phases in each of these sectors e.g. distorted s-wave in one sector and d-wave phase symmetry in the other,  is not so possible.

To know whether superconductivity in IBlG is single gap or multi gap, depending on what type of intercalant  is used,  one can  inspect numerically which sector will prevail. Hybridization of Carbon
  ($CP_z$)  and intercalant ($Is$) orbitals, electron or hole doping factor  (chemical potential), nearest neighbor hopping symmetry breaking ($\Gamma$ gap opening) and interlayer coupling ($H_{12}$) are important factors in specifying superconductivity pairing symmetry and dominant bands. 
 While just the even sector  is under the influence of C-I orbital hybridization and odd sector bands does not affected directly (except an small gap opening), but interlayer coupling  has a dual effects. It reduces the bandwidth of one sector 
(e.g. the even sector) but simultaneously it increases the bandwidth of the other sector(e.g. odd sector). It therefore plays a crucial  role in electronic correlation effects.  

Mathematical analysis shows that in the limit of strong interlayer hopping, so that inter- and 
intra- layer hoppings are the same (up to a minus sign), 
then bandwidths of the odd (even) sector become completely flat while the  bandwidths of the 
other sector are doubled. In this case superconductivity established in the flat band sector,
even with a small Cooper pairing  potential $g_0$, leads to a high critical temperature 
{\it i.e.}  $k_BT_c=g_0$. 
In this limit all of the possible superconducting phases, i.e. $d, p$ and $s$-wave symmetries,
 are degenerate. This observation suggests there may be  some aspect of 
unconventional superconductivity in  bilayer AA- graphene related to inter- versus 
intralayer hopping effects be available under high pressure.

The best conditions to induced superconductivity in IBlG are those that interlayer coupling be strong and band structure of even or odd sector or both slightly be deviated from pristine graphen-like structures. Under these circumstances if electron or hole doping causes  the nearly flat bands of odd or even sector meet the Fermi surface then chiral $d+id$ odd or even superconductivity  may induce. An inadequate inspecting of the band structure of different metal intercalated $C_6MC_6$ (M=Li, Na, K, Rb, Cs, Gr, Be, Mg, Ca,Sr and Ba)   such as that can be seen in the Fig.~3(a) of Ref. \cite{Tomoaki2017}, shows that odd sector flat band (not affected by I-C orbital coupling) always lay on top of even sector flat band, that  means the interlayer coupling $t_{12}$ in all of them is positive. The structures in which interlayer (IL) band is empty have negligible I-C coupling and so both even and odd sector have the structure similar to  shrunken graphene $C_6$.  The even band of Li intercalated BlG meet the Fermi Surface and interlayer coupling is stronger than the others. These is an  interesting structure that could host 
 even-superconducting beside normal odd-Dirac quasi-particles. But since IL band is empty, symmetry not allow out of plane phonon vibration to trigger superconductivity. However it may exhibit richer correlation effect  than single layer graphene under pressure, gating or proximity effects.   
 K and Rb intercalated BlG have potential to exhibit superconductivity, while in Gr intercalated BlG it seems that simultaneous odd and even Dirac cones coexist.
 
Experimental evidence for superconductivity in Ca- intercalated BlG has been reported. Motivated by this observation we performed numerical calculation for $C_6CaC_6$. 
Numerical calculations show that for both even and odd sector  gap equations,  $d$-wave phases, {\it i.e.} 
$\Psi_{{\pm},{d_{xy}}}^{2}$ and $\Psi_{{\pm}{d_{x^2-y^2}}}^{2}$, are dominant  
(smaller $g_0$ means less interaction energy for pairing) and slightly distorted 
by intercalation i.e. $\alpha _{d}^{2,\pm}\approx 1 $ while  $s$-wave symmetry 
{\it i.e.} $\Psi_{{\pm},s}^{2}$ require greater energy and are significantly distorted, 
$\alpha _{s}^{2,\pm}\neq 1$.    Phase Transition from d-wave single-gap to d-wave dual-gap Superconductivity in calcium intercalated bilayer graphene is possible. 
 Although $T_c$ experimentally around 4K and 
theoretically calculated near 6K are reported, 
 also  using Raman spectroscopy, possibility of distinguishing intralayer
and interlayer electron-phonon interactions in samples of twisted bilayer graphene has been reported by ref.[\cite{Eliel2018}] relying on these results, from  Fig. \ref{figure:twogap} it can 
be seen that both even and odd $d$-wave phases are nearly degenerate at 2K, consistent
with this system being a two gap superconductor around $T_c\approx 2K$. Our results support two d-wave gap superconductivity that has been proposed in 
ref.[\onlinecite{Margine2016}] (Fig.~2), although different sectors were not separated
in their studies.

Relying on pre-mentioned  properties, AA-stacked bilayer graphene may exhibit feature-rich electronic properties than singlelayer graphen. It seems that study of superconductivity in  pristine and intercalated AA-stacking BlG could tend to interesting experimental achievements- such as (even and odd) chiral superconducting  $d+id$ pairing which has been  predicted primarily in pristine single layer graphene at van Hove singularity point, and also simultaneous coexistence of different phases e.g.   superconductivity and normal Dirac  quasiparticles or two gap superconductivity (with different chirality) in different branches of their band-structures.  
\section{Acknowledgments}
R. Gholami acknowledges support that allowed an extended visit to the University of California Davis during part of this work. W.E.P. was supported by NSF grant DMR-1207622.

\section{Author contributions}
R.G. and R.M. conceived of the presented idea. R.G. developed the theory and performed the computations. R.M. and W. E. P verified the analytical methods. R.M. supervised the project. R.G, R. M. and W. E. P.  discussed the results and commented on the 
manuscript. S. M performed DFT calculation and 
designed the figures. 
\section{Additional information}
\textbf{ Competing interests:} The authors declare no competing interests.


\section{Appendix A: Accurate Tight Binding Model}
 In our previous work we used realistic multiband tight binding model for decorated monolayer graphene and obtained its band structure  analytically.\cite{Rouhollah2018} Here we follow and generalize that method and find analytic solutions for the intercalated bilayer graphene spectrum in general form. 
 We consider Bloch ket state of Eq.\ref{eq:H-normal} as 
\begin{eqnarray}
\left| \Psi_{\vec k} (\vec r) \right\rangle  = \frac{1}{\sqrt{N}}\sum\limits_{n = 1}^N {\sum\limits_{\alpha = 0}^{12}{\mathscr{C}_{{\alpha }}}} e^{i\vec k.{{\vec r}_{n{\alpha}}}}\left| \phi_{n\alpha} \right\rangle 
\label{eq:bloch-state-0}
\end{eqnarray}
in which ${\vec r}_{n\alpha}={\vec r}_{n}+{\vec d}_{\alpha}$ and ${\vec r}_{n}$ 
is $n$th Bravais lattice site vector position and ${\vec d}_{\alpha}$ is vector 
position of the $\alpha$-th subsite with respect to unit cell $n$. The  Ca sublattice is labeled by $\alpha=0$ also $A_{1}^{1}$, $A_{2}^{1}$, $A_{3}^{1}$, $B_{1}^{1}$, $B_{2}^{1}$, $B_{3}^{1}$, $A_{1}^{2}$, $A_{2}^{2}$, $A_{3}^{2}$, $B_{1}^{2}$,  $B_{2}^{2}$, $B_{3}^{2}$  subsites are labeled by $\alpha=1, ... ,12$ respectively.
$\left| \phi_{n\alpha} \right\rangle=
   \left| \phi_{n\alpha}({\vec r}-{\vec r}_{n}-{\vec d}_{\alpha}) \right\rangle$ 
is the atomic $\pi$ electron ket state of subsite $\alpha$ of site $n$. 
The Schr\"odinger equation for this system is 
\begin{eqnarray}
\sum^{12}_{\beta=0} \varepsilon_{\alpha\beta}({\vec k}) \mathscr{C}_{\beta }+(\epsilon_{\alpha}-\mu_{o})\mathscr{C}_{\alpha } = E({\vec k})\mathscr{C}_{\alpha } \;\;\mbox{where}\;\; \varepsilon_{\alpha\beta}({\vec k})=-\frac{1}{N}\sum_{ij} e^{i{\vec k}.({\vec r}_{i\alpha}-{\vec r}_{j\beta})}t^{\sigma\sigma}_{i\alpha j\beta}.
\label{eq:schro-bloch-state-1}
\end{eqnarray}
 Symmetries of this system imply that
  $
\mathscr{C}_{\alpha}(\vec k)=\pm\mathscr{C}_{\alpha+6 }(\vec k) 
.$
The Schr\"odinger equation Eq.\ref{eq:schro-bloch-state-1} can be written in the following $13\times13$ matrix form eigenvalue problem
\begin{eqnarray}
{H_N}(\vec k){\Psi _N}(\vec k) = \left[ {\begin{array}{*{20}{c}}
{{h_0}(\vec k)}&\vline& {{h_{01}}(\vec k)}&{{h_{02}}(\vec k)}\\
\hline
{{h_{10}}(\vec k)}&\vline& {{H_{11}}(\vec k)}&{{H_{12}}(\vec k)}\\
{{h_{20}}(\vec k)}&\vline& {{H_{21}}(\vec k)}&{{H_{22}}(\vec k)}
\end{array}} \right]\left( {\begin{array}{*{20}{c}}
{{\mathscr C}_{0}(\vec k)}\\
{\boldsymbol{\mathcal C}(\vec k)}\\
{ \pm \boldsymbol{\mathcal C}(\vec k)}
\end{array}} \right) = {E_ \pm }(\vec k)\left( {\begin{array}{*{20}{c}}
{{\mathscr C}_{0}(\vec k)}\\
{\boldsymbol{\mathcal C} }(\vec k)\\
{ \pm \boldsymbol{\mathcal C}(\vec k)}
\end{array}} \right)
\label{eq:k-schr-bigraphen1} 
\end{eqnarray}
where the column matrix $\boldsymbol{\mathcal C}({\vec k})$ is $\boldsymbol{\mathcal C}({\vec k}) 
=({\mathscr C}_{1}({\vec k})\; {\mathscr C}_{2}({\vec k})\;...\;{\mathscr C}_{6}({\vec k}))^{T}$ 
and the dispersion matrices satisfy $H_{11}=H_{22}$, $H_{12}=H_{21}$  
and $h_{01}=h_{02}=h_{10}^{\dagger}=h_{20}^{\dagger}.$ The Ca-C 
dispersion row matrices $h_{01}(\vec k)=h_{02}(\vec k)=(h_{CaA}(\vec k)~~h_{CaB}(\vec k))$ are given by 
\begin{eqnarray}
{h_{CaA}}(\vec k) &=& \left( {\begin{array}{*{20}{c}}
{{\varepsilon _{Ca{A_1}}}(\vec k)}&{{\varepsilon _{Ca{A_2}}}(\vec k)}&{{\varepsilon _{Ca{A_3}}}(\vec k)}
\end{array}} \right) =  - {t_{1}^{CaC}}\left( {\begin{array}{*{20}{c}}
{{e^{i\vec k.{{\vec \delta }_1}}}}~~&{{e^{i\vec k.{{\vec \delta }_3}}}}~~~&{{e^{i\vec k.{{\vec \delta }_2}}}}
\end{array}} \right)\nonumber \\
{h_{CaB}}(\vec k) &=& \left( {\begin{array}{*{20}{c}}
{{\varepsilon _{Ca{B_1}}}(\vec k)}&{{\varepsilon _{Ca{B_2}}}(\vec k)}&{{\varepsilon _{Ca{B_3}}}(\vec k)}
\end{array}} \right) =  - {t_{1}^{CaC}}\left( {\begin{array}{*{20}{c}}
{{e^{ - i\vec k.{{\vec \delta }_1}}}}&{{e^{ - i\vec k.{{\vec \delta }_3}}}}&{{e^{ - i\vec k.{{\vec \delta }_2}}}}
\end{array}} \right).
\label{eq:block-offdiag-Li}
\end{eqnarray}
Here $t^{CaC}_{i}$ is the hopping amplitude from Ca to $i$th neighbor C atom.  The Ca-Ca dispersion 
is $h_0(\vec k)={{\varepsilon_{CaCa}}(\vec k) + {\epsilon _{Ca}} - {\mu _{o}}}$
where 
\begin{eqnarray}
\varepsilon_{CaCa}(\vec{k}) &=& 2{t}_1^{CaCa}\left( {\cos \vec k.{{\vec \xi }_1} + \cos \vec k.{{\vec \xi }_2} + \cos \vec k.{{\vec \xi }_3}} \right) + 2{t}_2^{CaCa}\left( {\cos \vec k.({{\vec \xi }_1} - {{\vec \xi }_2}) + \cos  \vec k.({{\vec \xi }_1} - {{\vec \xi }_3}) + \cos \vec k.({{\vec \xi }_2} - {{\vec \xi }_3})} \right) \nonumber\\
&+& 2{t}_3^{CaCa}\left( {\cos 2\vec k.{{\vec \xi }_1} + \cos 2\vec k.{{\vec \xi }_2} + \cos 2\vec k.{{\vec \xi }_3}} \right) +  \ldots .
\label{eq:block-diag-Li}
\end{eqnarray}
The interlayer dispersion matrices $H_{11}$, $H_{22}$ and interlayer dispersion matrices $H_{12}$ and $H_{21}$  are given by
\begin{eqnarray}
{H_{mn}}(\vec k) = \left( {\begin{array}{*{20}{c}}
{h_{{A}{A}}^{mn}(\vec k) + (\varepsilon _{{A}}^{m} - {\mu _0}})\delta_{mn}I_{3\times3}&{h_{{A}{B}}^{mn}(\vec k)}\\
{h_{{B}{A}}^{mn}(\vec k)}&{h_{{B}{B}}^{mn}(\vec k) +(\varepsilon _{{B}}^{m} - {\mu _0}})\delta_{mn}I_{3\times3}
\end{array}} \right)
\end{eqnarray}
where the off-diagonal carbon-carbon dispersion matrices are
\begin{eqnarray}
h_{{A}{A}}^{mn}(\vec k) = h_{{B}{B}}^{*mn}(\vec k) = \left( {\begin{array}{*{20}{c}}
{{\varepsilon _{A_1^mA_1^n}}(\vec k)}&{{\varepsilon _{A_1^mA_2^n}}(\vec k)}&{{\varepsilon _{A_1^mA_3^n}}(\vec k)}\\
{{\varepsilon _{A_2^mA_1^n}}(\vec k)}&{{\varepsilon _{A_2^mA_2^n}}(\vec k)}&{{\varepsilon _{A_2^mA_3^n}}(\vec k)}\\
{{\varepsilon _{A_3^mA_1^n}}(\vec k)}&{{\varepsilon _{A_3^mA_2^n}}(\vec k)}&{{\varepsilon _{A_3^mA_3^n}}(\vec k)}
\end{array}} \right) =  \left( {\begin{array}{*{20}{c}}
{\alpha^{mn}(\vec k)}&{\beta^{mn}(\vec k)}&{\gamma^{mn}(\vec k)}\\
{\beta^{*mn}(\vec k)}&{\alpha^{mn}(\vec k)}&{\theta^{mn}(\vec k)}\\
{\gamma^{*mn}(\vec k)}&{\theta^{*mn}(\vec k)}&{\alpha^{mn}(\vec k)}
\end{array}} \right)
\end{eqnarray}
\begin{eqnarray}
h_{{A}{B}}^{mn}(\vec k) = h_{{B}{A}}^{\dag mn} (\vec k) = \left( {\begin{array}{*{20}{c}}
{{\varepsilon _{A_1^mB_1^n}}(\vec k)}&{{\varepsilon _{A_1^mB_2^n}}(\vec k)}&{{\varepsilon _{A_1^mB_3^n}}(\vec k)}\\
{{\varepsilon _{A_2^mB_1^n}}(\vec k)}&{{\varepsilon _{A_2^mB_2^n}}(\vec k)}&{{\varepsilon _{A_2^mB_3^n}}(\vec k)}\\
{{\varepsilon _{A_3^mB_1^n}}(\vec k)}&{{\varepsilon _{A_3^mB_2^n}}(\vec k)}&{{\varepsilon _{A_3^mB_3^n}}(\vec k)}
\end{array}} \right)=  \left( {\begin{array}{*{20}{c}}
{\tau _1^{mn}(\vec k)}&{d_2^{mn}(\vec k)}&{d_3^{mn}(\vec k)}\\
{d_2^{mn}(\vec k)}&{\tau _3^{mn}(\vec k)}&{d_1^{mn}(\vec k)}\\
{d_3^{mn}(\vec k)}&{d_1^{mn}(\vec k)}&{\tau _1^{mn}(\vec k)}
\end{array}} \right)
\end{eqnarray}
in which $m$ and $n$ are layer index. Shorthand notation has been introduced as follows:
\begin{eqnarray}
\begin{array}{l}
\beta^{mn}(\vec k)={\varepsilon _{A_1^mA_2^n}}(\vec k) = \varepsilon _{A_2^mA_1^n}^*(\vec k) = t_2^{mn}{e^{i\vec k.({{\vec \delta }_3} - {{\vec \delta }_1})}}\left[ {1 + w_t\left( {{e^{ - i\vec k.{{\vec \xi }_3}}} + {e^{i\vec k.{{\vec \xi }_1}}}} \right)} \right]\\
\gamma^{mn}(\vec k)={\varepsilon _{A_1^mA_3^n}}(\vec k) = \varepsilon _{A_3^mA_1^n}^*(\vec k) = t_2^{mn}{e^{i\vec k.({{\vec \delta }_2} - {{\vec \delta }_1})}}\left[ {1 + w_t\left( {{e^{ - i\vec k.{{\vec \xi }_2}}} + {e^{i\vec k.{{\vec \xi }_1}}}} \right)} \right]\\
\theta^{mn}(\vec k)={\varepsilon _{A_2^mA_3^n}}(\vec k) = \varepsilon _{A_3^mA_2^n}^*(\vec k) = t_2^{mn}{e^{i\vec k.({{\vec \delta }_2} - {{\vec \delta }_3})}}\left[ {1 + w_t\left( {{e^{ - i\vec k.{{\vec \xi }_2}}} + {e^{i\vec k.{{\vec \xi }_3}}}} \right)} \right]\\
 \alpha^{mn}(\vec k)={\varepsilon _{A_i^mA_i^n}}(\vec k) = {\varepsilon _{B_i^mB_i^n}}(\vec k)  =t_{0}^{mn}+ 2t_5^{mn}\left( {\cos \vec k.{{\vec \xi }_1} + \cos \vec k.{{\vec \xi }_2} + \cos \vec k.{{\vec \xi }_3}} \right)
\end{array}
\end{eqnarray}
and it has been supposed that $w_t=\frac{ t^{'mn}_{1}}{t_{1}^{mn}}=\frac{ t^{'mn}_{2}}{t_{2}^{mn}}=...$ and $\vec{\xi_{i}}=\vec{\tau_i}+2\vec{\delta_i}$. The $\tau$ and $d$-functions are given by
\begin{eqnarray}
{\tau _1^{mn}}(\vec k) = {e^{i\vec k.{{\vec \tau }_1}}}\left[ {t^{'mn}_{1} + t_3^{'mn}{e^{ - i\vec k.{{\vec \xi }_1}}} + t_4^{'mn}\left( {{e^{i\vec k.{{\vec \xi }_2}}} + {e^{i\vec k.{{\vec \xi }_3}}}} \right)} \right]; 
{d_1^{mn}}(\vec k) =  {e^{i\vec k.{{\vec \delta }_1}}}\left[ {t^{mn}_{1} + t_3^{mn}{e^{ - i\vec k.{{\vec \xi }_1}}} + t_4^{mn}\left( {{e^{i\vec k.{{\vec \xi }_2}}} + {e^{i\vec k.{{\vec \xi }_3}}}} \right)} \right]\nonumber \\
{\tau _2^{mn}}(\vec k) = {e^{i\vec k.{{\vec \tau }_2}}}\left[ {t^{'mn}_{1} + t_3^{'mn}{e^{ - i\vec k.{{\vec \xi }_2}}} + t_4^{'mn}\left( {{e^{i\vec k.{{\vec \xi }_3}}} + {e^{i\vec k.{{\vec \xi }_1}}}} \right)} \right];
 {d_2^{mn}}(\vec k) =  {e^{i\vec k.{{\vec \delta }_2}}}\left[ {t^{mn}_{1} + t_3^{mn}{e^{ - i\vec k.{{\vec \xi }_2}}} + t_4^{mn}\left( {{e^{i\vec k.{{\vec \xi }_3}}} + {e^{i\vec k.{{\vec \xi }_1}}}} \right)} \right]\nonumber \\
{\tau _3^{mn}}(\vec k) = {e^{i\vec k.{{\vec \tau }_3}}}\left[ {t^{'mn}_{1} + t_3^{'mn}{e^{ - i\vec k.{{\vec \xi }_3}}} + t_4^{'mn}\left( {{e^{i\vec k.{{\vec \xi }_1}}} + {e^{i\vec k.{{\vec \xi }_2}}}} \right)} \right];
{d_3^{mn}}(\vec k) =  {e^{i\vec k.{{\vec \delta }_3}}}\left[ {t^{mn}_{1} + t_3^{mn}{e^{ - i\vec k.{{\vec \xi }_3}}} +
 t_4^{mn}\left( {{e^{i\vec k.{{\vec \xi }_1}}} + {e^{i\vec k.{{\vec \xi }_2}}}} \right)} \right]\nonumber\\
\label{eq:epAB}
\end{eqnarray}
Using the following unitary transformation one can separate the bilayer graphene Hamiltonian 
Eq.~\ref{eq:k-schr-bigraphen1} 
 into two decoupled single layer pseudo-graphene Hamiltonians, where one of them is decorated with 
the intercalant layer 
\begin{eqnarray}
H_D=Q_T^\dag {H_N}(\vec k){Q_T} = \left[ {\begin{array}{*{20}{c}}
{{h_0}(\vec k)}&{\sqrt 2 {h_{01}}(\vec k)}&\vline& 0\\
{\sqrt 2 {h_{10}}(\vec k)}&{H^+}(\vec k)&\vline& 0\\
\hline
0&0&\vline& {H^-}(\vec k)
\end{array}} \right],~~~
{Q_T} = \frac{1}{{\sqrt 2 }}\left( {\begin{array}{*{20}{c}}
{\sqrt 2 }&\vline& 0&0\\
\hline
0&\vline& {{I_{6 \times 6}}}&{{I_{6 \times 6}}}\\
0&\vline& {{I_{6 \times 6}}}&{ - {I_{6 \times 6}}}
\end{array}} \right).
\label{eq:decoupled-k-schr-bigraphen-ap} 
\end{eqnarray}
Here $H^{\pm}=H_{11}(\vec k) \pm H_{12}$ in matrix notation is
\begin{eqnarray}
H ^{\pm} (\vec k) = {H_{11}}(\vec k) \pm {H_{12}}(\vec k) =   \left( {\begin{array}{*{20}{c}}
{\varepsilon _1^ \pm (\vec k)}& {\beta ^ \pm(\vec k)}&{ \gamma ^ \pm(\vec k)}&\vline& {\tau _1^ \pm (\vec k)}&{d_2^ \pm (\vec k)}&{d_3^ \pm (\vec k)}\\
{{\beta ^ {\pm *}}(\vec k)}&{\varepsilon _1^ \pm (\vec k)}&{\theta ^ \pm(\vec k)}&\vline& {d_2^ \pm (\vec k)}&{\tau _3^ \pm (\vec k)}&{d_1^ \pm (\vec k)}\\
{ {\gamma ^{\pm *}}(\vec k)}&{{\theta ^{\pm *}}(\vec k)}&{\varepsilon _1^ \pm (\vec k)}&\vline& {d_3^ \pm (\vec k)}&{d_1^ \pm (\vec k)}&{\tau _2^ \pm (\vec k)}\\
\hline
{\tau _1^{ \pm *}(\vec k)}&{d_2^{ \pm *}(\vec k)}&{d_3^{ \pm *}(\vec k)}&\vline& {\varepsilon _2^ \pm (\vec k)}&{ {\beta ^{\pm *}}(\vec k)}&{{\gamma ^{*\pm}}(\vec k)}\\
{d_2^{ \pm *}(\vec k)}&{\tau _3^{ \pm *}(\vec k)}&{d_1^{ \pm *}(\vec k)}&\vline& {\beta^ \pm (\vec k)}&{\varepsilon _2^ \pm (\vec k)}&{{\theta ^ {\pm *}}(\vec k)}\\
{d_3^{ \pm *}(\vec k)}&{d_1^{ \pm *}(\vec k)}&{\tau _2^{ \pm *}(\vec k)}&\vline& {\gamma ^ \pm(\vec k)}&{\theta ^ \pm(\vec k)}&{\varepsilon _2^ \pm (\vec k)}
\end{array}} \right)
\label{eq:def-H1}
\end{eqnarray}
in which $\vec{k}$-dependent on-site energies  have been defined as  
$\varepsilon_1^{\pm}({\vec{k}})= \epsilon_{A}-\mu_0+\alpha^{\pm}(\vec{k})$ and
$\varepsilon_2^{\pm}({\vec{k}})= \epsilon_{B}-\mu_0+\alpha^{\pm}(\vec{k}).$
Also  the following shorthand notation has been introduced 
\begin{eqnarray}
\alpha^{\pm}(\vec{k})&=&\left(\alpha^{11}({\vec k})\pm\alpha^{12}({\vec k})\right),~
\beta^{\pm}(\vec{k})=\left(\beta^{11}({\vec k})\pm\beta^{12}({\vec k})\right);~~~~\tau_i^{\pm}(\vec{k})=\left(\tau_i^{11}(\vec{k})\pm\tau_i^{12}({\vec k})\right)\nonumber\\
\gamma^{\pm}(\vec{k})&=&\left(\gamma^{11}({\vec k})\pm\gamma^{12}({\vec k})\right),~~
\theta^{\pm}(\vec{k})=\left(\theta^{11}({\vec k})\pm\theta^{12}({\vec k})\right);~~~~d_i^{\pm}(\vec{k})=\left(d_i^{11}({\vec k})\pm d_i^{12}({\vec k})\right)
\end{eqnarray}

Unitary transformation of Eq.~\ref{eq:decoupled-k-schr-bigraphen-ap} divides \textbf{thirteen} bands of intercalated bilayer graphene into, \textbf{six} and \textbf{seven} bands groups. Following the  approach\cite{Rouhollah2018} 
that  has been applied to monolayer decorated graphene, an exact analytical solution of the six-band group 
can be found in general case. These bands are eigenvalues of $H^-$ matrix and are not affected directly by 
the intercalant band. 

In the special case of pristine bilayer graphene in which $\gamma^{\pm*}(\vec{k})=\theta^{\pm}(\vec{k})=\beta^{\pm}(\vec{k})$, $\varepsilon_{1}^{\pm}(\vec{k})=\varepsilon_{2}^{\pm}(\vec{k})$ and $\tau_{i}^{\pm}(\vec{k})=d_{i}^{\pm}(\vec{k})$, Eq.~\ref{eq:def-H1} easily can be diagonalized to find eigenvalues and also eigenvectors. The eigenvalues are given by
\begin{eqnarray}
E^{\pm}_{\gamma,l}=\varepsilon_{1}^{\pm}(\vec{k})+\beta^{\pm}(\vec{k})+\beta^{*\pm}(\vec{k})
+(-1)^lt_1^{\pm}|\eta_{0}^{\pm}(\vec{k})|
\label{eq:gamma-branch}
\end{eqnarray}
\begin{eqnarray}
E^{\pm}_{\alpha,l}=\varepsilon_{1}^{\pm}(\vec{k})+e^{i2\pi/3}\beta^{\pm}(\vec{k})+e^{-i2\pi/3}\beta^{*\pm}(\vec{k})
+(-1)^lt_1^{\pm}|\eta_{1}^{\pm}(\vec{k})|
\label{eq:alpha-branch}
\end{eqnarray}
\begin{eqnarray}
E^{\pm}_{\beta,l}=\varepsilon_{1}^{\pm}(\vec{k})+e^{-i2\pi/3}\beta^{\pm}(\vec{k})+e^{i2\pi/3}\beta^{*\pm}(\vec{k})
+(-1)^lt_1^{\pm}|\eta_{2}^{\pm}(\vec{k})|
\label{eq:beta-branch}
\end{eqnarray}
with eigenvectors are given by replacing $m$ in the following equation with $m=0,  1, 2$ respectively 
\begin{eqnarray}
\phi_{m,l}^{\pm}(\vec{k})=\frac{1}{\sqrt{6}} [(u_m ~~ u_m^{*} ~~1)~~
(-1)^l\frac{\eta_m^{*\pm}}{|\eta_m^{\pm}|}(u_m^{*} ~~ u_m ~~1)]^T
\label{eq:pristine-eigenvectors}
\end{eqnarray}
wherein
\begin{eqnarray}
\eta_{m}^{\pm}(\vec{k})=d_{2}^{\pm}(\vec{k})+u_m d_{1}^{\pm}(\vec{k})+u_m^* d_{3}^{\pm}(\vec{k});~~~
u_m=e^{i2m\pi/3}
\label{eq:pristine-Bloch phase}
\end{eqnarray}
However, except at $\Gamma$ point it is challenging (and unhelpful) to obtain an exact forms of the seven-bands 
group analytically. These bands are eigenvalues of the $H^{+}_{c}$ matrix,  and analytic expressions for them 
can be obtained just in the particular case of no hopping between intercalant layer and graphene sheet, similar
to the case for lithium intercalated bilayer graphene where intercalant band is empty (no Li-C hopping). 
In these cases nontrivial  solutions are  eigenvalues of $H^+$ matrix. 
Eigenvalues of  $H^-$ and $H^+$ matrices are given by  \cite{Rouhollah2018}
\begin{eqnarray}
E_{sh;m,l}^{\pm}(t_{i},\vec{\xi_{i}},\vec{k}) = \mu_{m}^{\pm}(\vec k) - {\mu _{o}} + \frac{1}{2}\left[ {{\varepsilon _A} + {\varepsilon _B} + {{( - 1)}^l}\sqrt {{{\left( {{\varepsilon _A} - {\varepsilon _B}} \right)}^2} + 4{w_m^{\pm}}(t_{i},\vec{\xi_{i}},\vec{k})}   } \right],~~~m=1,2,3;~l=1,2 
\label{eq:full-shrun-eigen} 
\end{eqnarray}
in which $\vec{k}$ dependent chemical potentials are defined as,
\begin{eqnarray}
{\mu_{m}^\pm(\vec{k})} =\alpha^{\pm}(\vec{k})+ u_{m}\Pi_{0}^\pm(t_{2},\vec{\xi_{i}},\vec{k})  + 
  u^ *_{m} {\Pi}^{\pm*}_{0}(t_{2},\vec{\xi_{i}},\vec{k});~~ u_{m} = {e^{2 i m\pi /3}}.
\label{eq:diag-eigenvalues}
\end{eqnarray}
The $\Pi_{0}^{\pm}(t_{2},\vec{\xi_{i}},\vec{k})$ function is introduced as
\begin{eqnarray}
\Pi_{0}^{\pm}(t_{2},\vec{\xi_{i}},\vec{k})   = {\left( {{\textstyle{{c_{0}^{\pm}(t_{2},\vec{\xi_{i}},\vec{k})} \over 2}} + i\sqrt {{{\left( {{\textstyle{{{c_1^{\pm}(t_{2},\vec{\xi_{i}},\vec{k})}} \over 3}}} \right)}^3} - {{\left( {{\textstyle{{{c_0^{\pm}(t_{2},\vec{\xi_{i}},\vec{k})}} \over 2}}} \right)}^2}} } \right)^{{\raise0.5ex\hbox{$\scriptstyle 1$}
\kern-0.1em/\kern-0.15em
\lower0.25ex\hbox{$\scriptstyle 3$}}}}.
\label{eq:c0-c1-def}
\end{eqnarray}
and, ${c_0^{\pm}}(t_{2},\vec{\xi_{i}},\vec{k}) = 
{\gamma  ^{\pm*} }(\vec{k}) [{\beta^{\pm}}(\vec{k}) {\theta^{\pm}}(\vec{k})]  +{ \gamma^{\pm} }(\vec{k})[{\beta ^{\pm} }(\vec{k}){\theta ^{\pm } }(\vec{k})]^{*}$ and 
${c_1^{\pm}}(t_{2},\vec{\xi_{i}},\vec{k}) = {\left| \beta ^{\pm}(\vec{k}) \right|^2} + {\left| \theta ^{\pm}(\vec{k}) \right|^2} + {\left| \gamma ^{\pm}(\vec{k}) \right|^2}$. 
Also, $w_m^{\pm}(t_{i},\vec{\xi_{i}},\vec{k})$  are eigenvalues of the following matrices 
 \begin{eqnarray}
G^{\pm}= \left( {\begin{array}{*{20}{c}}
{\tau _1^{\pm}(\vec k)}&{d_2^{\pm}(\vec k)}&{d_3^{\pm}(\vec k)}\\
{d_2^{\pm}(\vec k)}&{\tau _3^{\pm}(\vec k)}&{d_1^{\pm}(\vec k)}\\
{d_3^{\pm}(\vec k)}&{d_1^{\pm}(\vec k)}&{\tau _1^{\pm}(\vec k)}
\end{array}} \right)
 \left( {\begin{array}{*{20}{c}}
{\tau _1^{\pm *}(\vec k)}&{d_2^{\pm *}(\vec k)}&{d_3^{\pm *}(\vec k)}\\
{d_2^{\pm *}(\vec k)}&{\tau _3^{\pm *}(\vec k)}&{d_1^{\pm *}(\vec k)}\\
{d_3^{\pm *}(\vec k)}&{d_1^{\pm *}(\vec k)}&{\tau _1^{\pm *}(\vec k)}
\end{array}} \right).
 \end{eqnarray}
The eigenvalues can be obtained as
\begin{eqnarray}
w_{m}^{\pm}(t_{i},\vec{\xi_{i}},\vec{k}) = \frac{C_2^{\pm}(t_{i},\vec{\xi_{i}},\vec{k})}{3}+ u_{m}\Pi_1^{\pm}(t_{i},\vec{\xi_{i}},\vec{k})  + u^ *_{m} {\Pi}_{1}^{\pm*}(t_{i},\vec{\xi_{i}},\vec{k}),~~~ u_{m} =  {e^{2 i m\pi /3}};~~~m=  1 , 2, 3.
\label{eq:cubic-w-solutions}
\end{eqnarray}
wherein 
\begin{eqnarray}
{C_2^{\pm}} (t_{i},\vec{\xi_{i}},\vec{k}) &=& {G_{11}^{\pm}} + {G_{22}^{\pm}} + {G_{33}^{\pm}}\nonumber\\
{C_1^{\pm}}(t_{i},\vec{\xi_{i}},\vec{k})  &=& {\left| {{G_{12}^{\pm}}} \right|^2} + {\left| {{G_{13}^{\pm}}} \right|^2} + {\left| {{G_{23}^{\pm}}} \right|^2} - \left( {{G_{11}^{\pm}}{G_{22}^{\pm}} + {G_{11}^{\pm}}{G_{33}^{\pm}} + {G_{22}^{\pm}}{G_{33}^{\pm}}} \right)\nonumber\\
{C_0^{\pm}}(t_{i},\vec{\xi_{i}},\vec{k}) & =& {G_{13}^{\pm}}{({G_{12}^{\pm}}{G_{23}^{\pm}})^*} + G_{13}^{\pm*}({G_{12}^{\pm}}{G_{23}^{\pm}}) - {G_{11}^{\pm}}{\left| {{G_{23}^{\pm}}} \right|^2} - {G_{22}^{\pm}}{\left| {{G_{13}^{\pm}}} \right|^2} - {G_{33}^{\pm}}{\left| {{G_{12}^{\pm}}} \right|^2} + {G_{11}^{\pm}}{G_{22}^{\pm}}{G_{33}^{\pm}}
\label{eq:Ci-coef-def}
\end{eqnarray}
\begin{eqnarray}
\Pi_1^{\pm}(t_{i},\vec{\xi_{i}},\vec{k}) &= &{\left( {Q^{\pm}(t_{i},\vec{\xi_{i}},\vec{k})+ i\sqrt {{P^{\pm}(t_{i},\vec{\xi_{i}},\vec{k})^3} - {Q^{\pm}(t_{i},\vec{\xi_{i}},\vec{k})^2}} } \right)^{\frac{1}{3}}}\nonumber\\
Q^{\pm}(t_{i},\vec{\xi_{i}},\vec{k}) &= &\frac{{{C_0^{\pm}(t_{i},\vec{\xi_{i}},\vec{k})}}}{2} + \frac{{{C_1^{\pm}(t_{i},\vec{\xi_{i}},\vec{k})}{C_2^{\pm}(t_{i},\vec{\xi_{i}},\vec{k})}}}{6} + \frac{{{C_2^{\pm}}^3(t_{i},\vec{\xi_{i}},\vec{k})}}{{27}} \nonumber\\
P^{\pm}(t_{i},\vec{\xi_{i}},\vec{k})& = &\frac{{{C_1^{\pm}(t_{i},\vec{\xi_{i}},\vec{k})}}}{3} + \frac{{{C_2^{\pm}}^2(t_{i},\vec{\xi_{i}},\vec{k})}}{9}.
\label{eq:delta0-def}
\end{eqnarray}

One can write the $H_D$ matrix in $H^{\pm}$ bases i. e. $H^{'}_{D}=U^{\dagger}H_D U$,
\begin{eqnarray}
H^{'}_{D}=\left( {\begin{array}{*{20}{c}}
h_0(\vec k) &\vline& \gamma _{1}(\vec k)&{{\gamma _2}(\vec k)}&{{\gamma _3}(\vec k)}&{{\gamma _4}(\vec k)}&{{\gamma _5}(\vec k)}&{{\gamma _6}(\vec k)}&\vline & 0&0&0&0&0&0\\
\hline
{\gamma _1^ * (\vec k)}&\vline& E_{1}^+(\vec k) &0&0&0&0&0&\vline  &0&0&0&0&0&0\\
{\gamma _2^ * (\vec k)}&\vline& 0&E_{2}^+(\vec k)  &0&0&0&0&\vline &0&0&0&0&0&0\\
{\gamma _3^ * (\vec k)}&\vline& 0&0&E_{3}^+(\vec k)  &0&0&0&\vline &0&0&0&0&0&0\\
{\gamma _4^ * (\vec k)}&\vline& 0&0&0&  E_{4}^+(\vec k)&0&0&\vline &0&0&0&0&0&0\\
{\gamma _5^ * (\vec k)}&\vline& 0&0&0&0&  E_{5}^+(\vec k)&0&\vline &0&0&0&0&0&0\\
{\gamma _6^ * (\vec k)}&\vline& 0&0&0&0&0&  E_{6}^+(\vec k)&\vline &0&0&0&0&0&0\\
\hline
0&\vline &0 &0&0&0&0&0&\vline & E_{1}^-(\vec k) &0&0&0&0&0\\
0&\vline &0 &0&0&0&0&0&\vline & 0&E_{2}^-(\vec k)  &0&0&0&0\\
0&\vline &0 &0&0&0&0&0&\vline & 0&0&E_{3}^-(\vec k)  &0&0&0\\
0&\vline &0 &0&0&0&0&0&\vline & 0&0&0&  E_{4}^-(\vec k)&0&0\\
0&\vline &0 &0&0&0&0&0&\vline & 0&0&0&0&  E_{5}^-(\vec k)&0\\
0&\vline &0 &0&0&0&0&0&\vline & 0&0&0&0&0&  E_{6}^-(\vec k)\\
\end{array}} \right)
\label{eq:h'}
\end{eqnarray}
The upper left portion of Eq.\ref{eq:h'} can be obtained sufficiently well by perturbation theory.

\section{Appendix B: Bogoliubov-de Gennes Transformation}
The interacting Hamiltonian 
$H_{su}$ in matrix representation is $14\times 14$ matrix,
\begin{eqnarray}
{H_{su}}(\vec k) =\sum_{\vec k} {{{\hat \Psi }^\dag }(\vec k)}\left( \begin{array}{*{20}{c}}
H_{N}(\vec k)&H_{P}(\vec k)\\
H^{\dag }_{P}(\vec k)&- H^{*}_N( -\vec k)
\end{array} \right)\hat \Psi (\vec k)
\label{eq:Hsu-matrix-text}
\end{eqnarray}
 where $ \hat \Psi^{\dagger} (\vec k)
=(\hat c^{\dagger}_{0\uparrow}(\vec k)\hat c^{\dagger}_{1\uparrow}(\vec k)\;\hat c^{\dagger}_{2\uparrow}(\vec k)\;...;\hat c^{\dagger}_{12\uparrow}(\vec k)\;\hat c_{0\downarrow}(-\vec k)\hat c_{1\downarrow}(-\vec k)\;\hat c_{2\downarrow}(-\vec k)\;...\;\hat c_{12\downarrow}(-\vec k))$.
$H_N$  is Hamiltonian of the  normal state and $H_p$ is the pair interaction matrix. 
The full matrix must be diagonalized to obtain the quasiparticle spectrum.

The  mean field superconducting Hamiltonian of  Eq.~\ref{eq:full-Hamiltonian} in  Nambu space is
$ {{\hat H}_{su}} = \sum_{\vec k} {{{\hat \Psi }^\dag }(\vec k)} {H_{su}}(\vec k)\hat \Psi (\vec k)$
 where $H_{su}$ in matrix representation is,
\begin{eqnarray}
{H_{su}}(\vec k) =\left( \begin{array}{*{20}{c}}
H_{N}(\vec k)&H_{P}(\vec k)\\
H^{\dag }_{P}(\vec k)&- H^{*}_N( -\vec k)
\end{array} \right)= \left( {\begin{array}{*{20}{c}}
{\left( {\begin{array}{*{20}{c}}
{{h_0}(\vec k)}&\vline& {{h_{01}}(\vec k)}&{{h_{02}}(\vec k)}\\
\hline
{{h_{10}}(\vec k)}&\vline& {{H_{11}}(\vec k)}&{{H_{12}}(\vec k)}\\
{{h_{20}}(\vec k)}&\vline& {{H_{21}}(\vec k)}&{{H_{22}}(\vec k)}
\end{array}} \right)}&{\left[ {\begin{array}{*{20}{c}}
0&\vline& 0&0\\
\hline
0&\vline& {H_{11}^P(\vec k)}&{H_{12}^P(\vec k)}\\
0&\vline& {H_{21}^P(\vec k)}&{H_{22}^P(\vec k)}
\end{array}} \right]}\\
{{{\left[ {\begin{array}{*{20}{c}}
0&\vline& 0&0\\
\hline
0&\vline& {H_{11}^P(\vec k)}&{H_{12}^P(\vec k)}\\
0&\vline& {H_{21}^P(\vec k)}&{H_{22}^P(\vec k)}
\end{array}} \right]}}}&{ - \left( {\begin{array}{*{20}{c}}
{{h_0}(\vec k)}&\vline& {{h_{01}}(\vec k)}&{{h_{02}}(\vec k)}\\
\hline
{{h_{10}}(\vec k)}&\vline& {{H_{11}}(\vec k)}&{{H_{12}}(\vec k)}\\
{{h_{20}}(\vec k)}&\vline& {{H_{21}}(\vec k)}&{{H_{22}}(\vec k)}
\end{array}} \right)}
\end{array}} \right)
\label{eq:Hsu-matrix}
\end{eqnarray}
and $ \hat \Psi^{\dagger} (\vec k)=(\hat c^{\dagger}_{0\uparrow}(\vec k)\hat c^{\dagger}_{1\uparrow}(\vec k)\;\hat c^{\dagger}_{2\uparrow}(\vec k)\;...;\hat c^{\dagger}_{12\uparrow}(\vec k)\;\hat c_{0\downarrow}(-\vec k)\hat c_{1\downarrow}(-\vec k)\;\hat c_{2\downarrow}(-\vec k)\;...\;\hat c_{12\downarrow}(-\vec k))$ where
  $H_{11}(\vec k)=H_{22}(\vec k)$. The interlayer pairing  matrices are  
${H_{11}^P(\vec k)}={H_{22}^P(\vec k)}$ and  interlayer pairing matrices are 
${H_{12}^P(\vec k)}={H_{21}^P(\vec k)}$. The pairing matrices are given by
\begin{eqnarray}
H_{mn}^P(\vec k) = \left( {\begin{array}{*{20}{c}}
0&0&0&\vline& {{\Sigma^{mn} _1}(\vec k)}&{{\Delta^{mn} _2}(\vec k)}&{{\Pi^{mn}_3}(\vec k)}\\
0&0&0&\vline& {{\Pi^{mn} _2}(\vec k)}&{{\Sigma^{mn} _3}(\vec k)}&{{\Delta^{mn} _1}(\vec k)}\\
0&0&0&\vline& {{\Delta^{mn} _3}(\vec k)}&{{\Pi^{mn} _1}(\vec k)}&{{\Sigma^{mn} _2}(\vec k)}\\
\hline
{\Sigma _1^{mn*}(\vec k)}&{\Pi _2^*(\vec k)}&{\Delta _3^{mn*}(\vec k)}&\vline& 0&0&0\\
{\Delta _2^{mn*}(\vec k)}&{\Sigma _3^{mn*}(\vec k)}&{\Pi _1^{mn*}(\vec k)}&\vline& 0&0&0\\
{\Pi _3^{mn*}(\vec k)}&{\Delta _1^{mn*}(\vec k)}&{\Sigma _2^{mn*}(\vec k)}&\vline& 0&0&0
\end{array}} \right)
\label{eq:pairing-sitespace-Hp}
\end{eqnarray}
where $m$ and $n$ are layer index which can take 1 or 2.  The order parameters accordingly in Fourier space  are 
\begin{eqnarray}
\begin{array}{l}
{\Sigma ^{11}_l}(\vec k) = {g_1}{\Sigma _{l<ij>}}{e^{i\vec k.{{\vec \tau }_l}}},~~{{\Sigma }^{12}_l}(\vec k) = {{g}^{'}_1}{\Sigma ^{'}_{l<ij>}}{e^{i\vec k.{{\vec \tau }_l}}}\\
{\Pi^{11}_l}(\vec k) = {g_0}{\Pi _{l<ij>}}{e^{i\vec k.{{\vec \delta }_l}}},~~{{\Pi}^{12}_l}(\vec k) = {{g}^{'}_0}{{\Pi}^{'}_{l<ij>}}{e^{i\vec k.{{\vec \delta }_l}}}\\
{\Delta^{11}_l}(\vec k) = {g_0}{\Delta _{l<ij>}}{e^{i\vec k{{\vec \delta }_l}}},~~{{\Delta}^{12}_l}(\vec k) = {g^{'}_0}{\Delta ^{'}_{l<ij>}}{e^{i\vec k.{{\vec \delta }_l}}},~~~~ l=1,2,3.
\end{array}
\label{eq:singlet-cond}
\end{eqnarray}
 where $<ij>$  subscript indicate nearest neighbor pairing amplitude in real space as illustrated  in Fig.\ref{figure:pairing amplitude}.  Introducing the following unitary transformation matrix, 
\begin{eqnarray}
{\hat H}_{su} = \sum_{\vec k} {\hat \Psi }^{\dag }(\vec k) Q\left[ {{Q^\dag }{H_{su}}(\vec k)Q} \right]{Q^\dag }\hat \Psi (\vec k)
 = \sum_{\vec k} \Lambda ^{\dag }(\vec k)H_{s}(\vec k)\Lambda (\vec k) ,~~~Q = \left( \begin{array}{*{20}{c}}
{Q}_{T}(\vec k)&0\\
0&{\hat Q_T^ * ( -\vec k)}
\end{array} \right),
\label{eq:trans-Hsu}
\end{eqnarray}
one can transform Eq.\ref{eq:trans-Hsu}.  Eq.\ref{eq:Hsu-matrix} can be transformed to  
the block diagonalize form
\begin{eqnarray}
{\hat H}_{su}=\sum_{\vec k} \Lambda ^{\dag }(\vec k)\left( {\begin{array}{*{20}{c}}
{{H_{su}^{+}}(\vec k)}&0\\
0&{  H_{su}^{-}( \vec k)}
\end{array}} \right)\Lambda (\vec k)
\label{eq:trans-Hsu-matrix}
\end{eqnarray}
in which
\begin{eqnarray}
H_{su}^{+}(\vec k)=\left(
\begin{array}{*{20}{c}}
H_{c}^{+}(\vec k)&H_{p}^{+}(\vec k)\\
H_{p}^{+}(\vec k)&-H_{c}^{+}(\vec k)
\end{array}\right)_{14\times14},~~~H_{su}^{-}(\vec k)=\left(
\begin{array}{*{20}{c}}
H^{-}(\vec k)&H_{p}^{-}(\vec k)\\
H_{p}^{-}(\vec k)&-H^{-}(\vec k)
\end{array}\right)_{12\times12}
\end{eqnarray}
and, $
 \Lambda^{\dagger} (\vec k)=\left([\hat c^{\dagger}_{0\uparrow}(\vec k)\hat c^{+\dagger}_{1\uparrow}(\vec k)...\hat c^{+\dagger}_{6\uparrow}(\vec k)\;\hat c_{0\downarrow}(-\vec k)\hat c_{1\downarrow}^{+}(-\vec k)...\hat c_{6\downarrow}^{+}(-\vec k)]~~~[\hat c^{-\dagger}_{1\uparrow}(\vec k)\;\hat c^{-\dagger}_{2\uparrow}(\vec k)...\hat c^{-\dagger}_{6\uparrow}(\vec k)\;\hat c_{1\downarrow}^{-}(-\vec k)\;\hat c_{2\downarrow}^{-}(-\vec k)...\hat c_{6\downarrow}^{-}(-\vec k)]\right),
$ where $\hat c^{(\pm) \dagger}_{m\sigma}(\vec k)=\frac{1}{\sqrt 2}(\hat c^{\dagger}_{m\sigma}(\vec k)\pm\hat c^{\dagger}_{m+6,\sigma}(\vec k)).$
New pairing matrices $H_{p}^{+}(\vec k)$ and $H_{p}^{-}(\vec k)$ are defined as,
\begin{eqnarray}
H_{p}^{+}(\vec k)=\left(
\begin{array}{*{20}{c}}
0&0\\
0&H_{p}^{11}(\vec k)+H_{p}^{12}(\vec k)
\end{array}\right),~~~H_{p}^{-}(\vec k)=\left(H_{p}^{11}(\vec k)-H_{p}^{12}(\vec k)\right).
\end{eqnarray}
 From Eq. \ref{eq:trans-Hsu-matrix} it can be seen that the superconducting Hamiltonian $H_{su}$ 
can be diagonalized into two decoupled new superconducting Hamiltonian $H^{+}_{su}$ and $H^{-}_{su}$. 
Thus electrons just can be paired within the seven bands sector $H_{c}^{+}$ or within the six bands sector  
$H^{-}$, without coupling between the sectors. Thus superconductivity in bilayer graphene cane be interpreted 
as two decoupled monolayer graphene-like systems with independent behaviors.

 \section{Appendix C: Two Superconducting Gap Equations}
  The linearized superconducting  gap equation are obtained by minimizing the quasiparticle free energy 
with respect to the nearest neighbor order parameter, or equivalently with respect to  $\Delta_{\pm}^{\alpha}$. 
The free energy of system is
\begin{eqnarray}
F = F^{+}+F^{-}+F_{0}= - \frac{2 }{\beta }\sum_{\vec k} \sum_{n = 1}^{13} 
        {\ln \left[ {2\cosh (\frac{ E_n^Q}{ 2 k_{B}T})} \right] } + {F_0},~~~  
   F_{0}=  -2N\sum_{\alpha  = 1}^{18} J_{\alpha}(\Delta^{\alpha})^{2}.    
\label{eq:quasi-su-free}
\end{eqnarray}
For $F^{+}$ the summation runs over $n=1,...,7$ giving $E_{n}^{Q}=E_{n,s}^{Q+}$; for $F^{-}$ the summation 
takes $n=8,...,13$ values giving $E_{n}^{Q}=E_{n,s}^{Q-}$, with $E_{n,s}^{Q\pm}$ introduced in 
Eqs.~\ref{eq:quasi-su1-spectrum} and \ref{eq:quasi-su2-spectrum}. 

   Minimization of the free energy    with respect to $\Delta_{+}^{\alpha}$ gives
\begin{equation}
\Delta^{\beta }_{<ij>} +\Delta^{'\beta }_{<ij>}=- \frac{1}{N}\sum_{\alpha =1}^9 \left[ \sum_{\vec k} 
   \sum_{n = 1}^7 \sum_{i = 1}^7 \frac{\tanh (\frac{{ E_n^{Q+}}}{ 2 k_{B}T})}{E_{n}^{+}(\vec k) 
  + E_{i}^{+}(\vec k) } \left( \Omega _{ni}^{+\alpha} (\vec k)\Omega _{ni}^{+ * \beta }(\vec k) 
  + \Omega _{ni}^{+\beta} (\vec k)\Omega _{ni}^{+ * \alpha }(\vec k) \right) \right] \Delta^{\alpha }_{+}
 \equiv - \sum_{\alpha  = 1}^9 \Gamma _{\beta \alpha }^{+} \Delta^{\alpha }_{+}.
\label{eq:gap-Equation-Hc}
\end{equation}
giving independent gap equations for the seven bands odd-symmetry  graphene-like Hamiltonian $H_{c}^{+}$. 
Minimizing the free energy with respect to  $\Delta_{-}^{\alpha}$ gives
\begin{equation}
\Delta^{\beta }_{<ij>} -\Delta^{'\beta }_{<ij>}=- \frac{1}{N}\sum_{\alpha =1}^9 \left[ \sum_{\vec k} 
   \sum_{n = 1}^6 \sum_{i = 1}^6 \frac{\tanh (\frac{{ E_n^{Q-}}}{ 2 k_{B}T})}{E_{n}^{-}(\vec k) 
  + E_{i}^{-}(\vec k) } \left( \Omega _{ni}^{-\alpha} (\vec k)\Omega _{ni}^{- * \beta }(\vec k) 
  + \Omega _{ni}^{-\beta} (\vec k)\Omega _{ni}^{- * \alpha }(\vec k) \right) \right] \Delta^{\alpha }_{-}
 \equiv - \sum_{\alpha  = 1}^9 \Gamma _{\beta \alpha }^{-} \Delta^{\alpha }_{-}.
\label{eq:gap-Equation-Ha}
\end{equation}
where $\Delta^\alpha$ as illustrated in Fig.\ref{figure:pairing amplitude}(b) covers all possible  
nearest neighbor inter- and intra-layer C-C pairing amplitudes. 
Eqs.~\ref{eq:gap-Equation-Hc} and Eq.~\ref{eq:gap-Equation-Ha}  in matrix form written as   
\begin{eqnarray}
\left[ \begin{array}{*{20}{c}}
A^{\pm}&B^{\pm}&B^{\pm}\\
B^{\pm}&C^{\pm}&D^{\pm}\\
B^{\pm}&D^{\pm}&C^{\pm}
\end{array} \right]\left( \begin{array}{*{20}{c}}
g_{1}\Sigma_{i}\pm g_{1}^{'}\Sigma_{i}^{'}\\
g_{0} \Pi_{i}\pm g_{0}^{'}\Pi_{i}^{'}\\
g_{0} \Delta_{i}\pm g_{0}^{'}\Delta_{i}^{'}
\end{array} \right) =-\left( \begin{array}{*{20}{c}}
\Sigma_{i}\pm \Sigma_{i}^{'}\\
 \Pi_{i}\pm\Pi_{i}^{'}\\
\Delta_{i}\pm \Delta_{i}^{'}
\end{array} \right).
\label{eq:matrix-form-gap-eq-six-seven-bands}
\end{eqnarray}
Equivalently, Eqs.~\ref{eq:gap-Equation-Hc} and \ref{eq:gap-Equation-Ha} can be combined in the following  non-Hermitian eigenvalue problem, 
\begin{eqnarray}
\left[ \begin{array}{*{20}{c}}
G^{+}&\kappa_0 G^{-}\\
G^{-}&\kappa_0 G^{+}\\
\end{array} \right]\left( \begin{array}{*{20}{c}}
\Psi_\Delta\\
\Psi_{\Delta}^{'}
\end{array} \right) =-\frac{1}{g_0}\left( \begin{array}{*{20}{c}}
\Psi_\Delta\\
 \Psi_{\Delta}^{'}
\end{array} \right)
\label{eq:gapequation-matrix-unhermit-form}
\end{eqnarray}
in which  $\kappa_0=\frac{g_0}{g_0^{'}}$ and
\begin{eqnarray}
G^{\pm}=\frac{1}{2}\left[ \begin{array}{*{20}{c}}
{\kappa (A^{+}\pm A^{-})}&{\kappa  (B^{+}\pm B^{-})}&{\kappa  (B^{+}\pm B^{-})}\\
(B^{+}\pm B^{-})&(C^{+}\pm C^{-})&(D^{+}\pm D^{-})\\
(B^{+}\pm B^{-})&(D^{+}\pm D^{-})&(C^{+}\pm C^{-})
\end{array} \right],~~~
 \Psi_{\Delta}=
\left( \begin{array}{*{20}{c}}
{{g_1}{V_1}}\\
{{g_0}{V_2}}\\
{{g_0}{V_3}}
\end{array} \right),~~~ \Psi_{\Delta}^{'}=
\left( \begin{array}{*{20}{c}}
{{g_1^{'}}{V_1^{'}}}\\
{{g_0}^{'}{V_2^{'}}}\\
{{g_0}^{'}{V_3^{'}}}
\end{array} \right).
\label{eq:gapequation-matrix-elements-def}
\end{eqnarray}
$\kappa=\frac{g_1}{g_0}=\frac{g_{1}^{'}}{g_{0}^{'}}$, 
$V_1^T=(\Sigma_1~\Sigma_2~\Sigma_3)$, $V_2^T=(\Pi_1~\Pi_2~\Pi_3)$, and 
$V_3^T=(\Delta_1~\Delta_2~\Delta_3)$. Also  
$V_1^{'T}=(\Sigma_1^{'}~\Sigma_2^{'}~\Sigma_3^{'})$, $V_2^{'T}=
(\Pi_1^{'}~\Pi_2^{'}~\Pi_3^{'})$, and $V_3^{'T}=(\Delta_1^{'}~\Delta_2^{'}~\Delta_3^{'})$. 
Equation~\ref{eq:gapequation-matrix-unhermit-form} is in fact the matrix representation of gap equation 
resulting from minimization of the free energy with respect to nearest neighbor order parameters 
instead of $\Delta_{\pm}^{\alpha}$, which can be solved to obtain the differing superconductivity 
phases and pairing interaction potentials.

In the limiting case $\kappa_0\rightarrow 1$, the inter- and intra-layer pairing amplitudes in real space 
are equal, {\it i.e.} $g_0=g_0^{'}$ and $g_1=g_1^{'}$.
 This restriction makes the matrix gap equation hermitian and implies that band order 
parameters $\Delta_{mn}^{\pm}(\vec{k})$ can be interpreted physically as pairing of electrons in different 
bands with pairing interaction $g_0^{\pm}$. Just in this limit $\Delta_{mn}^{\pm}(\vec{k})$ is equal 
to the product of band Green function and $g_{0}$, 
\begin{eqnarray}
\Delta_{mn}^{\pm}(\vec k)=g_{0}^{\pm}\langle \hat{d}^{\pm\uparrow}_{m}(\vec k)\hat{d}^{\pm\downarrow}_{n}(\vec k)\rangle.
\label{eq:order-green-app}
\end{eqnarray}
Here $\hat{d}^{\pm\sigma}_{i}(\vec k)=\sum^{7}_{m=1}{\mathscr{C}}^{\pm*}_{m} (E_{i}(\vec k)) 
  \hat{c}^{\sigma}_{m}(\vec k)$ annihilates an electron with  spin $\sigma$ in the $i$th six (odd) or 
seven (even) sector  bands with 
energy $E_{i}^{\pm}(\vec k)$. In this limit, Eq.~\ref{eq:gapequation-matrix-unhermit-form} has two 
solutions $\Psi_\Delta=\Psi_\Delta^{'}$ or $\Psi_\Delta=-\Psi_\Delta^{'}$. The corresponding gap equations 
 Eqs.~\ref{eq:matrix-form-gap-eq-six-seven-bands} and \ref{eq:gapequation-matrix-unhermit-form} become 
decoupled gap equations corresponding to the even or odd sector of the graphene-like systems.
\begin{eqnarray}
F^{+}\Psi_{\Delta}=g_{0}^{+}\Psi_{\Delta},~~~F^{-}\Psi_{\Delta}=g_{0}^{-}\Psi_{\Delta},~~~F^{\pm}=\left[ \begin{array}{*{20}{c}}
A^{\pm}&B^{\pm}&B^{\pm}\\
B^{\pm}&C^{\pm}&D^{\pm}\\
B^{\pm}&D^{\pm}&C^{\pm}
\end{array} \right]
\label{eq:decoupled-gap-equations}
\end{eqnarray} 
 \section{Appendix D: flat band(s) Superconductivity }
Mirror symmetry transformation rearranges the  noninteracting Hamiltonian 
Eq.~\ref{eq:H-normal} as the direct sum of 
two single layer pseudo-graphene structures $\hat{H}_{N}=\hat{H}^{+}_{N}\oplus\hat{H}^{-}_{N}$ (even sector (+ sign) and odd sector (- sign)) 
\begin{eqnarray}
\hat{H}^{+}_{N}=
\sum_{ij\sigma}\sum^{6}_{\alpha,\beta=0}{t^{+}_{i\alpha\sigma,j\beta\sigma}}\hat{c}^{+\dag}_{i\alpha\sigma}\hat{c}^{+}_{j\beta\sigma};~~~~
\hat{H}^{-}_{N}=
\sum_{ij\sigma}\sum^{6}_{\alpha,\beta=1}{t^{-}_{i\alpha\sigma,j\beta\sigma}}\hat{c}^{-\dag}_{i\alpha\sigma}\hat{c}^{-}_{j\beta\sigma}
\end{eqnarray} 
with renormalized hopping integrals of the form
$t^{\pm}_{i\alpha\sigma,j\beta\sigma}=t^{inter}_{i\alpha\sigma, j\beta\sigma}\pm t^{intra}_{i\alpha\sigma, j\beta+6\sigma}$. 
In the limit case of strong interlayer hopping wherein
 $t^{inter}_{i\alpha\sigma, j\beta\sigma}\to\pm t^{intra}_{i\alpha\sigma, j\beta+6\sigma}$ one can see that odd (or even) sector bandwidths completely become flat while the other sector bandwidth doubles. In this limit, thermal weight factor of Eq.~\ref{eq:gap-Equation-Hc-text} 
 $\frac{\tanh (\frac{{\beta E_i^{\pm}}}{ 2 })}{E_{j}^{\pm}(\vec k) 
  + E_{i}^{\pm}(\vec k) }\to \frac{\beta}{4}$ and so
$\Gamma$ matrix elements are given by
   \begin{equation}
\Gamma _{\beta \alpha }^{\pm} = \frac{\beta}{4N}\sum_{\vec k} 
   \sum_{ij } \left( \Omega _{ij}^{\pm\alpha} (\vec k)\Omega _{ji}^{\pm \beta }(\vec k) 
  + \Omega _{ji}^{\pm\alpha} (\vec k)\Omega _{ij}^{\pm \beta }(\vec k) \right).  
\label{eq:gap-Equation-Hc}
\end{equation}
 These elements are linked to the normal state Bloch coefficients via the
 $\Omega _{ij}^{\pm \beta }(\vec k)$ factors that given by Eq.~\ref{eq:band-orderparameter-coef}.
 For the case of pristine bilayer graphene, Eq.~\ref{eq:gap-Equation-Hc}
can be determined analytically. In this limit, normal state Bloch coefficients are given by
\begin{eqnarray}
[\mathscr{C}_1^{\pm}(E_{ml})...\mathscr{C}_6^{\pm}(E_{ml})]=\frac{1}{\sqrt{6}} [(u_m ~~ u_m^{*} ~~1)~~
(-1)^le^{i\phi_{m}^{\pm}(\vec{k})}(u_m^{*} ~~ u_m ~~1)];~~~~
u_m=e^{i2m\pi/3},~~~m=1,2,3~~ \&~l=1,2
\label{eq:pristine-eigenvectors-app}
\end{eqnarray}
wherein $e^{i\phi_{m}^{\pm}(\vec{k})}=\frac{\eta_m^{*\pm}}{|\eta_m^{\pm}|}$ and 
$\eta_{m}^{\pm}(\vec{k})=d_{2}^{\pm}(\vec{k})+u_m d_{1}^{\pm}(\vec{k})+u_m^* d_{3}^{\pm}(\vec{k})$.  $\Omega _{ij}^{\pm \beta }(\vec k)$ factors can be calculated by substituting 
Bloch coefficients Eq.~\ref{eq:pristine-eigenvectors-app} in the Eq.~\ref{eq:band-orderparameter-coef}. For instant one can show
\begin{eqnarray}
\Omega _{11}^{\pm 1}(\vec k)=\Omega _{11}^{\pm 4}(\vec k)
=\Omega _{11}^{\pm 7}(\vec k)=-\frac{1}{3}cos(\vec{k}.\vec{\delta_1}-\phi_1^{\pm}(\vec{k}))\nonumber\\
\Omega _{11}^{\pm 2}(\vec k)=\Omega _{11}^{\pm 5}(\vec k)
=\Omega _{11}^{\pm 8}(\vec k)=-\frac{1}{3}cos(\vec{k}.\vec{\delta_2}-\phi_1^{\pm}(\vec{k}))\nonumber\\
\Omega _{11}^{\pm 3}(\vec k)=\Omega _{11}^{\pm 6}(\vec k)
=\Omega _{11}^{\pm 9}(\vec k)=-\frac{1}{3}cos(\vec{k}.\vec{\delta_3}-\phi_1^{\pm}(\vec{k}))\nonumber\\
\Omega _{16}^{\pm 1}(\vec k)=\Omega _{16}^{\pm 4}(\vec k)
=\Omega _{16}^{\pm 7}(\vec k)=-\frac{i}{3}sin(\vec{k}.\vec{\delta_1}-\phi_1^{\pm}(\vec{k}))\nonumber\\
\Omega _{16}^{\pm 2}(\vec k)=\Omega _{16}^{\pm 5}(\vec k)
=\Omega _{16}^{\pm 8}(\vec k)=-\frac{i}{3}sin(\vec{k}.\vec{\delta_2}-\phi_1^{\pm}(\vec{k}))\nonumber\\
\Omega _{16}^{\pm 3}(\vec k)=\Omega _{16}^{\pm 6}(\vec k)
=\Omega _{16}^{\pm 9}(\vec k)=-\frac{i}{3}sin(\vec{k}.\vec{\delta_3}-\phi_1^{\pm}(\vec{k}))
\end{eqnarray}
By calculating a large number of these factors and replacing them in the  Eq.~\ref{eq:gap-Equation-Hc} one can obtain
$$\Gamma^{\pm}_{ij}=\beta_c\delta_{ij};~~~~~g_0=k_BT_c$$


\section{References}
\begin{thebibliography}{0}

\bibitem{brun2017}C. Brun, T. Cren, and D. Roditchev,
  Review of 2D superconductivity: the ultimate case of epitaxial monolayers,
  Supercond. Sci. \& Technol. {\bf 30}, 013003 (2017). 

\bibitem{Saito2016}
Y. Saito, T. Nojima \& Y. Iwasa “Highly crystalline 2D superconductors”
Nature Reviews Materials
2, 16094 (2016).

\bibitem{Nandkishore2012}
 Nandkishore, R., Levitov, L. S. \& Chubukov, A. V. Chiral superconductivity from repulsive interactions in doped graphene. Nat.
Phys. 8, 158–163 (2012).

\bibitem{Nandkishore2014}
Nandkishore, R., Thomale, R. \& Chubukov, A. V. Superconductivity from weak repulsions in hexagonal lattice systems. Phys. Rev. B
89, 144501 (2014).

\bibitem{Black-Schaffer2007}
Black-Schaffer, A. M. \& Doniach, S. Resonating valence bonds and mean field d-wave superconductivity in graphene. Phys. Rev. B
75, 134512 (2007).

\bibitem{Uchoa2007}Uchoa, B. \& Castro Neto, A. Superconducting States of Pure and Doped Graphene. Phy. Rev. Lett.98, 146801 (2007) 

\bibitem{Kiesel2012}
 Kiesel, M. L., Platt, C., Hanke, W., Abanin, D. A. \& Thomale, R. Competing many-body instabilities and unconventional
superconductivity in graphene. Phys. Rev. B 86, 020507R (2012).

\bibitem{Ma2014}
 Ma, T., Yang, F., Yao, H. \& Lin, H. Q. Possible triplet p + ip superconductivity in graphene at low filling. Phys. Rev. B 90, 245114
(2014).

\bibitem{Profeta2012} G. Profeta, M. Calandra, and F. Mauri, Phonon-mediated super-
conductivity in graphene by lithium deposition, Nat. Phys. 8,
131 (2012).

\bibitem{Ludbrook2015}B.M.  Ludbrook {\it et al.}, 
  Evidence for superconductivity in Li-decorated monolayer graphene, 
 Proc. Natl. Acad. Sci. (USA) {\bf 112}, 11795-11799 (2015).

\bibitem{Weller-naturephysics2005} T. E. Weller {\it et al.},
  Superconductivity in the intercalated graphite compounds C$_6$Yb and C$_6$Ca,
 Nat. Phys. {\bf 1}, 39 (2005).
 

 \bibitem{Tiwari2017}A. P. Tiwari, S. Shin, D. Hwang, S. G. Jung, T. Park, and H. Lee,
 Superconductivity at 7.4K in few layer graphene by Li intercalation,
 J. Phys.: Condens. Matt. {\bf 29}, 445701 (2017).
 
 \bibitem{Xue2012}
 Xue, M., Chen, G., Yang, H. \& Zhu, Y. Superconductivity in potassium-doped few-layer graphene. J. Am. Chem. Soc.134,6536–6539 (2012).
 
 \bibitem{Nori2016} AV Rozhkov, AO Sboychakov, AL Rakhmanov, F Nori,  Electronic properties of graphene-based bilayer systems,
Physics Reports 648, 1-104 (2016)

 \bibitem{Liu2009}
 Z. Liu, K. Suenaga, P. J. F. Harris, and S. Iijima, Open and Closed Edges of Graphene Layers, Phys. Rev. Lett.
102, 015501 (2009).

\bibitem{Borysiuk2011}
J. Borysiuk, J. Soltys, and J. Piechota, Stacking sequence dependence of graphene layers on SiC (0001)—Experimental and theoretical investigation, J. Appl. Phys. 109, 093523
(2011)

\bibitem{Lee2008}
J.-K. Lee, S.-Ch. Lee, J.-P. Ahn, S.-Ch. Kim, J. I. B. Wilson,
and P. John, The growth of AA
graphite on (111) diamond,  J. Chem. Phys. 129, 234709 (2008)

\bibitem{Cao-Mott2018}
Cao, Y. et al. Correlated insulator behaviour at half-filling in magic-angle graphene superlattices. Nature 556, https://doi.org/10.1038/nature26154 (2018)

\bibitem{Cao2018}Y. Cao, V Fatemi, S. Fang, K. Watanabe, T. Taniguchi, E. Kaxiras, and P. Jarillo-Herrero,
  Unconventional superconductivity in magic-angle graphene superlattices,
  Nature {\bf 556}, 43 (2018). doi:10.1038/nature26160.
  
\bibitem{MacDonald2019}
  MacDonald A. H., Trend: Bilayer Graphene’s Wicked, Twisted Road,Physics{\bf{12}} 12 (2019)   
  
 \bibitem{Chen2019}
 G. Chen et al., “Signatures of gate-tunable superconductivity in trilayer graphene/boron nitride moiré superlattices,” Nature 572, 215-219 (2019)
 
  
  \bibitem{Kanetani2012}Kanetani, K.; Sugawara, K.; Sato, T.; Shimizu, R.; Iwaya, K.;
Hitosugi, T.; Takahashi, T. Ca Intercalated Bilayer Graphene as A Thinnest Limit of Superconducting C6Ca.
Proc. Natl. Acad. Sci. U. S. A. 2012,109, 19610

\bibitem{Caffrey2016}Caffrey, N.M.; Johansson, L.I.; Xia, C.; Armiento, R.; Abrikosov, I.A.; Jacobi, C. Structural and electronic properties of Li-intercalated graphene on SiC(0001). Phys. Rev. B 2016, 93, 195421.

\bibitem{Satoru2016}  S. Ichinokura {\it et al.},  Superconducting calcium-intercalated bilayer
graphene, ACS Nano  {\bf 10}, 2761 (2016).


\bibitem{Chapman2016} J. Chapman {\it et al.}, 
 Superconductivity in Ca-doped graphene laminates, Sci. Rep. {\bf 6}, 23254 (2016). 
 
 \bibitem{Velasco2012}
J. Velasco Jr., L. Jing, W. Bao, Y. Lee, P. Kratz, V. Aji, M. Bockrath, C. Lau, C. Varma, R. Stillwell, et al., Transport spectroscopy of symmetry-broken
insulating states in bilayer graphene, Nat. Nanotechnol. 7 (2012) 156.
  
\bibitem{Vucicevic2012} J. Vucicevic, M. O. Goerbig, and M. V. Milovanovic, 
  $d$-wave superconductivity on the honeycomb bilayer, 
 Phys. Rev. B {\bf 86}, 214505 (2012).
 
\bibitem{James2014} J. M. Murray and O. Vafek, 
  Excitonic and superconducting orders from repulsive interaction on the doped 
    honeycomb bilayer, Phys. Rev. B {\bf 89}, 205119 (2014)
    
\bibitem{Hosseini2012}
Hosseini, M. V. \& Zareyan, M. Model of an exotic chiral superconducting phase in a graphene bilayer. Phys. Rev. Lett. 108, 147001 (2012).
    
\bibitem{Rakhmanov2012}
A.L. Rakhmanov, A.V. Rozhkov, A.O. Sboychakov, F. Nori, Instabilities of the AA-stacked graphene bilayer, Phys. Rev. Lett. 109 (2012) 206801.

\bibitem{Akzyanov2013}
R.S. Akzyanov, A.O. Sboychakov, A.V. Rozhkov, A.L. Rakhmanov, F. Nori, AA-stacked bilayer graphene in an applied electric field: Tunable
antiferromagnetism and coexisting exciton order parameter, Phys. Rev. B 90 (2014) 155415.

\bibitem{Sboychakov2013}
A.O. Sboychakov, A.V. Rozhkov, A.L. Rakhmanov, F. Nori, Antiferromagnetic states and phase separation in doped AA-stacked graphene bilayers, Phys.
Rev. B 88 (2013) 045409

\bibitem{SboychakovMetal2013}
A.O. Sboychakov, A.L. Rakhmanov, A.V. Rozhkov, F. Nori, Metal-insulator transition and phase separation in doped AA-stacked graphene bilayer,
Phys. Rev. B 87 (2013) 121401.

\bibitem{Alidoust2019}  
M. Alidoust, M. Willatzen \& A. Jauho,
Symmetry of superconducting correlations in displaced bilayers of graphene,
Phys. Rev. B 99, 155413 (2019)

\bibitem{Eliel2018}
G. S. N. Eliel \& et al.
Intralayer and interlayer electronphonon interactions in twisted
graphene heterostructures, Nat. Comm. 9, 1221 (2018).  

\bibitem{Fang2019}
 S. Fang {\it et al}	,Electric field-induced chiral $d+id$ superconducting state in AA-stacked bilayer graphene: A quantum Monte Carlo study,arXiv:1907.10236 
  
\bibitem{Endo2019}
Y. Endo {\it et al},Structure of Superconducting Ca-intercalated Bilayer Graphene/SiC studied using Total-Reflection High-Energy Positron Diffraction, 	arXiv:1906.11535 

\bibitem{Sanderson2013}
M. Sanderson, Y.S. Ang, C. Zhang, Klein tunneling and cone transport in AA-stacked bilayer graphene, Phys. Rev. B 88 (2013) 245404.

\bibitem{Putti2006}
M. Putti{\it et al},Observation of the crossover from two-gap to single-gap superconductivity through specific heat measurements in neutron-irradiated MgB2.
Phys. Rev. Lett. 96, 077003 (2006)

\bibitem{Margine2016} E. R. Margine, H. Lambert, and F. Giustino, 
  Electron-phonon interaction and pairing mechanism in
   superconducting Ca-intercalated bilayer graphene, 
  Sci. Rep. {\bf 6}, 21414 (2016).

 
\bibitem{Rouhollah2018} R. Gholami, R. Moradian, S. Moradian, W. E. Pickett, 
  Superconducting phases of lithium intercalated graphene, 
 Sci. Rep. {\bf 8}, 13795 (2018). doi: 10.1038/s41598-018-32050-9. 
 

\bibitem{Durajski2019} Durajski, A.; Skoczylas, K.; Szczesniak, R. Superconductivity in bilayer graphene intercalated with alkali
and alkaline earth metals. Phys. Chem. Chem. Phys. 2019, 21, 5925.
 





\bibitem{Tomoaki2017}  T. Kaneko and R. Saito, “First-Principles Study on Interlayer State in Alkali and 
Alkaline Earth Metal Atoms Intercalated Bilayer Graphene,”
Surface Science, vol. 665, 1–9, (2017).


\end{thebibliography}
\end{document}